\documentclass[prd,twocolumn,amsmath,amssymb,floatfix,superscriptaddress]{revtex4-1}

\usepackage{graphicx}
\usepackage{amssymb}
\usepackage{amsmath}
\usepackage{color}
\usepackage{ wasysym }
\usepackage{hyperref}
\usepackage{tabu}

\def\barray{\begin{array}}
	\def\earray{\end{array}}
\def\be{\begin{equation}}
\def\ee{\end{equation}}
\def\ben{\begin{equation} \nonumber}
\def\een{\end{equation}}
\def\ban{\begin{eqnarray*}}
	\def\ean{\end{eqnarray*}}
\def\ba{\begin{eqnarray}}
\def\ea{\end{eqnarray}}
\def\bs{\boldsymbol}

\def\({\left(}
\def\){\right)}

\def\aap{A\&A}
\def\apj{ApJ}

\def\apjl{ApJ}
\def\mnras{MNRAS}

\def\aj{AJ}

\def\prd{Phys. Rev. D}

\graphicspath{{./fig/}}

\begin{document}

\title{Primordial black hole detection through diffractive microlensing }

%\author[T. Nadery et~al.]{T. Nadery$^{1}$ ,
%	A. Mehrabi$^{2}$ \thanks{mehrabi@basu.ac.ir}, S. Rahvar$^{1,3}$,  \\
%	$^1$ Department of Physics, Sharif University of Technology, 11365–9161, Tehran, Iran\\
%	$^2$ Department of Physics, Bu-Ali Sina University, Hamedan 65178, 016016, Iran\\
  %       $^3$Perimeter Institute for Theoretical Physics, 31 Caroline Street North, Waterloo, Ontario N2L 2Y5, Canada
%}

\author{T. Naderi }
\affiliation{Department of Physics, Sharif University of Technology, 11365–9161, Tehran, Iran}

\author{A. Mehrabi }
\affiliation{Department of Physics, Bu-Ali Sina University, 65178, 016016, Hamedan, Iran}
\affiliation{School of Astronomy, Institute for Research in Fundamental Sciences (IPM), 19395-5531,  Tehran, Iran}

\author{S. Rahvar }
\affiliation{Department of Physics, Sharif University of Technology, 11365–9161, Tehran, Iran}
%\affiliation{Perimeter Institute for Theoretical Physics, 31 Caroline Street North, Waterloo, Ontario N2L 2Y5, Canada}

%\date{Accepted ?, Received ?; in original form \today}

%\pagerange{\pageref{firstpage}--\pageref{lastpage}} \pubyear{2015}

\begin{abstract}
Recent observations of gravitational waves motivate investigations for the existence of Primordial Black Holes (PBHs). We propose the 
observation of gravitational microlensing of distant quasars for  the range of 
infrared to the submillimeter wavelengths by sub-lunar PBHs as lenses. The advantage of observations in the longer wavelengths, comparable to the Schwarzschild radius of the lens (i.e. $R_{\rm sch}\simeq \lambda$) is the detection of the wave optics features of the gravitational microlensing.  The observation of diffraction pattern in the microlensing light curve of a quasar can break the degeneracy between the lens parameters and determine directly the lens mass as well as the distance of the lens from the observer.  We estimate the wave optics optical-depth, also calculate the rate of  $\sim 0.1$ to $\sim 0.3$ event per year per a quasar, 
 assuming that hundred percent of dark matter is made of sub-lunar PBHs. Also, we propose a long-term survey of quasars with the cadence of almost one hour to few days to resolve the wave optics features of the light curves to discover PBHs and determine the fraction of dark matter made of sub-lunar PBHs as well as their mass function.  
\end{abstract}

	\maketitle

\section{Introduction}\label{intro}
Observations of  type Ia supernova 
\citep{Perlmutter:1998np,Riess:2004nr,Astier:2005qq,Jha:2006fm}, cosmic microwave background (CMB) radiation 
\citep{Spergel:2006hy,Ade:2015rim} and baryon acoustic oscillation (BAO) 
	\citep{Seo:2005ys,Blake:2011en} indicate that around $25\%$ of matter content of the universe is made of dark matter (DM). There are many scenarios to explain the nature of DM and one of the models proposes 
DM might be composed totally or partially by the primordial black holes (PBHs) \citep{Blais:2002nd,Khlopov:2008qy,Frampton:2010sw}. 
	
There are several mechanisms to explain the formation of PBHs including sharp peaks in density fluctuations \citep{GarciaBellido:1996qt}, phase transitions \citep{Jedamzik:1999am}, resonant reheating \citep{Suyama:2004mz}, tachyonic preheating \citep{Suyama:2006sr} and curvaton scenarios \citep{Kohri:2012yw,Kawasaki:2012wr,Bugaev:2013vba}. PBHs smaller than about $10^{12}$kg should have already evaporated through the Hawking radiation \citep{2010RAA....10..495K,2010PhRvD..81j4019C}. However, the massive PBHs, are unaffected by the Hawking radiation might have various cosmological consequences, such as seeds for supermassive black holes \citep{2002PhRvD..66f3505B}, generation of the large-scale structures \citep{Afshordi:2003zb} and affects on the thermal and ionization history of the universe \citep{2008ApJ...680..829R}. 

The observations for searching the Massive Astrophysical Compact Halo Objects (MACHOs) by gravitational microlensing and femtolensing of $\gamma$-ray bursts excluded PBHs in the mass range of $\sim10^{-7}M\odot$ -$1 M\odot$ and $10^{14}$-$10^{17}$kg \citep{Tisserand:2006zx,Nemiroff:2001bp}.  However, assuming an extended mass function for the compact objects, it seems that various observational data  along with the dynamical constraints are consistent with PBHs as the dark matter candidate within the mass range of $10 M_\odot<M<10^3 M_\odot$ and/or $10^{17}~\text{kg} <M<10^{21}~\text{kg}$ \citep{carr,green}.  The later range for the PBHs is convenient to be written in terms of lunar
 mass, roughly as $10^{-6} <\bar{M}<10^{-2}$ (assuming the lunar mass of $M_{\rm m} \sim 7\times 10^{23}$ kg) where $\bar{M}={M}/{M_{\rm m}}$.
   
Gravitational lensing provides an exceptional tool for investigating the astrophysical phenomena including indirect detection of the compact objects \citep{rahvar:2015}. 
The light deflection in gravitational lensing depends on the configuration of the lens mass distribution and might produce several images. The term of gravitational microlensing is used when the images from the lensing cannot be resolved by the conventional telescopes. In this case, the result of lensing is the magnification of light receiving from the source star. Taking into account the relative motion of the lens, source and the observer results in a bell shape light curve, so-called Paczynski light curve \citep{pac86}. In recent years, microlensing has been used for discovering extra-solar planets \citep{Cottle:1991zza,Gould:1992aj,Gould:2008zu,batista,muraki}, investigating the properties of a distant source stars \citep{Afonso:2001gh,Gould:2001bg,Abe:2003xb,Fields:2003zx,sajadian} and studying the structure of the Milky Way galaxy \citep{moniez}.
Moreover, in the cosmological scales, the gravitational microlensing provides a useful method for studying compact objects. 

The quasar microlensing in the cosmological scales uses the caustic crossing of an ensemble of lenses \cite{2001ASPC..237..185W} for studying the distribution of stars and micro-halos around the galaxies \cite{2007A&A...475..453Z,rahvar2014}. The possibility of detection of PBHs have been studied through observation of Quasars in X-ray \citep{pbh2017} and they found no-candidate in the mass range of   $0.05 M_\odot<M<0.45 M_\odot$. Also, some microlensing observation of Quasars has been done in the 
survey mode where the aim was the detection of the caustic crossing of the lenses in the halo of the strong lensed galaxies \citep{meld,Giannini}. All these observations have been done in the optical and shorter wavelengths. Here, we propose extending observations to the infrared and millimeter wavelengths where effects of wave optics is important in the light curve of sub-lunar mass lenses. The advantage of wave optics is that we can obtain more information about the parameters of lenses compared to the geometric microlensing. From the observational point of view, Spitzer  space-based telescope and the Atacama Large Millimeter/Submillimeter
Array (ALMA)  are ideal tools for studying the light curve of quasars in our desired wavelengths  \citep{2015MNRAS.452...88K, Venemans:2015hyr,2016ApJ...824..132N} .

In section (\ref{geometric}) we introduce the gravitational microlensing in the geometric optics regime. In section (\ref{WO}), we introduce the wave optics feature of gravitational microlensing and calculate the diffraction pattern from the scattering of the electromagnetic wave from a PBH on the observer plane. In section (\ref{obs}), we study the observational features of the diffraction of light from a PBH as a lens as well as we calculate the optical depth for the observation of this event. Conclusions are given in (\ref{sec:conclude}). 

\section{Geometric microlensing}
\label{geometric}
The standard gravitational lensing formalism uses the geometric optics for the limit of 
$\lambda \ll R_{\rm sch}$, where $\lambda$ is the wavelength of the light and $R_{\rm sch}$ is the Schwarzschild radius of the lens. However, when $ \lambda \approx R_{\rm sch}$, the wave optics features of lensing such as interference of the light from different images produces the interference pattern on the observer plane. The relative 
motion of the observer with respect to the fringes results in a time variation in the intensity of light where the time-scale and the amplitude of these fringes provide more information compared to that of geometric optics. 
The diffractive gravitational lensing has been studied for a system with a galaxy as a lens and a point radio source \cite{ohanian:1983}. Following this work, the caustic-crossing of quasars in the wave optics regime has been investigated, where they put a limit on the size of quasars \cite{Jaroszynski:1995cd}. The wave optics aspect of microlensing with a sub-stellar mass lens, like a free-floating planet in the galaxy, have been studied in a series of papers \citep{Heyl:2009av,Heyl:2011b,Heyl:2010hm}. Also, \citet{Mehrabi:2012dy} investigated the wave optics features for a binary lensing system near the caustic lines. 

The gravity of a lens deflects light ray from a distant source and this deflection produces multiple images from a single source.
It is more convenient to write the lens equation in terms of angular scales \citep{Schneider1985}:
\begin{equation}\label{eq:lens}
\bs{\beta}=\bs{\theta}-\bs{\alpha}(\bs{\theta})\;,
\end{equation}
where $\bs{\beta}$ and $\bs{\theta}$ are the source and the image positions and $\bs{\alpha}(\bs{\theta})$ is the deflection angle. Notice that all angles are normalized to the Einstein-angle
\begin{equation}\label{eq:theta-e}
\theta_{E}=\sqrt{\frac{2R_{\rm sch}}{D_s}\frac{1-x}{x}},
\end{equation}
where $x={D_l}/{D_s}$ is the ratio of the comoving distance of the lens to the comoving distance of the source
%, $R_{\rm sch}$ is the Schwarzschild radius
 and $D(z)$ in $\Lambda$CDM model is given by 
      \begin{equation}\label{eq:dis-cos}
      D(z)=\frac{c}{H_0}\int_0^z \frac{dz}{\sqrt{\Omega_m(1+z)^3+\Omega_{\Lambda}}}\;.
      \end{equation}     
In the geometric optics limit, after 
 solving equation (\ref{eq:lens}), one can find the corresponding map between the source position to the image positions. The Jacobian of transformation in the equation of $\theta = \theta(\beta)$, provides the ratio of areas in the image space to the source space which is equivalent to the magnification in the gravitational microlensing.
  
 In what follows, we concern the wave optics regime of the gravitational microlensing by considering the interferometry of the light rays. It is convenient to use the Fermat potential for the light ray which is proportional to the time delay between a given trajectory and a straight path. The Fermat potential for a generic lens is given by 
    \begin{equation}\label{eq:ferma}
    \phi(\bs{\theta},\bs{\beta})=\frac{1}{2}(\bs{\theta}-\bs{\beta})^2-\psi(\bs{\theta})\;,
    \end{equation} 
    where $\psi(\bs{\theta})$ is the gravitational potential in 2D and is defined on the lens plane as    
    \begin{equation}\label{eq:psi}
    \psi(\bs{\theta})=\frac{1}{\pi}\int\frac{\Sigma(\bs{\theta'})}{\Sigma_{\rm cr}}\ln\vert\bs{\theta}-\bs{\theta}^{\prime}\vert d^2\bs{\theta}^{\prime},\;
    \end{equation}  
    and $\Sigma(\bs{\theta})$ is the surface mass density of the lens and the critical mass density is given by
 \begin{equation}
 \nonumber
    \Sigma_{\rm cr}=\frac{c^2}{4\pi GD_{\rm l}(1-x)}. 
    \end{equation}
    The lens equation gives rise from the Fermat principle,  
   $ \nabla_{\bs{\theta}}\phi(\bs{\theta},\bs{\beta})=0,$
  and in terms of the Fermat potential, the deflection angle in equation (\ref{eq:lens}) is given by
     $\bs{\alpha(\bs{\theta})}=\bs{\nabla}\psi(\bs{\theta})$.
        
 \section{Diffractive microlensing}   
 \label{WO}
In wave optics limit where the time delay between trajectories from a source to 
observer is less than a period of light, the light rays can be considered temporally coherent and 
the result is the production of the interference pattern on the observer plane. Under condition where the source and deflector are far from the observer, we can use the Huygens-Fresnel   
principle for analyzing gravitational lensing. Then, every point on the
lens plane can be taken as a secondary source and the amplitude of the electromagnetic wave at each point on the observer plane is
the superposition of the light from various sources on the lens plane. This analysis can be done for a point source, however, for an extended realistic source the amplification is calculated by the superposition of the infinitesimal 
incoherent sources. Finally, multiplying the superposition
of the electromagnetic wave by its complex conjugate results in
the magnification on the observer plane. The magnification for a point 
source \citep{Falco:1992} is given by 
\begin{equation}\label{eq:mag-wave}
\mu(\bs{\beta};k)=\frac{f^2}{4\pi^2}\vert\int e^{if\phi(\bs{\theta},\bs{\beta})}d^{2}\theta\vert^2,\;
\end{equation} 
where $f=2kR_{\rm sch}$ and $k$ is the wave-number.
$\phi(\bs{\theta},\bs{\beta})$ is the 
Fermat potential for a single lens and is given by:
\begin{equation}\label{eq:ferma:point}
\phi(\bs{\theta},\bs{\beta})=\frac{1}{2}(\bs{\theta}-\bs{\beta})^2-\ln\vert\bs{\theta}\vert.\;
\end{equation}

\begin{figure}
	\centering
	\includegraphics[width=.45\textwidth]{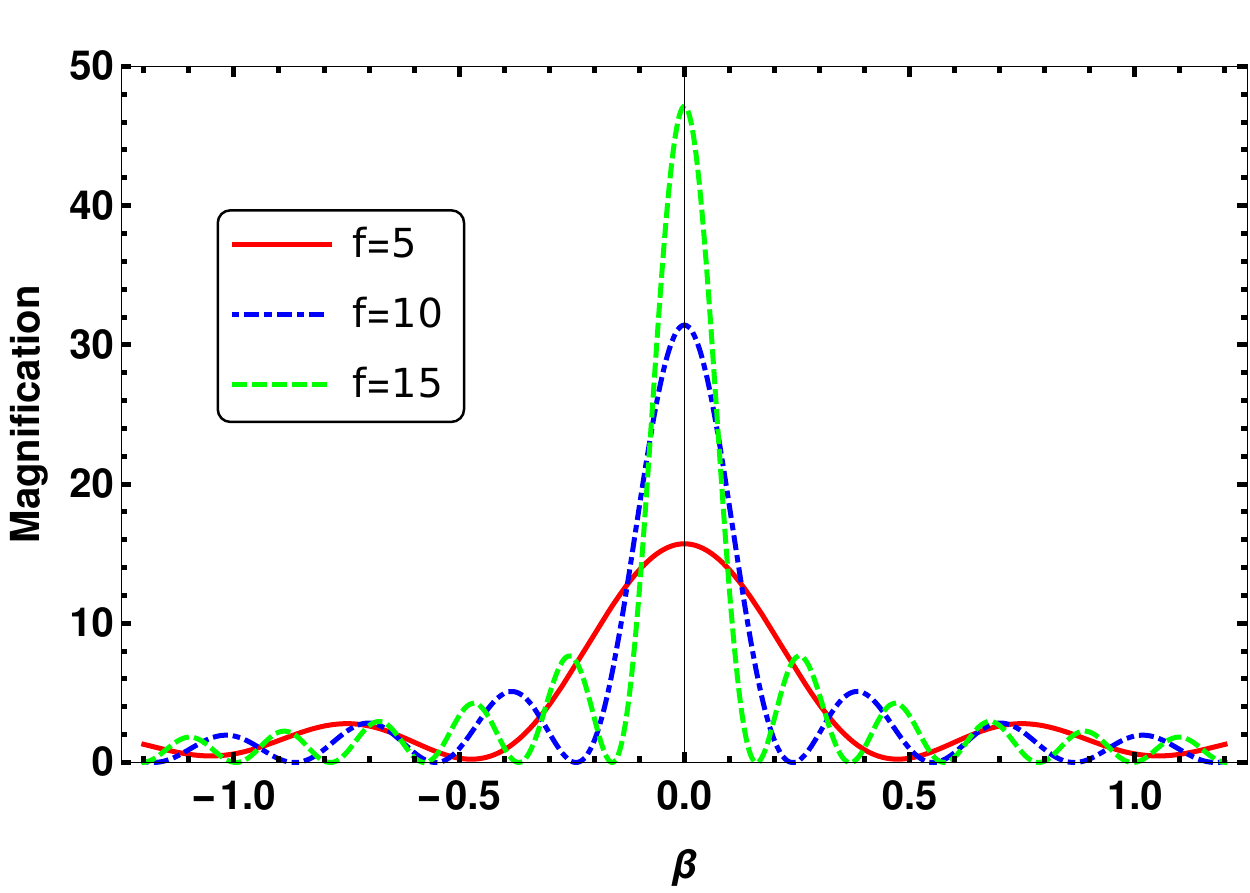}
	\caption{Magnification of a point source for varies values of $f$. The solid red, dot-dashed blue and dashed green lines represent the magnification for $f=5, 10$ and $f=15$, respectively. }
	\label{fig:mag-point}
\end{figure}

Substituting equation~(\ref{eq:ferma:point}) in equation~(\ref{eq:mag-wave}), the  magnification is given 
as follows
\begin{equation}\label{eq:mag-poi-main}
\mu(\beta;f)=\frac{\pi \frac{f}{2}}{\sinh(\pi \frac{f}{2})}e^{\pi \frac{f}{2}}\vert 1F1(1-i\frac{f}{2},1,i\frac{f\beta^2}{2})\vert^2, \;
\end{equation}  
where $1F1(a,b,x)$ is the confluent hypergeometric function \citep{Falco:1992}. Fig.~(\ref{fig:mag-point}) presents the magnification in terms of $\beta$ for different values of $f$.
The diffraction pattern is observable when $\lambda\approx R_{\rm sch}$ (i.e. $f \simeq \mathcal{O}(1)$). 
By increasing $f$ (smaller wavelength or massive lens), the fringes shrink and the diffraction pattern converge to the geometric optics magnification. 

%the peak of magnification increases by the factor of $\pi f$ 

To simplify equation (\ref{eq:mag-poi-main}), we expand the  Fermat potential in equation~(\ref{eq:ferma:point})
around the critical point of ($\bs{\theta}=1$, $\bs{\beta}=0$) where according to Fig. (\ref{fig:mag-point}), the light curve has peak around it, as follows:
\begin{equation}\label{eq:fer-expand}
\phi(\bs{\theta},\bs{\beta})=\theta^2 -2\theta -\theta\beta\cos\gamma,\;
\end{equation}
where  polar coordinate $(\theta,\gamma)$ is used on the lens plane.  
Then, from equation~(\ref{eq:mag-wave}) the magnification simplifies to
\begin{equation}\label{eq:mag-appr}
\mu(\beta;k)=\pi f J_0^2(f\beta),
\end{equation}
where $J_0$ is the Bessel function of the first kind.  We note that the relative difference of magnification  from Eq. (\ref{eq:mag-appr}) compare to the exact equation is less than $1\%$ for sources with $\bs{\beta}<0.5$ and this difference decreases rapidly when the source moves toward the lens position (i.e. $\beta \rightarrow 0$). 
In practice, it is possible to observe the light curve in two different wavelengths say $\lambda_1$ and $\lambda_2$. In this case, the relative magnification is given by
  \begin{equation}\label{eq:mag-rel}
  \frac{\mu_1}{\mu_2}=\frac{\lambda_2}{\lambda_1}\frac{J_0^2(f_1\beta)}{J_0^2(f_2\beta)},
  \end{equation}
  where close to the maximum magnification, using the series of $J_0(x)\approx 1 - \frac{x^2}{4}+\frac{x^4}{64}$, equation (\ref{eq:mag-rel}) simplifies as 
     \begin{equation}\label{eq:mag-rel2}
     \frac{\mu_1}{\mu_2}=\frac{\lambda_2}{\lambda_1}\frac{1-\frac{1}{2}(f_1\beta)^2+\frac{3}{32}(f_1\beta)^4}{1-\frac{1}{2}(\frac{\lambda_2}{\lambda_1})^2(f_1\beta)^2+\frac{3}{32}(\frac{\lambda_2}{\lambda_1})^4 (f_1\beta)^4}.
     \end{equation}
 In this case the right-hand side of this equation is a function of $f_1\beta(t)$ where from the measurement of $\mu_1$ and $\mu_2$  as a function of time, we can extract $f_1\beta(t)$. This parameter depends on 
 the lensing parameters as follows:  
   \begin{equation}\label{eq:fy}
     f_1\beta(t)= f_1 \left(u_0^2 + (\frac{t}{t_E})^2\right)^{\frac{1}{2}},
     \end{equation}  
 where $u_0$ is the minimum impact parameter 
 and $t_E$ is the Einstein-crossing time. From the observation of a microlensing event in the regime of geometric optics (i.e. $\lambda\ll R_{\rm sch}$), we can extract $u_0$ and $t_E$. On the other hand, knowing the left-hand side of equation (\ref{eq:fy}) from the wave optics and right-hand side from the geometric optics at $t=0$, we  determine directly $f_1$ or mass of the lens. We note that unlike to the geometric optics, the mass of lens determine from this method is independent of the distance of lens and source as well as their relative velocities with respect to the observer.

In reality, quasars as the source in the lensing have finite sizes and this effect should be taken into account in the wave-optics calculation. For a given source, the total magnification is calculated by integrating over all individual elements on a source where these elements are independent and incoherent. Then the magnification of an extended source \citep{Mehrabi:2012dy} is given as 
\begin{equation}\label{eq:mag-finite}
\mu(\bs{\beta},\rho;k)=\int_{s<\rho}\frac{I_{\rm w}(\bs{\beta})\mu(\bs{\beta};k)d^2 s}{I_{\rm w}(\bs{\beta})d^2 s},\;
\end{equation}  
where $I_{\rm w}(\bs{\beta})$ is the surface brightness of the source (that might depends on the wavelength) and  $\rho={\theta_s}/{\theta_E}$ is the angular size of source normalized to the Einstein angle.  Here the integration in equation (\ref{eq:mag-finite}) is taken over the source area (i.e. $s<\rho$). 

%------

\begin{figure}
 	\centering
 		\includegraphics[width=.45\textwidth]{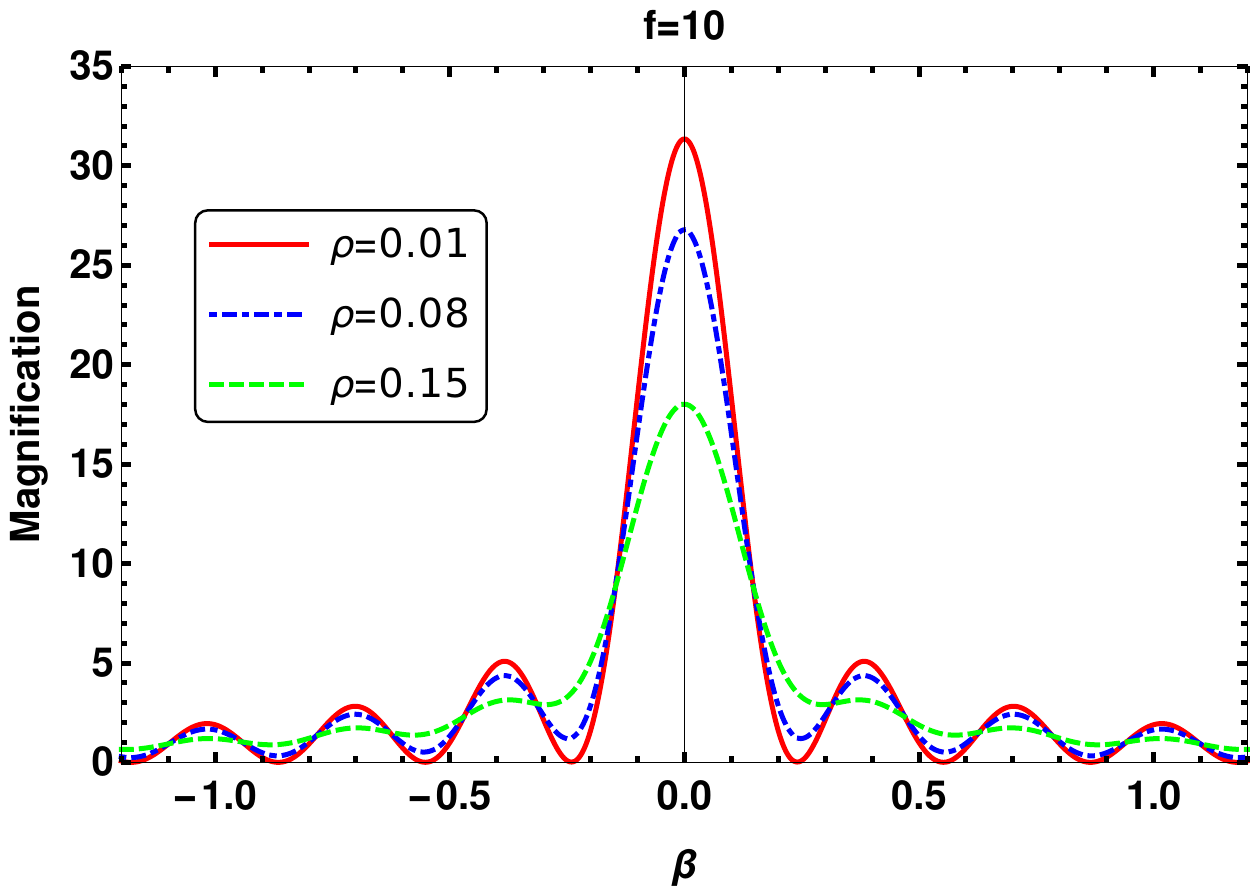}
\caption{Magnification of a uniform luminous source as a function of $\beta$ for $f=10$. The solid red, dot-dashed blue and dashed green lines show the magnification for $\rho=0.01$, $\rho = 0.08$ and $\rho=0.15$, respectively. By increasing the size of source, the incoherent light of the source results in magnification pattern converge to the geometric optics limit profile.}
\label{fig:mag-finite}
 \end{figure}

In Fig.~(\ref{fig:mag-finite}), we depict the magnification in terms of $\beta$ for three sources with different sizes in the wave-optics regime. 
For the small sources, the magnification resembles a point source in the wave-optics regime and by increasing the source size, the fringes are smeared out and the magnification looks like the geometric optics. 
The distance between the fringes for a point-lens is $\Delta \beta={2\pi}/{f}$ \citep{Falco:1992} and for a typical extended source with $\rho>\Delta \beta$, fringes smear out due to integration over a highly oscillating function. Hence, the fringes are observable only for  the extended source that satisfies the condition of $\rho<\Delta \beta$. Summarizing this part,  in the wave optics regime fringes for an extended source can be produced under the following condition of
  \begin{equation}\label{eq:sour-size1}
  \theta_{\rm s} <\frac{1}{2}\frac{\lambda}{R_{\rm sch}}\theta_{E},\;
  \end{equation}
where taking into account the redshift of deflector at $z_d$, this condition can be written as
\begin{equation}\label{eq:sour-size2}
	\theta_{E}>2\theta_s\left(1+z_d\right){R_{\rm sch}}/{\lambda_{obs}},\;
\end{equation}
  where $\lambda_{obs}$ is the wavelength of observation. 
Also, since $\theta_E\propto\sqrt{M}$ and $R_{\rm sch}\propto M$, the detection of fringes is in favour of small mass PBHs.

 Now, let us assume the lens mass to be in the range of $10^{-6}\lesssim \bar{M}\lesssim 10^{-2}$.
Then we rewrite the wave optics parameter of microlensing $f$, as follows
\begin{equation}\label{eq:f-pbh}
f=4\pi(1+z_{\rm d})\frac{R_{\rm sch}}{\lambda_{obs}}=4\pi(1+z_{\rm d})(\frac{\lambda_{obs}}{0.1\text{mm}})^{-1}{\bar{M}}.
\end{equation}
In the case of strong lensing of a quasar by a galaxy, it is more likely that PBH resides in the halo of the lensed galaxy, which allows us to measure the redshift of lens and from 
equation (\ref{eq:f-pbh}) directly obtain the mass of PBH. We note that we had also another method 
of mass measurement from equation (\ref{eq:mag-rel2}),  if we use at least two different wavelengths for the observation.

There are other observables in the geometric optics that can be used for breaking the degeneracy between the lens parameters. Let us take the finite size effect in the geometric optics which smoothes the peak of a light curve \citep{2011AJ....141..105P,Paris:2016xdm,2017PKAS...32..305S}. 
Knowing the physical size of a quasar as a source from the astrophysical informations, from the finite-size effect, we can extract the projected Einstein radius on the source plane (i.e. $R_E^{(s)}=\theta_{E}D_s$) as
 \begin{equation}
 \label{eq1}
 R_E^{(s)} = 1.65 \text{A.U.}(\frac{D_{\rm s}}{6\text{Gpc}})^{1/2}(\frac{1-x}{x})^{1/2}{\bar M}^{1/2},
 \end{equation}
 where in this case we assume a source at redshift $z\approx 3$. We can combine equation (\ref{eq1}) with the parameter $f = 2kR_{\rm sch}$ from the wave optics observations to obtain the distance of the lens as well as its mass.

\section{Observational prospect and optical depth}
\label{obs}
From observational point of view, the cadence between the data points in the light curve of a quasar should be small enough to measure the oscillations due to the diffraction pattern in the light curve. In order to estimate the time scale between the fringes, we use 
 $\Delta \beta$ where in terms of $f$ in equation (\ref{eq:f-pbh}) is given by 
\begin{equation}\label{eq:mag-osc} 
\Delta \beta = \frac{2\pi}{f}=\frac12\frac{\lambda _{obs}}{0.1\text{mm}}\frac{1}{1+z_d}\frac{1}{\bar M}.\;
\end{equation} 
 By multiplying $\Delta \beta$ to the Einstein crossing time of lens, $t_{E}$, the time-scale for the transit of fringes can be obtained. The Einstein crossing time for typical parameters of a lens with lunar mass at the cosmological distances is 
 \begin{equation}\label{eq:te-geo}
 t_{E} = 5.7{\text d}~(\frac{D_{\rm s}}{6\text{Gpc}})^{1/2}(\frac{1-x}{x})^{1/2}{\bar M}^{1/2}(\frac{500}{v_t}),
 \end{equation}
where $v_t$ is the relative transverse velocity of the lens-source-observer, which is $\sim 1000$ km/s for a rich cluster and  $\sim 200$ km/s in the galactic scales. Hence, the time-scale for transit of fringes (i.e. $\Delta t = t_E \Delta\beta $)  is given by 

\begin{equation}\label{eq:dely-time}
\Delta t =  2.9\text{d}~ (\frac{\lambda_{obs}}{0.1\text{mm}})\frac{1}{1+z_d}(\frac{D_{\rm s}}{6\text{Gpc}})^{1/2}(\frac{1-x}{x})^{1/2}(\frac{500}{v_t}){\bar M}^{-1/2}.
\end{equation}

%\begin{equation}\label{eq:dely-time}
%\Delta t = \frac{2\pi}{f}t_E = \frac{\lambda_{obs}}{0.1mm}\frac{1}{1+z_d}\frac{1}{\bar M} \frac{t_E}{2}.
%\end{equation}
 
%We note the ratio of two time scales of $\Delta t$ to $t_E$ is proportional to $ {\lambda_{obs}}/{\bar M}$, (i.e. $\Delta t/t_E \propto {\lambda_{obs}}/{\bar M}$).
 For a lunar mass PBH located at the distance $z_d\sim1$ and wavelength of $\lambda_{obs}=100\mu m$, we have $\Delta\beta\sim 10^{-1}$ and the time scale of fringe-transit  is of the order of $\sim 1.5$ days. According to (\ref{eq:dely-time}), the transit time is proportional to the $\lambda$ and decreases for shorter wavelengths. For PBHs in the mass range of $10^{-6} <\bar{M}<10^{-2}$ the time scale of fringe-transit is within the range of $15 \text{d}<\Delta t< 4 \text{yr}$.

One of the important technical issues in the observations of quasars is the filtering of intrinsic variabilities compare to the diffraction signals. For a 
quasar with the variability time scale shorter or in the same order of fringe transit time-scale, it is difficult to filter out the background signals.  Some of the quasars with very rapid variabilities have been detected in the timescales shorter than hours to minutes, so-called micro-variability \citep{Whiting:2001vj}. 
One solution is to survey those quasars with the low variabilities.
The other possibility is to study the quasars in the strong lensing systems and remove any intrinsic variations in the light curve by shifting the light curves according to the time delay between the images \citep{2001ASPC..237..185W}. This method in recent years is used for detecting microlensing signals in the geometric optics regime \citep{Ricci2011,Ricci,Giannini}.

%As we noted in the previous section, the time scale of fringes is about a few days. 
%Among all the observed quasars we select the quasar "3C 273", which is a bright quasar at sub-millimeter wavelengths to check the intrinsic time scale variability.

% This part is temporary removed from the text ----------------------------------------------

%As an example, Figure (\ref{fig:quasar})  represents the light curve of quasar "3C 273" in two sub-millimeter wavelengths for the duration of ten days. The intrinsic variability of light curve in this case is less than $20\%$, however, the cadence of light curve data is not sufficient to see the short time scale variability.  
% \begin{figure}
% 	\centering
% 	\includegraphics[width=.48\textwidth]{light-curve}
% 	\caption{The light curve of quasar 3C 273 in two different wavelengths for the duration of $10$ days in year 2003.}
% 	\label{fig:quasar}
 % \end{figure} 

%  -------------------------------- End temporary from the text ---------------------------------------

 \begin{figure}
  	\centering
  	\includegraphics[width=.48\textwidth]{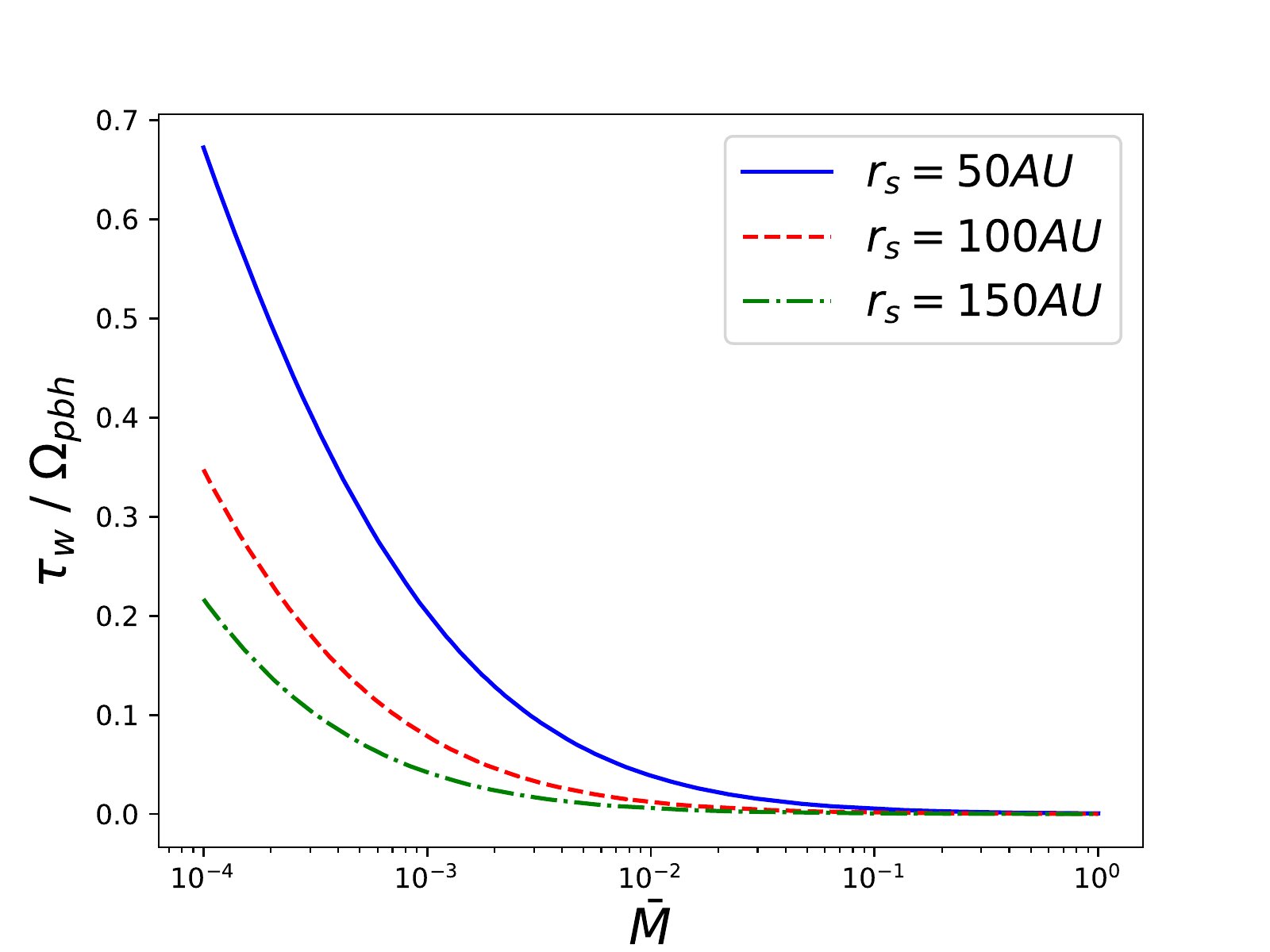}
  	\caption{The wave optics optical depth as a function of PBH mass for sources with the size of 50 AU (solid blue line), 100AU (dashed red line) and 150AU (dot-dashed green line). Here we consider concordance $\Lambda$CDM with $\Omega_m=0.3$, $h=0.7$ and use $\lambda_{\rm obs}=100 \mu m$. While for a lunar mass PBH the optical depth is negligible, for the smaller PBH the optical depth is larger. In addition, the optical depth increases by decreasing the size of source.}
  	\label{fig:opt-dep}
  \end{figure} 
  
    \begin{figure}
  	\centering
  	\includegraphics[width=.48\textwidth]{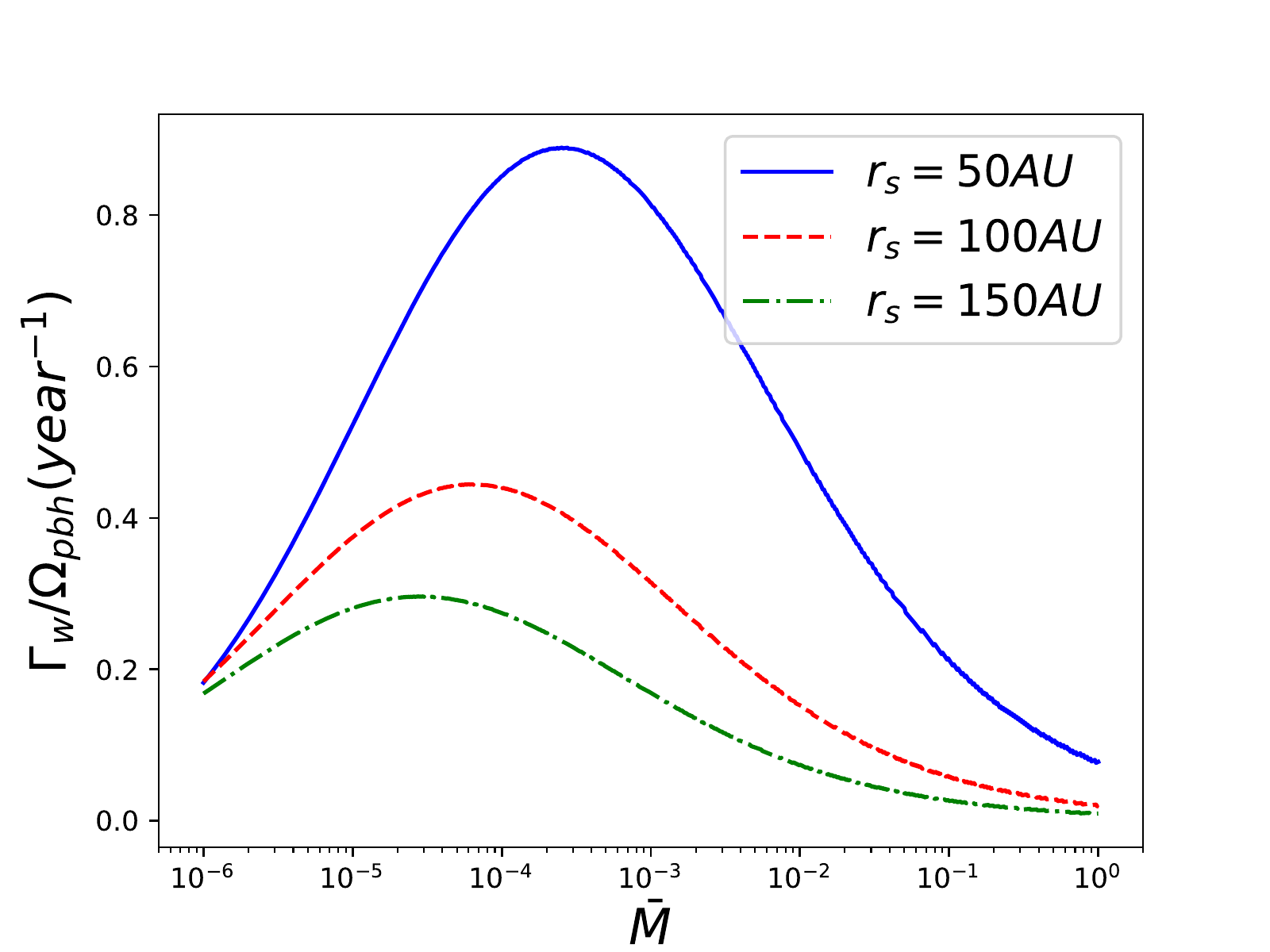}
  	\caption{The number of expected events per year as a function of PBH mass. Here the wave length of the observation is adapted $\lambda_{\rm obs}=100 \mu m$ and the rate is plotted for three different size of sources as 50 AU (solid blue line), 100AU (dashed red line) and 150AU (dot-dashed green line).  We note that $\tau_{\text{w}}$ is larger for the smaller PBHs as $z_{max}$ is getting larger and the optical depth is larger for the smaller mass from Fig. (\ref{fig:opt-dep}). On the other hand $\Delta t$ also is getting larger to the smaller PBHs. So the ratio of these two terms in (\ref{eq:rate-geo}) is a function of mass and  has a peak for the rate of number of events as depicted in this figure. 
	%The number decreases due to small value of the optical depth for $\bar{M}\sim 1$ and due to a large value of $\Delta t$ for tiny PBHs.  
	}
  	\label{fig:rate}
  \end{figure}

 In order to estimate the number of detectable events, we calculate the microlensing optical depth for detection of PBH. The optical depth  \citep{rahvar:2015,2007A&A...475..453Z} is defined by
  \begin{equation}\label{eq:opt-dep1}
  \tau=\int\pi R_E^2n(M,z)c\frac{dt}{dz}dz,\;
  \end{equation}
  where $n(M,z)$ is the number density of PBHs and it follows the spatial clustering of the cold dark matter. For the lower redshifts, the optical depth is a function of the direction of line of sight, depending on the cosmic density perturbations that cross the line of sight. However, for quasars at the higher redshifts, we can take almost a uniform number density of PBHs that is proportional to the dark matter density of the universe.  This assumption has been carefully investigated in \citep{Zackrisson}, where by considering 6 different models for halos and sub-halos, for quasars at higher redshifts (i.e. $z>0.25$) the optical depth from the clustered and uniform distribution of lenses converge. Here, we assume a uniform distribution for the density of PBHs in the optical depth calculation.  
  
  For a uniform distribution of PBH, we define a new optical depth, so-called the wave optics optical depth by $\tau_{\text w}$ where in $\Lambda$CDM, it is given by    
  \begin{equation}\label{eq:opt-dep2}
    \tau_{\text{w}}=\frac{3}{2}\frac{D_H}{D_s}\Omega_{pbh}\int_0^{z_{max}}\frac{(1+z)^2(D_s-D_l)D_ldz}{\sqrt{\Omega_m(1+z)^3+\Omega_{\Lambda}}},\;
    \end{equation}
  where $\Omega_{pbh}$ is the density parameter of PBHs and $D_H$ is the present horizon size of universe.  The difference between this definition with the conventional optical depth is that in this equation, $z_{max}$ is not assigned to the position of the quasar while that is the largest distance for a lens that satisfies the detection of the wave optics condition ($\rho<\Delta y$).  Moreover, in the geometric optics $\tau$ is independent of the mass function of the lenses and it depends on the overall mass density of the lenses. However, for the wave optics regime, the optical depth depends 
on the mass of lens as well as $z_{max}$. In Fig.(\ref{fig:opt-dep}) we plot the optical depth in unite of $\Omega_{pbh}$ for three different values of the source sizes. In this plot, we consider the concordance cosmology model of $\Omega_{m}=0.3~,~h=0.7$ and put the source at the redshift $z=3$. As it is expected, the small mass PBH and small size sources are in favor of wave optics microlensing detection.
  
 For a lunar mass PBH, the optical depth is very small, however it grows rapidly by decreasing the mass of lenses. For the case of $\bar{M}\ll 1$, from equation (\ref{eq:dely-time}) and equation (\ref{eq:te-geo}), $t_E \ll \Delta t$ and we take $\Delta t$ as the corresponding time-scale for the microlensing events in the wave optics regime instead of $t_E$. 
 %In this case $\Delta t$ is proportional to $\bar{M}^{-0.5}$ which increases for small PBHs.
 Now we define the rate of events in the regime of wave optics microlensing as 
 \begin{equation}\label{eq:rate-geo}
 \Gamma_{\text w}=\frac{2}{\pi}\frac{\tau_{\text{w}}}{\Delta t},\;
 \end{equation} 
where both $\tau_{\text{w}}$ and $\Delta t $ depend on the mass of PBH. Assuming Dirac-Delta function for the mass function of PBHs, in Fig. (\ref{fig:rate}) the rate of events per year is depicted as a function of $\bar{M}$. In equation (\ref{eq:rate-geo}), the optical depth increases for the smaller masses. On the other hand $\Delta t$ also increase for the smaller masses with the factor of ${\bar M}^{-1/2}$, so the ratio of these two terms in  (\ref{eq:rate-geo}) results in a peak as depicted in Fig. (\ref{fig:rate}).
%The rate is small for $\bar{M}\sim 1$ due to negligible value of the optical depth and due to large value of $\Delta t$ for $\bar{M}<<1$. 
For a given quasar, the number of detectable events is $N_{obs}=\Gamma T_{obs}$, where $T_{obs}$ is the duration of observation. Let us take a quasar with the size of $50$ AU, then from Fig. (\ref{fig:rate}), we expect to detect $N_{obs} \simeq 0.9\Omega_{\rm pbh}/{\text yr}$ for PBHs with $\bar{M}\sim 10^{-3}$ and $N_{obs} = 0.3\Omega_{\rm pbh}/{\text yr}$ for the sources with the radius $r_s=100AU$. Now if hundred percent of the dark matter is made of PBHs  (i.e. $\Omega_{\rm pbh} = 0.3$), all with the mass of $\bar{M}\sim 10^{-3}$, we expect to detect $0.27$ and $0.09$ event per year, respectively. Also for this mass, equation (\ref{eq:dely-time}) provides $\Delta t \simeq 46$~d and a cadence in the observation of light curve with one day can reveal the oscillation mode of diffraction pattern.
 %As an example, for detection of a lens with the mass of $\bar{M}=10^{-3}$PBH, a cadence of order of $\sim$ day and smaller is needed at the observation wavelength $\sim 100 \mu m$

\begin{figure}
 	\centering
 	\includegraphics[width=.48\textwidth]{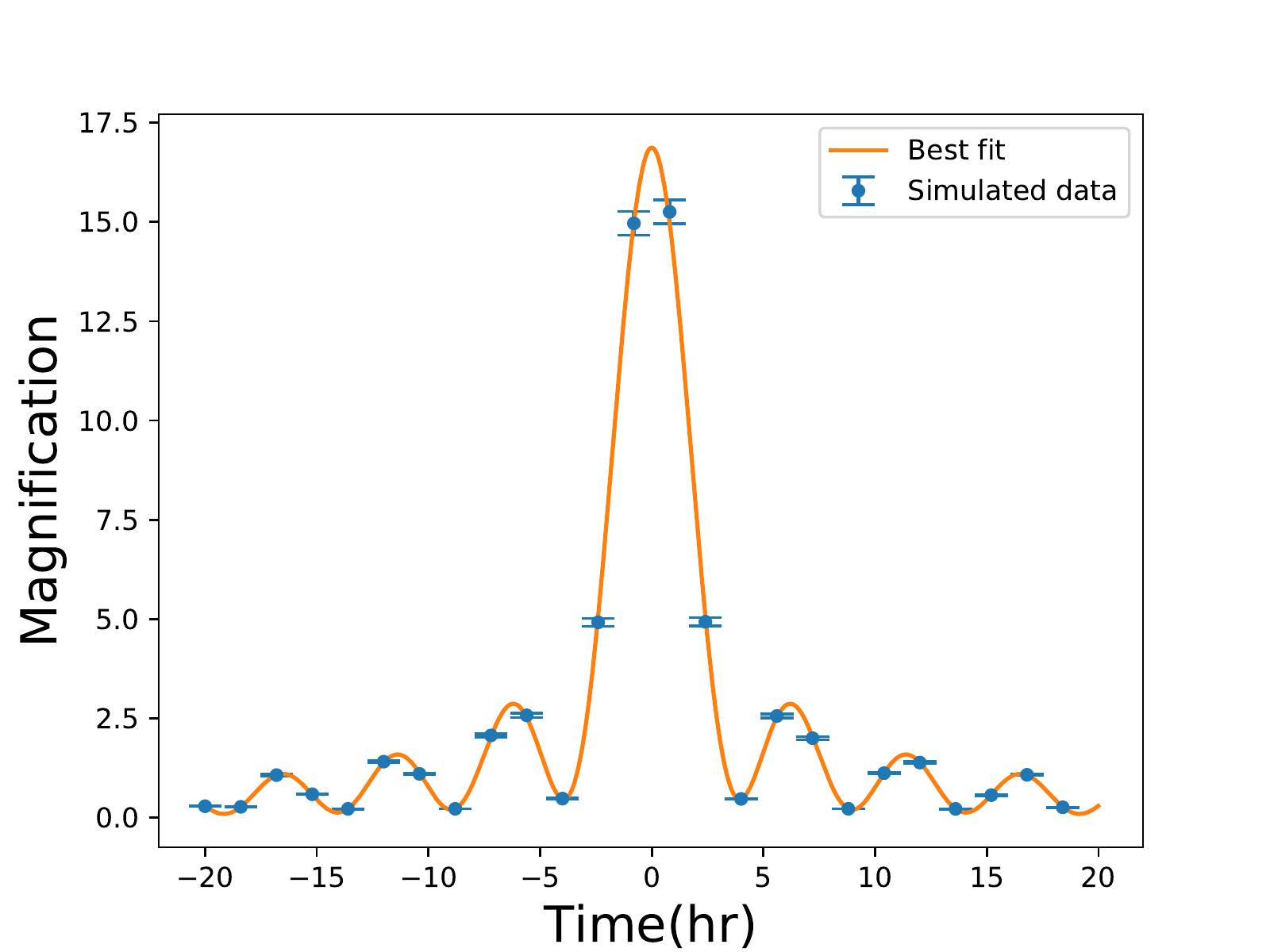}
 	\caption{Simulated data points and the best fit model for a typical event. Here we use $t_E=10$ hr, $\rho=0.1$, $u_0=0.05$, $\bar{M}=7\times 10^{-3}$ and $\lambda=2\mu m$ to simulate data points.  
		}
 	\label{fig:sampl-data}
 \end{figure}

% For example if $\tau_w \approx 10^{-1}\Omega_{pbh}$, the number of expected events per quasar (with the average value of $\Delta t\approx 3$ days) is around $7.7\Omega_{pbh}$ per year. 
Let us define the contribution of PBH on the density of dark matter as $f_{pbh} =  \Omega_{pbh}/\Omega_{m}$.
Then we also define the parameter of $df_{pbh}/d\bar{M}$ which provide the fraction of dark matter in form of PBHs within the range of $(\bar{M}, \bar{M} + d\bar{M})$. This function can be measured by a long-term survey of quasars with cadence of order of $\sim$ days. 
 Fig.(\ref{fig:sampl-data}) demonstrates the simulation of data points for a microlensing event with the parameters of $t_{E}=10$~hr, $\rho=0.1$, $u_0=0.05$ and $\bar{M}=0.007$.  Here we adapt the wavelength of $\lambda=2\mu m$ which results in the transit time scale of fringes of order of a few hours.  We assumed a photometric signal to noise ratio of $S/N=50$ and recover the parameters of light curve, using the Markov chain Monte Carlo method. 
The best values of parameters with 1-$\sigma$ uncertainty for the light curve in Fig. (\ref{fig:sampl-data}) are given in Tab.(\ref{tab:res}). 
The maximum magnification for this light curve is around $\sim 17$ and increases rapidly by decreasing $u_0$ and $\rho$. We note that the ratio of light in the anti-nodes to the nodes  (where $A_{node}\rightarrow 0$) of the interference pattern in Fig. (\ref{fig:sampl-data}) is a large number,  much larger than the intrinsic variations of a quasar which is about  $\sim 50\%$ \citep{Soldi:2008ev}. So, once we have enough photometric accuracy and the cadence shorter than the interference crossing time-scale, we can detect our desired signals.

%where compare to the intrinsic variations of a typical quasar in the order of $\sim 50\%$ \citep{Soldi:2008ev}, the diffraction pattern produces larger contrast in the light curve.
%The variability due to magnification is around $\sim 60-70\%$ near the central peak and relatively larger for the central peak.  %The time scale of these variability in Fig. (\ref{fig:sampl-data}) is of order of $\sim$ hour and if we use the longer wavelengths for the observations that would be event larger.. On the other hand the intrinsic fractional variability for a typical quasar (3C 273) at our desired wavelengths is $\sim 50\%$ \citep{Soldi:2008ev} but for cadence much larger than a few hours. In our case, we need to know the intrinsic variability of quasars at $\sim$ hour cadence which is not currently available.  

\begin{table}
 	\centering
\begin{tabular} { l  c}	
	Parameter &  68\% limits\\
	\hline
	{\boldmath$t_E            $} & $10.035\pm 0.036           $\\
	
	{\boldmath$u_0            $} & $0.0557\pm 0.0033          $\\
	
	{\boldmath$\rho           $} & $0.09995\pm 0.00070        $\\
	
	{\boldmath$\bar{M}              $} & $0.007027\pm 0.000025      $\\
	\hline
\end{tabular}
  	\caption{The best fit parameters with 1-$\sigma$ confidence level recovered from the fitting to the light curve in Fig. (\ref{fig:sampl-data}), using the Markov chain Monte Carlo method.}
  	\label{tab:res}
  \end{table}

% This part is removed temporary -------------------------------

%In addition, the results in 1,2 and 3$\sigma$ confidence levels are shown in Fig.(\ref{fig:like}). As it is clear, there is a tight correlation between $M$ and $t_E$ an uncertainty in $M$ is very small.   
% \begin{figure}
 %	\centering
 %	\includegraphics[width=.48\textwidth]{like}
 %	\caption{The likelihood and 1,2 and 3 $\sigma$ confidence level for a typical microlensing event. for details see the text.}
 %	\label{fig:like}
 %\end{figure}

% ------------------------------------------------------------------------- 

\section{Conclusion}\label{sec:conclude}

%\textcolor{blue}{I will work on the conclusion after finalizing the body of paper} \\

%=================================================================
%\acknowledgements

Summarizing this work, we have proposed a new method for the microlensing observation of quasars from the far infrared to the millimeter wavelengths. For the small mass lenses where the Schwarzschild radius of the lens is of the order of the wavelength of observation, the gravitational lens can produce distortions on the wavefront of the light in the order of one wavelength. Since the lens and the source are far enough from the observer, this situation is similar to the Huygens-Fresnel approximation and the result is the diffraction pattern from the phase-shifted electromagnetic wave on the lens plane. A relative motion of the observer through the diffraction pattern on the observer plane produces a modulation in the light curve of the quasar. One of the problems with this wave-optics microlensing observation would be the intrinsic variations of quasars that might be mixed with our desired signals. In order to solve this problem, we suggested the observation of quasars with the multiple images from the strong lensing.  The advantage of using these quasars is that by shifting the time delay between the images, we can remove the intrinsic variations of the quasar and extract our desired signals.

We suggested the observation of wave-optics microlensing in two different wavelengths. This technic enable us to measure directly the mass of a lens. Also with single wavelength observation, from the measuring the redshift of the strong lensing galaxy and the redshift of the source, we can break the degeneracy between the lens parameters and extract the mass of lenses. One of the possible candidates for the dark matter is the PBHs in the mass range smaller than the lunar mass. In this work, we proposed the observation of quasars with suitable cadence and photometric accuracy to observe the transit of the fringes in the diffraction pattern by the observer. The optical depth and the rate of events depend on the fraction of dark matter made of PBHs as well as the mass of the PBHs. Assuming the mass of PBHs in the order of $10^{-3}$ lunar mass and hundred percent of dark matter is made of PBHs, we obtained the rate of event detection per year for a given quasar within the range of $\sim 0.1$ to $\sim 0.3$. The wave-optics quasar microlensing can put a constraint on the fraction of dark matter made of PBHs as well as the mass function of PBHs. A long term survey of quasars by the infrared telescopes such as Spitzer space-based telescope or millimeter and submillimeter wavelength ground-based telescopes such as ALMA was suggested for this project.

 \acknowledgments
We thank David Spergel for the useful discussions. Also we thank anonymous 
referee for his/her useful comments and suggestions. S. Rahvar was supported by Sharif
University of Technology's Office of Vice President for Research
under Grant No. G950214.

 \bibliographystyle{apsrev4-1}
  \bibliography{ref}

\begin{thebibliography}{65}%
\makeatletter
\providecommand \@ifxundefined [1]{%
 \@ifx{#1\undefined}
}%
\providecommand \@ifnum [1]{%
 \ifnum #1\expandafter \@firstoftwo
 \else \expandafter \@secondoftwo
 \fi
}%
\providecommand \@ifx [1]{%
 \ifx #1\expandafter \@firstoftwo
 \else \expandafter \@secondoftwo
 \fi
}%
\providecommand \natexlab [1]{#1}%
\providecommand \enquote  [1]{``#1''}%
\providecommand \bibnamefont  [1]{#1}%
\providecommand \bibfnamefont [1]{#1}%
\providecommand \citenamefont [1]{#1}%
\providecommand \href@noop [0]{\@secondoftwo}%
\providecommand \href [0]{\begingroup \@sanitize@url \@href}%
\providecommand \@href[1]{\@@startlink{#1}\@@href}%
\providecommand \@@href[1]{\endgroup#1\@@endlink}%
\providecommand \@sanitize@url [0]{\catcode `\\12\catcode `\$12\catcode
  `\&12\catcode `\#12\catcode `\^12\catcode `\_12\catcode `\%12\relax}%
\providecommand \@@startlink[1]{}%
\providecommand \@@endlink[0]{}%
\providecommand \url  [0]{\begingroup\@sanitize@url \@url }%
\providecommand \@url [1]{\endgroup\@href {#1}{\urlprefix }}%
\providecommand \urlprefix  [0]{URL }%
\providecommand \Eprint [0]{\href }%
\providecommand \doibase [0]{http://dx.doi.org/}%
\providecommand \selectlanguage [0]{\@gobble}%
\providecommand \bibinfo  [0]{\@secondoftwo}%
\providecommand \bibfield  [0]{\@secondoftwo}%
\providecommand \translation [1]{[#1]}%
\providecommand \BibitemOpen [0]{}%
\providecommand \bibitemStop [0]{}%
\providecommand \bibitemNoStop [0]{.\EOS\space}%
\providecommand \EOS [0]{\spacefactor3000\relax}%
\providecommand \BibitemShut  [1]{\csname bibitem#1\endcsname}%
\let\auto@bib@innerbib\@empty
%</preamble>
\bibitem [{\citenamefont {Perlmutter}\ \emph {et~al.}(1999)\citenamefont
  {Perlmutter} \emph {et~al.}}]{Perlmutter:1998np}%
  \BibitemOpen
  \bibfield  {author} {\bibinfo {author} {\bibfnamefont {S.}~\bibnamefont
  {Perlmutter}} \emph {et~al.} (\bibinfo {collaboration} {Supernova Cosmology
  Project}),\ }\href {\doibase 10.1086/307221} {\bibfield  {journal} {\bibinfo
  {journal} {Astrophys.J.}\ }\textbf {\bibinfo {volume} {517}},\ \bibinfo
  {pages} {565} (\bibinfo {year} {1999})},\ \Eprint
  {http://arxiv.org/abs/astro-ph/9812133} {arXiv:astro-ph/9812133 [astro-ph]}
  \BibitemShut {NoStop}%
%%CITATION = ASTRO-PH/9812133;%%
\bibitem [{\citenamefont {Riess}\ \emph {et~al.}(2004)\citenamefont {Riess}
  \emph {et~al.}}]{Riess:2004nr}%
  \BibitemOpen
  \bibfield  {author} {\bibinfo {author} {\bibfnamefont {A.~G.}\ \bibnamefont
  {Riess}} \emph {et~al.} (\bibinfo {collaboration} {Supernova Search Team}),\
  }\href {\doibase 10.1086/383612} {\bibfield  {journal} {\bibinfo  {journal}
  {Astrophys.J.}\ }\textbf {\bibinfo {volume} {607}},\ \bibinfo {pages} {665}
  (\bibinfo {year} {2004})},\ \Eprint {http://arxiv.org/abs/astro-ph/0402512}
  {arXiv:astro-ph/0402512 [astro-ph]} \BibitemShut {NoStop}%
%%CITATION = ASTRO-PH/0402512;%%
\bibitem [{\citenamefont {Astier}\ \emph {et~al.}(2006)\citenamefont {Astier}
  \emph {et~al.}}]{Astier:2005qq}%
  \BibitemOpen
  \bibfield  {author} {\bibinfo {author} {\bibfnamefont {P.}~\bibnamefont
  {Astier}} \emph {et~al.} (\bibinfo {collaboration} {SNLS Collaboration}),\
  }\href {\doibase 10.1051/0004-6361:20054185} {\bibfield  {journal} {\bibinfo
  {journal} {Astron.Astrophys.}\ }\textbf {\bibinfo {volume} {447}},\ \bibinfo
  {pages} {31} (\bibinfo {year} {2006})},\ \Eprint
  {http://arxiv.org/abs/astro-ph/0510447} {arXiv:astro-ph/0510447 [astro-ph]}
  \BibitemShut {NoStop}%
%%CITATION = ASTRO-PH/0510447;%%
\bibitem [{\citenamefont {Jha}\ \emph {et~al.}(2007)\citenamefont {Jha},
  \citenamefont {Riess},\ and\ \citenamefont {Kirshner}}]{Jha:2006fm}%
  \BibitemOpen
  \bibfield  {author} {\bibinfo {author} {\bibfnamefont {S.}~\bibnamefont
  {Jha}}, \bibinfo {author} {\bibfnamefont {A.~G.}\ \bibnamefont {Riess}}, \
  and\ \bibinfo {author} {\bibfnamefont {R.~P.}\ \bibnamefont {Kirshner}},\
  }\href {\doibase 10.1086/512054} {\bibfield  {journal} {\bibinfo  {journal}
  {Astrophys.J.}\ }\textbf {\bibinfo {volume} {659}},\ \bibinfo {pages} {122}
  (\bibinfo {year} {2007})},\ \Eprint {http://arxiv.org/abs/astro-ph/0612666}
  {arXiv:astro-ph/0612666 [astro-ph]} \BibitemShut {NoStop}%
%%CITATION = ASTRO-PH/0612666;%%
\bibitem [{\citenamefont {Spergel}\ \emph {et~al.}(2007)\citenamefont {Spergel}
  \emph {et~al.}}]{Spergel:2006hy}%
  \BibitemOpen
  \bibfield  {author} {\bibinfo {author} {\bibfnamefont {D.}~\bibnamefont
  {Spergel}} \emph {et~al.} (\bibinfo {collaboration} {WMAP Collaboration}),\
  }\href {\doibase 10.1086/513700} {\bibfield  {journal} {\bibinfo  {journal}
  {Astrophys.J.Suppl.}\ }\textbf {\bibinfo {volume} {170}},\ \bibinfo {pages}
  {377} (\bibinfo {year} {2007})},\ \Eprint
  {http://arxiv.org/abs/astro-ph/0603449} {arXiv:astro-ph/0603449 [astro-ph]}
  \BibitemShut {NoStop}%
%%CITATION = ASTRO-PH/0603449;%%
\bibitem [{\citenamefont {Ade}\ \emph {et~al.}(2016)\citenamefont {Ade} \emph
  {et~al.}}]{Ade:2015rim}%
  \BibitemOpen
  \bibfield  {author} {\bibinfo {author} {\bibfnamefont {P.~A.~R.}\
  \bibnamefont {Ade}} \emph {et~al.} (\bibinfo {collaboration} {Planck}),\
  }\href {\doibase 10.1051/0004-6361/201525814} {\bibfield  {journal} {\bibinfo
   {journal} {Astron. Astrophys.}\ }\textbf {\bibinfo {volume} {594}},\
  \bibinfo {pages} {A14} (\bibinfo {year} {2016})},\ \Eprint
  {http://arxiv.org/abs/1502.01590} {arXiv:1502.01590 [astro-ph.CO]}
  \BibitemShut {NoStop}%
%%CITATION = ARXIV:1502.01590;%%
\bibitem [{\citenamefont {Seo}\ and\ \citenamefont
  {Eisenstein}(2005)}]{Seo:2005ys}%
  \BibitemOpen
  \bibfield  {author} {\bibinfo {author} {\bibfnamefont {H.-J.}\ \bibnamefont
  {Seo}}\ and\ \bibinfo {author} {\bibfnamefont {D.~J.}\ \bibnamefont
  {Eisenstein}},\ }\href {\doibase 10.1086/491599} {\bibfield  {journal}
  {\bibinfo  {journal} {Astrophys.J.}\ }\textbf {\bibinfo {volume} {633}},\
  \bibinfo {pages} {575} (\bibinfo {year} {2005})},\ \Eprint
  {http://arxiv.org/abs/astro-ph/0507338} {arXiv:astro-ph/0507338 [astro-ph]}
  \BibitemShut {NoStop}%
%%CITATION = ASTRO-PH/0507338;%%
\bibitem [{\citenamefont {Blake}\ \emph {et~al.}(2011)\citenamefont {Blake},
  \citenamefont {Kazin}, \citenamefont {Beutler}, \citenamefont {Davis},
  \citenamefont {Parkinson} \emph {et~al.}}]{Blake:2011en}%
  \BibitemOpen
  \bibfield  {author} {\bibinfo {author} {\bibfnamefont {C.}~\bibnamefont
  {Blake}}, \bibinfo {author} {\bibfnamefont {E.}~\bibnamefont {Kazin}},
  \bibinfo {author} {\bibfnamefont {F.}~\bibnamefont {Beutler}}, \bibinfo
  {author} {\bibfnamefont {T.}~\bibnamefont {Davis}}, \bibinfo {author}
  {\bibfnamefont {D.}~\bibnamefont {Parkinson}},  \emph {et~al.},\ }\href
  {\doibase 10.1111/j.1365-2966.2011.19592.x} {\bibfield  {journal} {\bibinfo
  {journal} {Mon.Not.Roy.Astron.Soc.}\ }\textbf {\bibinfo {volume} {418}},\
  \bibinfo {pages} {1707} (\bibinfo {year} {2011})},\ \Eprint
  {http://arxiv.org/abs/1108.2635} {arXiv:1108.2635 [astro-ph.CO]} \BibitemShut
  {NoStop}%
%%CITATION = ARXIV:1108.2635;%%
\bibitem [{\citenamefont {Blais}\ \emph {et~al.}(2002)\citenamefont {Blais},
  \citenamefont {Kiefer},\ and\ \citenamefont {Polarski}}]{Blais:2002nd}%
  \BibitemOpen
  \bibfield  {author} {\bibinfo {author} {\bibfnamefont {D.}~\bibnamefont
  {Blais}}, \bibinfo {author} {\bibfnamefont {C.}~\bibnamefont {Kiefer}}, \
  and\ \bibinfo {author} {\bibfnamefont {D.}~\bibnamefont {Polarski}},\ }\href
  {\doibase 10.1016/S0370-2693(02)01803-8} {\bibfield  {journal} {\bibinfo
  {journal} {Phys.Lett.}\ }\textbf {\bibinfo {volume} {B535}},\ \bibinfo
  {pages} {11} (\bibinfo {year} {2002})},\ \Eprint
  {http://arxiv.org/abs/astro-ph/0203520} {arXiv:astro-ph/0203520 [astro-ph]}
  \BibitemShut {NoStop}%
%%CITATION = ASTRO-PH/0203520;%%
\bibitem [{\citenamefont {Khlopov}(2010)}]{Khlopov:2008qy}%
  \BibitemOpen
  \bibfield  {author} {\bibinfo {author} {\bibfnamefont {M.~Y.}\ \bibnamefont
  {Khlopov}},\ }\href {\doibase 10.1088/1674-4527/10/6/001} {\bibfield
  {journal} {\bibinfo  {journal} {Res.Astron.Astrophys.}\ }\textbf {\bibinfo
  {volume} {10}},\ \bibinfo {pages} {495} (\bibinfo {year} {2010})},\ \Eprint
  {http://arxiv.org/abs/0801.0116} {arXiv:0801.0116 [astro-ph]} \BibitemShut
  {NoStop}%
%%CITATION = ARXIV:0801.0116;%%
\bibitem [{\citenamefont {Frampton}\ \emph {et~al.}(2010)\citenamefont
  {Frampton}, \citenamefont {Kawasaki}, \citenamefont {Takahashi},\ and\
  \citenamefont {Yanagida}}]{Frampton:2010sw}%
  \BibitemOpen
  \bibfield  {author} {\bibinfo {author} {\bibfnamefont {P.~H.}\ \bibnamefont
  {Frampton}}, \bibinfo {author} {\bibfnamefont {M.}~\bibnamefont {Kawasaki}},
  \bibinfo {author} {\bibfnamefont {F.}~\bibnamefont {Takahashi}}, \ and\
  \bibinfo {author} {\bibfnamefont {T.~T.}\ \bibnamefont {Yanagida}},\ }\href
  {\doibase 10.1088/1475-7516/2010/04/023} {\bibfield  {journal} {\bibinfo
  {journal} {JCAP}\ }\textbf {\bibinfo {volume} {1004}},\ \bibinfo {pages}
  {023} (\bibinfo {year} {2010})},\ \Eprint {http://arxiv.org/abs/1001.2308}
  {arXiv:1001.2308 [hep-ph]} \BibitemShut {NoStop}%
%%CITATION = ARXIV:1001.2308;%%
\bibitem [{\citenamefont {Garcia-Bellido}\ \emph {et~al.}(1996)\citenamefont
  {Garcia-Bellido}, \citenamefont {Linde},\ and\ \citenamefont
  {Wands}}]{GarciaBellido:1996qt}%
  \BibitemOpen
  \bibfield  {author} {\bibinfo {author} {\bibfnamefont {J.}~\bibnamefont
  {Garcia-Bellido}}, \bibinfo {author} {\bibfnamefont {A.~D.}\ \bibnamefont
  {Linde}}, \ and\ \bibinfo {author} {\bibfnamefont {D.}~\bibnamefont
  {Wands}},\ }\href {\doibase 10.1103/PhysRevD.54.6040} {\bibfield  {journal}
  {\bibinfo  {journal} {Phys.Rev.}\ }\textbf {\bibinfo {volume} {D54}},\
  \bibinfo {pages} {6040} (\bibinfo {year} {1996})},\ \Eprint
  {http://arxiv.org/abs/astro-ph/9605094} {arXiv:astro-ph/9605094 [astro-ph]}
  \BibitemShut {NoStop}%
%%CITATION = ASTRO-PH/9605094;%%
\bibitem [{\citenamefont {Jedamzik}\ and\ \citenamefont
  {Niemeyer}(1999)}]{Jedamzik:1999am}%
  \BibitemOpen
  \bibfield  {author} {\bibinfo {author} {\bibfnamefont {K.}~\bibnamefont
  {Jedamzik}}\ and\ \bibinfo {author} {\bibfnamefont {J.~C.}\ \bibnamefont
  {Niemeyer}},\ }\href {\doibase 10.1103/PhysRevD.59.124014} {\bibfield
  {journal} {\bibinfo  {journal} {Phys.Rev.}\ }\textbf {\bibinfo {volume}
  {D59}},\ \bibinfo {pages} {124014} (\bibinfo {year} {1999})},\ \Eprint
  {http://arxiv.org/abs/astro-ph/9901293} {arXiv:astro-ph/9901293 [astro-ph]}
  \BibitemShut {NoStop}%
%%CITATION = ASTRO-PH/9901293;%%
\bibitem [{\citenamefont {Suyama}\ \emph {et~al.}(2005)\citenamefont {Suyama},
  \citenamefont {Tanaka}, \citenamefont {Bassett},\ and\ \citenamefont
  {Kudoh}}]{Suyama:2004mz}%
  \BibitemOpen
  \bibfield  {author} {\bibinfo {author} {\bibfnamefont {T.}~\bibnamefont
  {Suyama}}, \bibinfo {author} {\bibfnamefont {T.}~\bibnamefont {Tanaka}},
  \bibinfo {author} {\bibfnamefont {B.}~\bibnamefont {Bassett}}, \ and\
  \bibinfo {author} {\bibfnamefont {H.}~\bibnamefont {Kudoh}},\ }\href
  {\doibase 10.1103/PhysRevD.71.063507} {\bibfield  {journal} {\bibinfo
  {journal} {Phys.Rev.}\ }\textbf {\bibinfo {volume} {D71}},\ \bibinfo {pages}
  {063507} (\bibinfo {year} {2005})},\ \Eprint
  {http://arxiv.org/abs/hep-ph/0410247} {arXiv:hep-ph/0410247 [hep-ph]}
  \BibitemShut {NoStop}%
%%CITATION = HEP-PH/0410247;%%
\bibitem [{\citenamefont {Suyama}\ \emph {et~al.}(2006)\citenamefont {Suyama},
  \citenamefont {Tanaka}, \citenamefont {Bassett},\ and\ \citenamefont
  {Kudoh}}]{Suyama:2006sr}%
  \BibitemOpen
  \bibfield  {author} {\bibinfo {author} {\bibfnamefont {T.}~\bibnamefont
  {Suyama}}, \bibinfo {author} {\bibfnamefont {T.}~\bibnamefont {Tanaka}},
  \bibinfo {author} {\bibfnamefont {B.}~\bibnamefont {Bassett}}, \ and\
  \bibinfo {author} {\bibfnamefont {H.}~\bibnamefont {Kudoh}},\ }\href
  {\doibase 10.1088/1475-7516/2006/04/001} {\bibfield  {journal} {\bibinfo
  {journal} {JCAP}\ }\textbf {\bibinfo {volume} {0604}},\ \bibinfo {pages}
  {001} (\bibinfo {year} {2006})},\ \Eprint
  {http://arxiv.org/abs/hep-ph/0601108} {arXiv:hep-ph/0601108 [hep-ph]}
  \BibitemShut {NoStop}%
%%CITATION = HEP-PH/0601108;%%
\bibitem [{\citenamefont {Kohri}\ \emph {et~al.}(2013)\citenamefont {Kohri},
  \citenamefont {Lin},\ and\ \citenamefont {Matsuda}}]{Kohri:2012yw}%
  \BibitemOpen
  \bibfield  {author} {\bibinfo {author} {\bibfnamefont {K.}~\bibnamefont
  {Kohri}}, \bibinfo {author} {\bibfnamefont {C.-M.}\ \bibnamefont {Lin}}, \
  and\ \bibinfo {author} {\bibfnamefont {T.}~\bibnamefont {Matsuda}},\ }\href
  {\doibase 10.1103/PhysRevD.87.103527} {\bibfield  {journal} {\bibinfo
  {journal} {Phys.Rev.}\ }\textbf {\bibinfo {volume} {D87}},\ \bibinfo {pages}
  {103527} (\bibinfo {year} {2013})},\ \Eprint {http://arxiv.org/abs/1211.2371}
  {arXiv:1211.2371 [hep-ph]} \BibitemShut {NoStop}%
%%CITATION = ARXIV:1211.2371;%%
\bibitem [{\citenamefont {Kawasaki}\ \emph {et~al.}(2013)\citenamefont
  {Kawasaki}, \citenamefont {Kitajima},\ and\ \citenamefont
  {Yanagida}}]{Kawasaki:2012wr}%
  \BibitemOpen
  \bibfield  {author} {\bibinfo {author} {\bibfnamefont {M.}~\bibnamefont
  {Kawasaki}}, \bibinfo {author} {\bibfnamefont {N.}~\bibnamefont {Kitajima}},
  \ and\ \bibinfo {author} {\bibfnamefont {T.~T.}\ \bibnamefont {Yanagida}},\
  }\href {\doibase 10.1103/PhysRevD.87.063519} {\bibfield  {journal} {\bibinfo
  {journal} {Phys.Rev.}\ }\textbf {\bibinfo {volume} {D87}},\ \bibinfo {pages}
  {063519} (\bibinfo {year} {2013})},\ \Eprint {http://arxiv.org/abs/1207.2550}
  {arXiv:1207.2550 [hep-ph]} \BibitemShut {NoStop}%
%%CITATION = ARXIV:1207.2550;%%
\bibitem [{\citenamefont {Bugaev}\ and\ \citenamefont
  {Klimai}(2013)}]{Bugaev:2013vba}%
  \BibitemOpen
  \bibfield  {author} {\bibinfo {author} {\bibfnamefont {E.}~\bibnamefont
  {Bugaev}}\ and\ \bibinfo {author} {\bibfnamefont {P.}~\bibnamefont
  {Klimai}},\ }\href {\doibase 10.1142/S021827181350034X} {\bibfield  {journal}
  {\bibinfo  {journal} {Int.J.Mod.Phys.}\ }\textbf {\bibinfo {volume} {D22}},\
  \bibinfo {pages} {1350034} (\bibinfo {year} {2013})},\ \Eprint
  {http://arxiv.org/abs/1303.3146} {arXiv:1303.3146 [astro-ph.CO]} \BibitemShut
  {NoStop}%
%%CITATION = ARXIV:1303.3146;%%
\bibitem [{\citenamefont {{Khlopov}}(2010)}]{2010RAA....10..495K}%
  \BibitemOpen
  \bibfield  {author} {\bibinfo {author} {\bibfnamefont {M.~Y.}\ \bibnamefont
  {{Khlopov}}},\ }\href {\doibase 10.1088/1674-4527/10/6/001} {\bibfield
  {journal} {\bibinfo  {journal} {Research in Astronomy and Astrophysics}\
  }\textbf {\bibinfo {volume} {10}},\ \bibinfo {pages} {495} (\bibinfo {year}
  {2010})},\ \Eprint {http://arxiv.org/abs/0801.0116} {arXiv:0801.0116}
  \BibitemShut {NoStop}%
\bibitem [{\citenamefont {{Carr}}\ \emph {et~al.}(2010)\citenamefont {{Carr}},
  \citenamefont {{Kohri}}, \citenamefont {{Sendouda}},\ and\ \citenamefont
  {{Yokoyama}}}]{2010PhRvD..81j4019C}%
  \BibitemOpen
  \bibfield  {author} {\bibinfo {author} {\bibfnamefont {B.~J.}\ \bibnamefont
  {{Carr}}}, \bibinfo {author} {\bibfnamefont {K.}~\bibnamefont {{Kohri}}},
  \bibinfo {author} {\bibfnamefont {Y.}~\bibnamefont {{Sendouda}}}, \ and\
  \bibinfo {author} {\bibfnamefont {J.}~\bibnamefont {{Yokoyama}}},\ }\href
  {\doibase 10.1103/PhysRevD.81.104019} {\bibfield  {journal} {\bibinfo
  {journal} {\prd}\ }\textbf {\bibinfo {volume} {81}},\ \bibinfo {eid} {104019}
  (\bibinfo {year} {2010})},\ \Eprint {http://arxiv.org/abs/0912.5297}
  {arXiv:0912.5297 [astro-ph.CO]} \BibitemShut {NoStop}%
\bibitem [{\citenamefont {{Bean}}\ and\ \citenamefont
  {{Magueijo}}(2002)}]{2002PhRvD..66f3505B}%
  \BibitemOpen
  \bibfield  {author} {\bibinfo {author} {\bibfnamefont {R.}~\bibnamefont
  {{Bean}}}\ and\ \bibinfo {author} {\bibfnamefont {J.}~\bibnamefont
  {{Magueijo}}},\ }\href {\doibase 10.1103/PhysRevD.66.063505} {\bibfield
  {journal} {\bibinfo  {journal} {\prd}\ }\textbf {\bibinfo {volume} {66}},\
  \bibinfo {eid} {063505} (\bibinfo {year} {2002})},\ \Eprint
  {http://arxiv.org/abs/astro-ph/0204486} {astro-ph/0204486} \BibitemShut
  {NoStop}%
\bibitem [{\citenamefont {Afshordi}\ \emph {et~al.}(2003)\citenamefont
  {Afshordi}, \citenamefont {McDonald},\ and\ \citenamefont
  {Spergel}}]{Afshordi:2003zb}%
  \BibitemOpen
  \bibfield  {author} {\bibinfo {author} {\bibfnamefont {N.}~\bibnamefont
  {Afshordi}}, \bibinfo {author} {\bibfnamefont {P.}~\bibnamefont {McDonald}},
  \ and\ \bibinfo {author} {\bibfnamefont {D.~N.}\ \bibnamefont {Spergel}},\
  }\href {\doibase 10.1086/378763} {\bibfield  {journal} {\bibinfo  {journal}
  {Astrophys. J.}\ }\textbf {\bibinfo {volume} {594}},\ \bibinfo {pages} {L71}
  (\bibinfo {year} {2003})},\ \Eprint {http://arxiv.org/abs/astro-ph/0302035}
  {arXiv:astro-ph/0302035 [astro-ph]} \BibitemShut {NoStop}%
%%CITATION = ASTRO-PH/0302035;%%
\bibitem [{\citenamefont {{Ricotti}}\ \emph {et~al.}(2008)\citenamefont
  {{Ricotti}}, \citenamefont {{Ostriker}},\ and\ \citenamefont
  {{Mack}}}]{2008ApJ...680..829R}%
  \BibitemOpen
  \bibfield  {author} {\bibinfo {author} {\bibfnamefont {M.}~\bibnamefont
  {{Ricotti}}}, \bibinfo {author} {\bibfnamefont {J.~P.}\ \bibnamefont
  {{Ostriker}}}, \ and\ \bibinfo {author} {\bibfnamefont {K.~J.}\ \bibnamefont
  {{Mack}}},\ }\href {\doibase 10.1086/587831} {\bibfield  {journal} {\bibinfo
  {journal} {\apj}\ }\textbf {\bibinfo {volume} {680}},\ \bibinfo {eid}
  {829-845} (\bibinfo {year} {2008})},\ \Eprint
  {http://arxiv.org/abs/0709.0524} {arXiv:0709.0524} \BibitemShut {NoStop}%
\bibitem [{\citenamefont {Tisserand}\ \emph {et~al.}(2007)\citenamefont
  {Tisserand} \emph {et~al.}}]{Tisserand:2006zx}%
  \BibitemOpen
  \bibfield  {author} {\bibinfo {author} {\bibfnamefont {P.}~\bibnamefont
  {Tisserand}} \emph {et~al.} (\bibinfo {collaboration} {EROS-2}),\ }\href
  {\doibase 10.1051/0004-6361:20066017} {\bibfield  {journal} {\bibinfo
  {journal} {Astron. Astrophys.}\ }\textbf {\bibinfo {volume} {469}},\ \bibinfo
  {pages} {387} (\bibinfo {year} {2007})},\ \Eprint
  {http://arxiv.org/abs/astro-ph/0607207} {arXiv:astro-ph/0607207 [astro-ph]}
  \BibitemShut {NoStop}%
%%CITATION = ASTRO-PH/0607207;%%
\bibitem [{\citenamefont {Nemiroff}\ \emph {et~al.}(2001)\citenamefont
  {Nemiroff}, \citenamefont {Marani}, \citenamefont {Norris},\ and\
  \citenamefont {Bonnell}}]{Nemiroff:2001bp}%
  \BibitemOpen
  \bibfield  {author} {\bibinfo {author} {\bibfnamefont {R.~J.}\ \bibnamefont
  {Nemiroff}}, \bibinfo {author} {\bibfnamefont {G.~F.}\ \bibnamefont
  {Marani}}, \bibinfo {author} {\bibfnamefont {J.~P.}\ \bibnamefont {Norris}},
  \ and\ \bibinfo {author} {\bibfnamefont {J.~T.}\ \bibnamefont {Bonnell}},\
  }\href {\doibase 10.1103/PhysRevLett.86.580} {\bibfield  {journal} {\bibinfo
  {journal} {Phys. Rev. Lett.}\ }\textbf {\bibinfo {volume} {86}},\ \bibinfo
  {pages} {580} (\bibinfo {year} {2001})},\ \Eprint
  {http://arxiv.org/abs/astro-ph/0101488} {arXiv:astro-ph/0101488 [astro-ph]}
  \BibitemShut {NoStop}%
%%CITATION = ASTRO-PH/0101488;%%
\bibitem [{\citenamefont {{Carr}}\ \emph {et~al.}(2016)\citenamefont {{Carr}},
  \citenamefont {{K{\"u}hnel}},\ and\ \citenamefont {{Sandstad}}}]{carr}%
  \BibitemOpen
  \bibfield  {author} {\bibinfo {author} {\bibfnamefont {B.}~\bibnamefont
  {{Carr}}}, \bibinfo {author} {\bibfnamefont {F.}~\bibnamefont
  {{K{\"u}hnel}}}, \ and\ \bibinfo {author} {\bibfnamefont {M.}~\bibnamefont
  {{Sandstad}}},\ }\href {\doibase 10.1103/PhysRevD.94.083504} {\bibfield
  {journal} {\bibinfo  {journal} {\prd}\ }\textbf {\bibinfo {volume} {94}},\
  \bibinfo {eid} {083504} (\bibinfo {year} {2016})},\ \Eprint
  {http://arxiv.org/abs/1607.06077} {arXiv:1607.06077} \BibitemShut {NoStop}%
\bibitem [{\citenamefont {{Green}}(2016)}]{green}%
  \BibitemOpen
  \bibfield  {author} {\bibinfo {author} {\bibfnamefont {A.~M.}\ \bibnamefont
  {{Green}}},\ }\href {\doibase 10.1103/PhysRevD.94.063530} {\bibfield
  {journal} {\bibinfo  {journal} {\prd}\ }\textbf {\bibinfo {volume} {94}},\
  \bibinfo {eid} {063530} (\bibinfo {year} {2016})},\ \Eprint
  {http://arxiv.org/abs/1609.01143} {arXiv:1609.01143} \BibitemShut {NoStop}%
\bibitem [{\citenamefont {{Rahvar}}(2015)}]{rahvar:2015}%
  \BibitemOpen
  \bibfield  {author} {\bibinfo {author} {\bibfnamefont {S.}~\bibnamefont
  {{Rahvar}}},\ }\href {\doibase 10.1142/S0218271815300207} {\bibfield
  {journal} {\bibinfo  {journal} {International Journal of Modern Physics D}\
  }\textbf {\bibinfo {volume} {24}},\ \bibinfo {eid} {1530020} (\bibinfo {year}
  {2015})},\ \Eprint {http://arxiv.org/abs/1503.04271} {arXiv:1503.04271
  [astro-ph.IM]} \BibitemShut {NoStop}%
\bibitem [{\citenamefont {{Paczynski}}(1986)}]{pac86}%
  \BibitemOpen
  \bibfield  {author} {\bibinfo {author} {\bibfnamefont {B.}~\bibnamefont
  {{Paczynski}}},\ }\href {\doibase 10.1086/164140} {\bibfield  {journal}
  {\bibinfo  {journal} {\apj}\ }\textbf {\bibinfo {volume} {304}},\ \bibinfo
  {pages} {1} (\bibinfo {year} {1986})}\BibitemShut {NoStop}%
\bibitem [{\citenamefont {Cottle}\ \emph {et~al.}(1991)\citenamefont {Cottle},
  \citenamefont {Kennedy}, \citenamefont {Kemper}, \citenamefont {Brown},
  \citenamefont {Jacobsen} \emph {et~al.}}]{Cottle:1991zza}%
  \BibitemOpen
  \bibfield  {author} {\bibinfo {author} {\bibfnamefont {P.}~\bibnamefont
  {Cottle}}, \bibinfo {author} {\bibfnamefont {M.}~\bibnamefont {Kennedy}},
  \bibinfo {author} {\bibfnamefont {K.}~\bibnamefont {Kemper}}, \bibinfo
  {author} {\bibfnamefont {J.~D.}\ \bibnamefont {Brown}}, \bibinfo {author}
  {\bibfnamefont {E.}~\bibnamefont {Jacobsen}},  \emph {et~al.},\ }\href
  {\doibase 10.1103/PhysRevC.44.1668} {\bibfield  {journal} {\bibinfo
  {journal} {Phys.Rev.}\ }\textbf {\bibinfo {volume} {C44}},\ \bibinfo {pages}
  {1668} (\bibinfo {year} {1991})}\BibitemShut {NoStop}%
%%CITATION = PHRVA,C44,1668;%%
\bibitem [{\citenamefont {Gould}\ and\ \citenamefont
  {Loeb}(1992)}]{Gould:1992aj}%
  \BibitemOpen
  \bibfield  {author} {\bibinfo {author} {\bibfnamefont {A.}~\bibnamefont
  {Gould}}\ and\ \bibinfo {author} {\bibfnamefont {A.}~\bibnamefont {Loeb}},\
  }\href {\doibase 10.1086/171700} {\bibfield  {journal} {\bibinfo  {journal}
  {Astrophys.J.}\ }\textbf {\bibinfo {volume} {396}},\ \bibinfo {pages} {104}
  (\bibinfo {year} {1992})}\BibitemShut {NoStop}%
%%CITATION = ASJOA,396,104;%%
\bibitem [{\citenamefont {{Gould}}\ \emph {et~al.}(2010)\citenamefont
  {{Gould}}, \citenamefont {{Dong}}, \citenamefont {{Gaudi}}, \citenamefont
  {{Udalski}}, \citenamefont {{Bond}}, \citenamefont {{Greenhill}},
  \citenamefont {{Street}}, \citenamefont {{Dominik}}, \citenamefont {{Sumi}},
  \citenamefont {{Szyma{\'n}ski}}, \citenamefont {{Han}}, \citenamefont
  {{Allen}}, \citenamefont {{Bolt}}, \citenamefont {{Bos}}, \citenamefont
  {{Christie}}, \citenamefont {{DePoy}}, \citenamefont {{Drummond}},
  \citenamefont {{Eastman}}, \citenamefont {{Gal-Yam}}, \citenamefont
  {{Higgins}}, \citenamefont {{Janczak}}, \citenamefont {{Kaspi}},
  \citenamefont {{Koz{\l}owski}}, \citenamefont {{Lee}}, \citenamefont
  {{Mallia}}, \citenamefont {{Maury}}, \citenamefont {{Maoz}}, \citenamefont
  {{McCormick}}, \citenamefont {{Monard}}, \citenamefont {{Moorhouse}},
  \citenamefont {{Morgan}}, \citenamefont {{Natusch}}, \citenamefont {{Ofek}},
  \citenamefont {{Park}}, \citenamefont {{Pogge}}, \citenamefont {{Polishook}},
  \citenamefont {{Santallo}}, \citenamefont {{Shporer}}, \citenamefont
  {{Spector}}, \citenamefont {{Thornley}}, \citenamefont {{Yee}}, \citenamefont
  {{{$\mu$}FUN Collaboration}}, \citenamefont {{Kubiak}}, \citenamefont
  {{Pietrzy{\'n}ski}}, \citenamefont {{Soszy{\'n}ski}}, \citenamefont
  {{Szewczyk}}, \citenamefont {{Wyrzykowski}}, \citenamefont {{Ulaczyk}},
  \citenamefont {{Poleski}}, \citenamefont {{OGLE Collaboration}},
  \citenamefont {{Abe}}, \citenamefont {{Bennett}}, \citenamefont {{Botzler}},
  \citenamefont {{Douchin}}, \citenamefont {{Freeman}}, \citenamefont
  {{Fukui}}, \citenamefont {{Furusawa}}, \citenamefont {{Hearnshaw}},
  \citenamefont {{Hosaka}}, \citenamefont {{Itow}}, \citenamefont {{Kamiya}},
  \citenamefont {{Kilmartin}}, \citenamefont {{Korpela}}, \citenamefont
  {{Lin}}, \citenamefont {{Ling}}, \citenamefont {{Makita}}, \citenamefont
  {{Masuda}}, \citenamefont {{Matsubara}}, \citenamefont {{Miyake}},
  \citenamefont {{Muraki}}, \citenamefont {{Nagaya}}, \citenamefont
  {{Nishimoto}}, \citenamefont {{Ohnishi}}, \citenamefont {{Okumura}},
  \citenamefont {{Perrott}}, \citenamefont {{Philpott}}, \citenamefont
  {{Rattenbury}}, \citenamefont {{Saito}}, \citenamefont {{Sako}},
  \citenamefont {{Sullivan}}, \citenamefont {{Sweatman}}, \citenamefont
  {{Tristram}}, \citenamefont {{von Seggern}}, \citenamefont {{Yock}},
  \citenamefont {{MOA Collaboration}}, \citenamefont {{Albrow}}, \citenamefont
  {{Batista}}, \citenamefont {{Beaulieu}}, \citenamefont {{Brillant}},
  \citenamefont {{Caldwell}}, \citenamefont {{Calitz}}, \citenamefont
  {{Cassan}}, \citenamefont {{Cole}}, \citenamefont {{Cook}}, \citenamefont
  {{Coutures}}, \citenamefont {{Dieters}}, \citenamefont {{Dominis Prester}},
  \citenamefont {{Donatowicz}}, \citenamefont {{Fouqu{\'e}}}, \citenamefont
  {{Hill}}, \citenamefont {{Hoffman}}, \citenamefont {{Jablonski}},
  \citenamefont {{Kane}}, \citenamefont {{Kains}}, \citenamefont {{Kubas}},
  \citenamefont {{Marquette}}, \citenamefont {{Martin}}, \citenamefont
  {{Martioli}}, \citenamefont {{Meintjes}}, \citenamefont {{Menzies}},
  \citenamefont {{Pedretti}}, \citenamefont {{Pollard}}, \citenamefont
  {{Sahu}}, \citenamefont {{Vinter}}, \citenamefont {{Wambsganss}},
  \citenamefont {{Watson}}, \citenamefont {{Williams}}, \citenamefont {{Zub}},
  \citenamefont {{PLANET Collaboration}}, \citenamefont {{Allan}},
  \citenamefont {{Bode}}, \citenamefont {{Bramich}}, \citenamefont
  {{Burgdorf}}, \citenamefont {{Clay}}, \citenamefont {{Fraser}}, \citenamefont
  {{Hawkins}}, \citenamefont {{Horne}}, \citenamefont {{Kerins}}, \citenamefont
  {{Lister}}, \citenamefont {{Mottram}}, \citenamefont {{Saunders}},
  \citenamefont {{Snodgrass}}, \citenamefont {{Steele}}, \citenamefont
  {{Tsapras}}, \citenamefont {{RoboNet Collaboration}}, \citenamefont
  {{J{\o}rgensen}}, \citenamefont {{Anguita}}, \citenamefont {{Bozza}},
  \citenamefont {{Calchi Novati}}, \citenamefont {{Harps{\o}e}}, \citenamefont
  {{Hinse}}, \citenamefont {{Hundertmark}}, \citenamefont {{Kj{\ae}rgaard}},
  \citenamefont {{Liebig}}, \citenamefont {{Mancini}}, \citenamefont {{Masi}},
  \citenamefont {{Mathiasen}}, \citenamefont {{Rahvar}}, \citenamefont
  {{Ricci}}, \citenamefont {{Scarpetta}}, \citenamefont {{Southworth}},
  \citenamefont {{Surdej}}, \citenamefont {{Th{\"o}ne}},\ and\ \citenamefont
  {{MiNDSTEp Consortium}}}]{Gould:2008zu}%
  \BibitemOpen
  \bibfield  {author} {\bibinfo {author} {\bibfnamefont {A.}~\bibnamefont
  {{Gould}}}, \bibinfo {author} {\bibfnamefont {S.}~\bibnamefont {{Dong}}},
  \bibinfo {author} {\bibfnamefont {B.~S.}\ \bibnamefont {{Gaudi}}}, \bibinfo
  {author} {\bibfnamefont {A.}~\bibnamefont {{Udalski}}}, \bibinfo {author}
  {\bibfnamefont {I.~A.}\ \bibnamefont {{Bond}}}, \bibinfo {author}
  {\bibfnamefont {J.}~\bibnamefont {{Greenhill}}}, \bibinfo {author}
  {\bibfnamefont {R.~A.}\ \bibnamefont {{Street}}}, \bibinfo {author}
  {\bibfnamefont {M.}~\bibnamefont {{Dominik}}}, \bibinfo {author}
  {\bibfnamefont {T.}~\bibnamefont {{Sumi}}}, \bibinfo {author} {\bibfnamefont
  {M.~K.}\ \bibnamefont {{Szyma{\'n}ski}}}, \bibinfo {author} {\bibfnamefont
  {C.}~\bibnamefont {{Han}}}, \bibinfo {author} {\bibfnamefont
  {W.}~\bibnamefont {{Allen}}}, \bibinfo {author} {\bibfnamefont
  {G.}~\bibnamefont {{Bolt}}}, \bibinfo {author} {\bibfnamefont
  {M.}~\bibnamefont {{Bos}}}, \bibinfo {author} {\bibfnamefont {G.~W.}\
  \bibnamefont {{Christie}}}, \bibinfo {author} {\bibfnamefont {D.~L.}\
  \bibnamefont {{DePoy}}}, \bibinfo {author} {\bibfnamefont {J.}~\bibnamefont
  {{Drummond}}}, \bibinfo {author} {\bibfnamefont {J.~D.}\ \bibnamefont
  {{Eastman}}}, \bibinfo {author} {\bibfnamefont {A.}~\bibnamefont
  {{Gal-Yam}}}, \bibinfo {author} {\bibfnamefont {D.}~\bibnamefont
  {{Higgins}}}, \bibinfo {author} {\bibfnamefont {J.}~\bibnamefont
  {{Janczak}}}, \bibinfo {author} {\bibfnamefont {S.}~\bibnamefont {{Kaspi}}},
  \bibinfo {author} {\bibfnamefont {S.}~\bibnamefont {{Koz{\l}owski}}},
  \bibinfo {author} {\bibfnamefont {C.-U.}\ \bibnamefont {{Lee}}}, \bibinfo
  {author} {\bibfnamefont {F.}~\bibnamefont {{Mallia}}}, \bibinfo {author}
  {\bibfnamefont {A.}~\bibnamefont {{Maury}}}, \bibinfo {author} {\bibfnamefont
  {D.}~\bibnamefont {{Maoz}}}, \bibinfo {author} {\bibfnamefont
  {J.}~\bibnamefont {{McCormick}}}, \bibinfo {author} {\bibfnamefont
  {L.~A.~G.}\ \bibnamefont {{Monard}}}, \bibinfo {author} {\bibfnamefont
  {D.}~\bibnamefont {{Moorhouse}}}, \bibinfo {author} {\bibfnamefont
  {N.}~\bibnamefont {{Morgan}}}, \bibinfo {author} {\bibfnamefont
  {T.}~\bibnamefont {{Natusch}}}, \bibinfo {author} {\bibfnamefont {E.~O.}\
  \bibnamefont {{Ofek}}}, \bibinfo {author} {\bibfnamefont {B.-G.}\
  \bibnamefont {{Park}}}, \bibinfo {author} {\bibfnamefont {R.~W.}\
  \bibnamefont {{Pogge}}}, \bibinfo {author} {\bibfnamefont {D.}~\bibnamefont
  {{Polishook}}}, \bibinfo {author} {\bibfnamefont {R.}~\bibnamefont
  {{Santallo}}}, \bibinfo {author} {\bibfnamefont {A.}~\bibnamefont
  {{Shporer}}}, \bibinfo {author} {\bibfnamefont {O.}~\bibnamefont
  {{Spector}}}, \bibinfo {author} {\bibfnamefont {G.}~\bibnamefont
  {{Thornley}}}, \bibinfo {author} {\bibfnamefont {J.~C.}\ \bibnamefont
  {{Yee}}}, \bibinfo {author} {\bibnamefont {{{$\mu$}FUN Collaboration}}},
  \bibinfo {author} {\bibfnamefont {M.}~\bibnamefont {{Kubiak}}}, \bibinfo
  {author} {\bibfnamefont {G.}~\bibnamefont {{Pietrzy{\'n}ski}}}, \bibinfo
  {author} {\bibfnamefont {I.}~\bibnamefont {{Soszy{\'n}ski}}}, \bibinfo
  {author} {\bibfnamefont {O.}~\bibnamefont {{Szewczyk}}}, \bibinfo {author}
  {\bibfnamefont {{\L}.}~\bibnamefont {{Wyrzykowski}}}, \bibinfo {author}
  {\bibfnamefont {K.}~\bibnamefont {{Ulaczyk}}}, \bibinfo {author}
  {\bibfnamefont {R.}~\bibnamefont {{Poleski}}}, \bibinfo {author}
  {\bibnamefont {{OGLE Collaboration}}}, \bibinfo {author} {\bibfnamefont
  {F.}~\bibnamefont {{Abe}}}, \bibinfo {author} {\bibfnamefont {D.~P.}\
  \bibnamefont {{Bennett}}}, \bibinfo {author} {\bibfnamefont {C.~S.}\
  \bibnamefont {{Botzler}}}, \bibinfo {author} {\bibfnamefont {D.}~\bibnamefont
  {{Douchin}}}, \bibinfo {author} {\bibfnamefont {M.}~\bibnamefont
  {{Freeman}}}, \bibinfo {author} {\bibfnamefont {A.}~\bibnamefont {{Fukui}}},
  \bibinfo {author} {\bibfnamefont {K.}~\bibnamefont {{Furusawa}}}, \bibinfo
  {author} {\bibfnamefont {J.~B.}\ \bibnamefont {{Hearnshaw}}}, \bibinfo
  {author} {\bibfnamefont {S.}~\bibnamefont {{Hosaka}}}, \bibinfo {author}
  {\bibfnamefont {Y.}~\bibnamefont {{Itow}}}, \bibinfo {author} {\bibfnamefont
  {K.}~\bibnamefont {{Kamiya}}}, \bibinfo {author} {\bibfnamefont {P.~M.}\
  \bibnamefont {{Kilmartin}}}, \bibinfo {author} {\bibfnamefont
  {A.}~\bibnamefont {{Korpela}}}, \bibinfo {author} {\bibfnamefont
  {W.}~\bibnamefont {{Lin}}}, \bibinfo {author} {\bibfnamefont {C.~H.}\
  \bibnamefont {{Ling}}}, \bibinfo {author} {\bibfnamefont {S.}~\bibnamefont
  {{Makita}}}, \bibinfo {author} {\bibfnamefont {K.}~\bibnamefont {{Masuda}}},
  \bibinfo {author} {\bibfnamefont {Y.}~\bibnamefont {{Matsubara}}}, \bibinfo
  {author} {\bibfnamefont {N.}~\bibnamefont {{Miyake}}}, \bibinfo {author}
  {\bibfnamefont {Y.}~\bibnamefont {{Muraki}}}, \bibinfo {author}
  {\bibfnamefont {M.}~\bibnamefont {{Nagaya}}}, \bibinfo {author}
  {\bibfnamefont {K.}~\bibnamefont {{Nishimoto}}}, \bibinfo {author}
  {\bibfnamefont {K.}~\bibnamefont {{Ohnishi}}}, \bibinfo {author}
  {\bibfnamefont {T.}~\bibnamefont {{Okumura}}}, \bibinfo {author}
  {\bibfnamefont {Y.~C.}\ \bibnamefont {{Perrott}}}, \bibinfo {author}
  {\bibfnamefont {L.}~\bibnamefont {{Philpott}}}, \bibinfo {author}
  {\bibfnamefont {N.}~\bibnamefont {{Rattenbury}}}, \bibinfo {author}
  {\bibfnamefont {T.}~\bibnamefont {{Saito}}}, \bibinfo {author} {\bibfnamefont
  {T.}~\bibnamefont {{Sako}}}, \bibinfo {author} {\bibfnamefont {D.~J.}\
  \bibnamefont {{Sullivan}}}, \bibinfo {author} {\bibfnamefont {W.~L.}\
  \bibnamefont {{Sweatman}}}, \bibinfo {author} {\bibfnamefont {P.~J.}\
  \bibnamefont {{Tristram}}}, \bibinfo {author} {\bibfnamefont
  {E.}~\bibnamefont {{von Seggern}}}, \bibinfo {author} {\bibfnamefont
  {P.~C.~M.}\ \bibnamefont {{Yock}}}, \bibinfo {author} {\bibnamefont {{MOA
  Collaboration}}}, \bibinfo {author} {\bibfnamefont {M.}~\bibnamefont
  {{Albrow}}}, \bibinfo {author} {\bibfnamefont {V.}~\bibnamefont {{Batista}}},
  \bibinfo {author} {\bibfnamefont {J.~P.}\ \bibnamefont {{Beaulieu}}},
  \bibinfo {author} {\bibfnamefont {S.}~\bibnamefont {{Brillant}}}, \bibinfo
  {author} {\bibfnamefont {J.}~\bibnamefont {{Caldwell}}}, \bibinfo {author}
  {\bibfnamefont {J.~J.}\ \bibnamefont {{Calitz}}}, \bibinfo {author}
  {\bibfnamefont {A.}~\bibnamefont {{Cassan}}}, \bibinfo {author}
  {\bibfnamefont {A.}~\bibnamefont {{Cole}}}, \bibinfo {author} {\bibfnamefont
  {K.}~\bibnamefont {{Cook}}}, \bibinfo {author} {\bibfnamefont
  {C.}~\bibnamefont {{Coutures}}}, \bibinfo {author} {\bibfnamefont
  {S.}~\bibnamefont {{Dieters}}}, \bibinfo {author} {\bibfnamefont
  {D.}~\bibnamefont {{Dominis Prester}}}, \bibinfo {author} {\bibfnamefont
  {J.}~\bibnamefont {{Donatowicz}}}, \bibinfo {author} {\bibfnamefont
  {P.}~\bibnamefont {{Fouqu{\'e}}}}, \bibinfo {author} {\bibfnamefont
  {K.}~\bibnamefont {{Hill}}}, \bibinfo {author} {\bibfnamefont
  {M.}~\bibnamefont {{Hoffman}}}, \bibinfo {author} {\bibfnamefont
  {F.}~\bibnamefont {{Jablonski}}}, \bibinfo {author} {\bibfnamefont {S.~R.}\
  \bibnamefont {{Kane}}}, \bibinfo {author} {\bibfnamefont {N.}~\bibnamefont
  {{Kains}}}, \bibinfo {author} {\bibfnamefont {D.}~\bibnamefont {{Kubas}}},
  \bibinfo {author} {\bibfnamefont {J.-B.}\ \bibnamefont {{Marquette}}},
  \bibinfo {author} {\bibfnamefont {R.}~\bibnamefont {{Martin}}}, \bibinfo
  {author} {\bibfnamefont {E.}~\bibnamefont {{Martioli}}}, \bibinfo {author}
  {\bibfnamefont {P.}~\bibnamefont {{Meintjes}}}, \bibinfo {author}
  {\bibfnamefont {J.}~\bibnamefont {{Menzies}}}, \bibinfo {author}
  {\bibfnamefont {E.}~\bibnamefont {{Pedretti}}}, \bibinfo {author}
  {\bibfnamefont {K.}~\bibnamefont {{Pollard}}}, \bibinfo {author}
  {\bibfnamefont {K.~C.}\ \bibnamefont {{Sahu}}}, \bibinfo {author}
  {\bibfnamefont {C.}~\bibnamefont {{Vinter}}}, \bibinfo {author}
  {\bibfnamefont {J.}~\bibnamefont {{Wambsganss}}}, \bibinfo {author}
  {\bibfnamefont {R.}~\bibnamefont {{Watson}}}, \bibinfo {author}
  {\bibfnamefont {A.}~\bibnamefont {{Williams}}}, \bibinfo {author}
  {\bibfnamefont {M.}~\bibnamefont {{Zub}}}, \bibinfo {author} {\bibnamefont
  {{PLANET Collaboration}}}, \bibinfo {author} {\bibfnamefont {A.}~\bibnamefont
  {{Allan}}}, \bibinfo {author} {\bibfnamefont {M.~F.}\ \bibnamefont {{Bode}}},
  \bibinfo {author} {\bibfnamefont {D.~M.}\ \bibnamefont {{Bramich}}}, \bibinfo
  {author} {\bibfnamefont {M.~J.}\ \bibnamefont {{Burgdorf}}}, \bibinfo
  {author} {\bibfnamefont {N.}~\bibnamefont {{Clay}}}, \bibinfo {author}
  {\bibfnamefont {S.}~\bibnamefont {{Fraser}}}, \bibinfo {author}
  {\bibfnamefont {E.}~\bibnamefont {{Hawkins}}}, \bibinfo {author}
  {\bibfnamefont {K.}~\bibnamefont {{Horne}}}, \bibinfo {author} {\bibfnamefont
  {E.}~\bibnamefont {{Kerins}}}, \bibinfo {author} {\bibfnamefont {T.~A.}\
  \bibnamefont {{Lister}}}, \bibinfo {author} {\bibfnamefont {C.}~\bibnamefont
  {{Mottram}}}, \bibinfo {author} {\bibfnamefont {E.~S.}\ \bibnamefont
  {{Saunders}}}, \bibinfo {author} {\bibfnamefont {C.}~\bibnamefont
  {{Snodgrass}}}, \bibinfo {author} {\bibfnamefont {I.~A.}\ \bibnamefont
  {{Steele}}}, \bibinfo {author} {\bibfnamefont {Y.}~\bibnamefont {{Tsapras}}},
  \bibinfo {author} {\bibnamefont {{RoboNet Collaboration}}}, \bibinfo {author}
  {\bibfnamefont {U.~G.}\ \bibnamefont {{J{\o}rgensen}}}, \bibinfo {author}
  {\bibfnamefont {T.}~\bibnamefont {{Anguita}}}, \bibinfo {author}
  {\bibfnamefont {V.}~\bibnamefont {{Bozza}}}, \bibinfo {author} {\bibfnamefont
  {S.}~\bibnamefont {{Calchi Novati}}}, \bibinfo {author} {\bibfnamefont
  {K.}~\bibnamefont {{Harps{\o}e}}}, \bibinfo {author} {\bibfnamefont {T.~C.}\
  \bibnamefont {{Hinse}}}, \bibinfo {author} {\bibfnamefont {M.}~\bibnamefont
  {{Hundertmark}}}, \bibinfo {author} {\bibfnamefont {P.}~\bibnamefont
  {{Kj{\ae}rgaard}}}, \bibinfo {author} {\bibfnamefont {C.}~\bibnamefont
  {{Liebig}}}, \bibinfo {author} {\bibfnamefont {L.}~\bibnamefont {{Mancini}}},
  \bibinfo {author} {\bibfnamefont {G.}~\bibnamefont {{Masi}}}, \bibinfo
  {author} {\bibfnamefont {M.}~\bibnamefont {{Mathiasen}}}, \bibinfo {author}
  {\bibfnamefont {S.}~\bibnamefont {{Rahvar}}}, \bibinfo {author}
  {\bibfnamefont {D.}~\bibnamefont {{Ricci}}}, \bibinfo {author} {\bibfnamefont
  {G.}~\bibnamefont {{Scarpetta}}}, \bibinfo {author} {\bibfnamefont
  {J.}~\bibnamefont {{Southworth}}}, \bibinfo {author} {\bibfnamefont
  {J.}~\bibnamefont {{Surdej}}}, \bibinfo {author} {\bibfnamefont {C.~C.}\
  \bibnamefont {{Th{\"o}ne}}}, \ and\ \bibinfo {author} {\bibnamefont
  {{MiNDSTEp Consortium}}},\ }\href {\doibase 10.1088/0004-637X/720/2/1073}
  {\bibfield  {journal} {\bibinfo  {journal} {\apj}\ }\textbf {\bibinfo
  {volume} {720}},\ \bibinfo {pages} {1073} (\bibinfo {year} {2010})},\ \Eprint
  {http://arxiv.org/abs/1001.0572} {arXiv:1001.0572 [astro-ph.EP]} \BibitemShut
  {NoStop}%
\bibitem [{\citenamefont {{Batista}}\ \emph {et~al.}(2011)\citenamefont
  {{Batista}}, \citenamefont {{Gould}}, \citenamefont {{Dieters}},
  \citenamefont {{Dong}}, \citenamefont {{Bond}}, \citenamefont {{Beaulieu}},
  \citenamefont {{Maoz}}, \citenamefont {{Monard}}, \citenamefont {{Christie}},
  \citenamefont {{McCormick}}, \citenamefont {{Albrow}}, \citenamefont
  {{Horne}}, \citenamefont {{Tsapras}}, \citenamefont {{Burgdorf}},
  \citenamefont {{Calchi Novati}}, \citenamefont {{Skottfelt}}, \citenamefont
  {{Caldwell}}, \citenamefont {{Koz{\l}owski}}, \citenamefont {{Kubas}},
  \citenamefont {{Gaudi}}, \citenamefont {{Han}}, \citenamefont {{Bennett}},
  \citenamefont {{An}}, \citenamefont {{MOA Collaboration}}, \citenamefont
  {{Abe}}, \citenamefont {{Botzler}}, \citenamefont {{Douchin}}, \citenamefont
  {{Freeman}}, \citenamefont {{Fukui}}, \citenamefont {{Furusawa}},
  \citenamefont {{Hearnshaw}}, \citenamefont {{Hosaka}}, \citenamefont
  {{Itow}}, \citenamefont {{Kamiya}}, \citenamefont {{Kilmartin}},
  \citenamefont {{Korpela}}, \citenamefont {{Lin}}, \citenamefont {{Ling}},
  \citenamefont {{Makita}}, \citenamefont {{Masuda}}, \citenamefont
  {{Matsubara}}, \citenamefont {{Miyake}}, \citenamefont {{Muraki}},
  \citenamefont {{Nagaya}}, \citenamefont {{Nishimoto}}, \citenamefont
  {{Ohnishi}}, \citenamefont {{Okumura}}, \citenamefont {{Perrott}},
  \citenamefont {{Rattenbury}}, \citenamefont {{Saito}}, \citenamefont
  {{Sullivan}}, \citenamefont {{Sumi}}, \citenamefont {{Sweatman}},
  \citenamefont {{Tristram}}, \citenamefont {{von Seggern}}, \citenamefont
  {{Yock}}, \citenamefont {{PLANET Collaboration}}, \citenamefont {{Brillant}},
  \citenamefont {{Calitz}}, \citenamefont {{Cassan}}, \citenamefont {{Cole}},
  \citenamefont {{Cook}}, \citenamefont {{Coutures}}, \citenamefont {{Dominis
  Prester}}, \citenamefont {{Donatowicz}}, \citenamefont {{Greenhill}},
  \citenamefont {{Hoffman}}, \citenamefont {{Jablonski}}, \citenamefont
  {{Kane}}, \citenamefont {{Kains}}, \citenamefont {{Marquette}}, \citenamefont
  {{Martin}}, \citenamefont {{Martioli}}, \citenamefont {{Meintjes}},
  \citenamefont {{Menzies}}, \citenamefont {{Pedretti}}, \citenamefont
  {{Pollard}}, \citenamefont {{Sahu}}, \citenamefont {{Vinter}}, \citenamefont
  {{Wambsganss}}, \citenamefont {{Watson}}, \citenamefont {{Williams}},
  \citenamefont {{Zub}}, \citenamefont {{FUN Collaboration}}, \citenamefont
  {{Allen}}, \citenamefont {{Bolt}}, \citenamefont {{Bos}}, \citenamefont
  {{DePoy}}, \citenamefont {{Drummond}}, \citenamefont {{Eastman}},
  \citenamefont {{Gal-Yam}}, \citenamefont {{Gorbikov}}, \citenamefont
  {{Higgins}}, \citenamefont {{Janczak}}, \citenamefont {{Kaspi}},
  \citenamefont {{Lee}}, \citenamefont {{Mallia}}, \citenamefont {{Maury}},
  \citenamefont {{Monard}}, \citenamefont {{Moorhouse}}, \citenamefont
  {{Morgan}}, \citenamefont {{Natusch}}, \citenamefont {{Ofek}}, \citenamefont
  {{Park}}, \citenamefont {{Pogge}}, \citenamefont {{Polishook}}, \citenamefont
  {{Santallo}}, \citenamefont {{Shporer}}, \citenamefont {{Spector}},
  \citenamefont {{Thornley}}, \citenamefont {{Yee}}, \citenamefont {{MiNDSTEp
  Consortium}}, \citenamefont {{Bozza}}, \citenamefont {{Browne}},
  \citenamefont {{Dominik}}, \citenamefont {{Dreizler}}, \citenamefont
  {{Finet}}, \citenamefont {{Glitrup}}, \citenamefont {{Grundahl}},
  \citenamefont {{Harps{\o}e}}, \citenamefont {{Hessman}}, \citenamefont
  {{Hinse}}, \citenamefont {{Hundertmark}}, \citenamefont {{J{\o}rgensen}},
  \citenamefont {{Liebig}}, \citenamefont {{Maier}}, \citenamefont {{Mancini}},
  \citenamefont {{Mathiasen}}, \citenamefont {{Rahvar}}, \citenamefont
  {{Ricci}}, \citenamefont {{Scarpetta}}, \citenamefont {{Southworth}},
  \citenamefont {{Surdej}}, \citenamefont {{Zimmer}}, \citenamefont {{RoboNet
  Collaboration}}, \citenamefont {{Allan}}, \citenamefont {{Bramich}},
  \citenamefont {{Snodgrass}}, \citenamefont {{Steele}},\ and\ \citenamefont
  {{Street}}}]{batista}%
  \BibitemOpen
  \bibfield  {author} {\bibinfo {author} {\bibfnamefont {V.}~\bibnamefont
  {{Batista}}}, \bibinfo {author} {\bibfnamefont {A.}~\bibnamefont {{Gould}}},
  \bibinfo {author} {\bibfnamefont {S.}~\bibnamefont {{Dieters}}}, \bibinfo
  {author} {\bibfnamefont {S.}~\bibnamefont {{Dong}}}, \bibinfo {author}
  {\bibfnamefont {I.}~\bibnamefont {{Bond}}}, \bibinfo {author} {\bibfnamefont
  {J.~P.}\ \bibnamefont {{Beaulieu}}}, \bibinfo {author} {\bibfnamefont
  {D.}~\bibnamefont {{Maoz}}}, \bibinfo {author} {\bibfnamefont
  {B.}~\bibnamefont {{Monard}}}, \bibinfo {author} {\bibfnamefont {G.~W.}\
  \bibnamefont {{Christie}}}, \bibinfo {author} {\bibfnamefont
  {J.}~\bibnamefont {{McCormick}}}, \bibinfo {author} {\bibfnamefont {M.~D.}\
  \bibnamefont {{Albrow}}}, \bibinfo {author} {\bibfnamefont {K.}~\bibnamefont
  {{Horne}}}, \bibinfo {author} {\bibfnamefont {Y.}~\bibnamefont {{Tsapras}}},
  \bibinfo {author} {\bibfnamefont {M.~J.}\ \bibnamefont {{Burgdorf}}},
  \bibinfo {author} {\bibfnamefont {S.}~\bibnamefont {{Calchi Novati}}},
  \bibinfo {author} {\bibfnamefont {J.}~\bibnamefont {{Skottfelt}}}, \bibinfo
  {author} {\bibfnamefont {J.}~\bibnamefont {{Caldwell}}}, \bibinfo {author}
  {\bibfnamefont {S.}~\bibnamefont {{Koz{\l}owski}}}, \bibinfo {author}
  {\bibfnamefont {D.}~\bibnamefont {{Kubas}}}, \bibinfo {author} {\bibfnamefont
  {B.~S.}\ \bibnamefont {{Gaudi}}}, \bibinfo {author} {\bibfnamefont
  {C.}~\bibnamefont {{Han}}}, \bibinfo {author} {\bibfnamefont {D.~P.}\
  \bibnamefont {{Bennett}}}, \bibinfo {author} {\bibfnamefont {J.}~\bibnamefont
  {{An}}}, \bibinfo {author} {\bibnamefont {{MOA Collaboration}}}, \bibinfo
  {author} {\bibfnamefont {F.}~\bibnamefont {{Abe}}}, \bibinfo {author}
  {\bibfnamefont {C.~S.}\ \bibnamefont {{Botzler}}}, \bibinfo {author}
  {\bibfnamefont {D.}~\bibnamefont {{Douchin}}}, \bibinfo {author}
  {\bibfnamefont {M.}~\bibnamefont {{Freeman}}}, \bibinfo {author}
  {\bibfnamefont {A.}~\bibnamefont {{Fukui}}}, \bibinfo {author} {\bibfnamefont
  {K.}~\bibnamefont {{Furusawa}}}, \bibinfo {author} {\bibfnamefont {J.~B.}\
  \bibnamefont {{Hearnshaw}}}, \bibinfo {author} {\bibfnamefont
  {S.}~\bibnamefont {{Hosaka}}}, \bibinfo {author} {\bibfnamefont
  {Y.}~\bibnamefont {{Itow}}}, \bibinfo {author} {\bibfnamefont
  {K.}~\bibnamefont {{Kamiya}}}, \bibinfo {author} {\bibfnamefont {P.~M.}\
  \bibnamefont {{Kilmartin}}}, \bibinfo {author} {\bibfnamefont
  {A.}~\bibnamefont {{Korpela}}}, \bibinfo {author} {\bibfnamefont
  {W.}~\bibnamefont {{Lin}}}, \bibinfo {author} {\bibfnamefont {C.~H.}\
  \bibnamefont {{Ling}}}, \bibinfo {author} {\bibfnamefont {B.~S.}\
  \bibnamefont {{Makita}}}, \bibinfo {author} {\bibfnamefont {K.}~\bibnamefont
  {{Masuda}}}, \bibinfo {author} {\bibfnamefont {Y.}~\bibnamefont
  {{Matsubara}}}, \bibinfo {author} {\bibfnamefont {N.}~\bibnamefont
  {{Miyake}}}, \bibinfo {author} {\bibfnamefont {Y.}~\bibnamefont {{Muraki}}},
  \bibinfo {author} {\bibfnamefont {M.}~\bibnamefont {{Nagaya}}}, \bibinfo
  {author} {\bibfnamefont {K.}~\bibnamefont {{Nishimoto}}}, \bibinfo {author}
  {\bibfnamefont {K.}~\bibnamefont {{Ohnishi}}}, \bibinfo {author}
  {\bibfnamefont {T.}~\bibnamefont {{Okumura}}}, \bibinfo {author}
  {\bibfnamefont {Y.~C.}\ \bibnamefont {{Perrott}}}, \bibinfo {author}
  {\bibfnamefont {N.}~\bibnamefont {{Rattenbury}}}, \bibinfo {author}
  {\bibfnamefont {T.}~\bibnamefont {{Saito}}}, \bibinfo {author} {\bibfnamefont
  {D.~J.}\ \bibnamefont {{Sullivan}}}, \bibinfo {author} {\bibfnamefont
  {T.}~\bibnamefont {{Sumi}}}, \bibinfo {author} {\bibfnamefont {W.~L.}\
  \bibnamefont {{Sweatman}}}, \bibinfo {author} {\bibfnamefont {P.~J.}\
  \bibnamefont {{Tristram}}}, \bibinfo {author} {\bibfnamefont
  {E.}~\bibnamefont {{von Seggern}}}, \bibinfo {author} {\bibfnamefont
  {P.~C.~M.}\ \bibnamefont {{Yock}}}, \bibinfo {author} {\bibnamefont {{PLANET
  Collaboration}}}, \bibinfo {author} {\bibfnamefont {S.}~\bibnamefont
  {{Brillant}}}, \bibinfo {author} {\bibfnamefont {J.~J.}\ \bibnamefont
  {{Calitz}}}, \bibinfo {author} {\bibfnamefont {A.}~\bibnamefont {{Cassan}}},
  \bibinfo {author} {\bibfnamefont {A.}~\bibnamefont {{Cole}}}, \bibinfo
  {author} {\bibfnamefont {K.}~\bibnamefont {{Cook}}}, \bibinfo {author}
  {\bibfnamefont {C.}~\bibnamefont {{Coutures}}}, \bibinfo {author}
  {\bibfnamefont {D.}~\bibnamefont {{Dominis Prester}}}, \bibinfo {author}
  {\bibfnamefont {J.}~\bibnamefont {{Donatowicz}}}, \bibinfo {author}
  {\bibfnamefont {J.}~\bibnamefont {{Greenhill}}}, \bibinfo {author}
  {\bibfnamefont {M.}~\bibnamefont {{Hoffman}}}, \bibinfo {author}
  {\bibfnamefont {F.}~\bibnamefont {{Jablonski}}}, \bibinfo {author}
  {\bibfnamefont {S.~R.}\ \bibnamefont {{Kane}}}, \bibinfo {author}
  {\bibfnamefont {N.}~\bibnamefont {{Kains}}}, \bibinfo {author} {\bibfnamefont
  {J.-B.}\ \bibnamefont {{Marquette}}}, \bibinfo {author} {\bibfnamefont
  {R.}~\bibnamefont {{Martin}}}, \bibinfo {author} {\bibfnamefont
  {E.}~\bibnamefont {{Martioli}}}, \bibinfo {author} {\bibfnamefont
  {P.}~\bibnamefont {{Meintjes}}}, \bibinfo {author} {\bibfnamefont
  {J.}~\bibnamefont {{Menzies}}}, \bibinfo {author} {\bibfnamefont
  {E.}~\bibnamefont {{Pedretti}}}, \bibinfo {author} {\bibfnamefont
  {K.}~\bibnamefont {{Pollard}}}, \bibinfo {author} {\bibfnamefont {K.~C.}\
  \bibnamefont {{Sahu}}}, \bibinfo {author} {\bibfnamefont {C.}~\bibnamefont
  {{Vinter}}}, \bibinfo {author} {\bibfnamefont {J.}~\bibnamefont
  {{Wambsganss}}}, \bibinfo {author} {\bibfnamefont {R.}~\bibnamefont
  {{Watson}}}, \bibinfo {author} {\bibfnamefont {A.}~\bibnamefont
  {{Williams}}}, \bibinfo {author} {\bibfnamefont {M.}~\bibnamefont {{Zub}}},
  \bibinfo {author} {\bibnamefont {{FUN Collaboration}}}, \bibinfo {author}
  {\bibfnamefont {W.}~\bibnamefont {{Allen}}}, \bibinfo {author} {\bibfnamefont
  {G.}~\bibnamefont {{Bolt}}}, \bibinfo {author} {\bibfnamefont
  {M.}~\bibnamefont {{Bos}}}, \bibinfo {author} {\bibfnamefont {D.~L.}\
  \bibnamefont {{DePoy}}}, \bibinfo {author} {\bibfnamefont {J.}~\bibnamefont
  {{Drummond}}}, \bibinfo {author} {\bibfnamefont {J.~D.}\ \bibnamefont
  {{Eastman}}}, \bibinfo {author} {\bibfnamefont {A.}~\bibnamefont
  {{Gal-Yam}}}, \bibinfo {author} {\bibfnamefont {E.}~\bibnamefont
  {{Gorbikov}}}, \bibinfo {author} {\bibfnamefont {D.}~\bibnamefont
  {{Higgins}}}, \bibinfo {author} {\bibfnamefont {J.}~\bibnamefont
  {{Janczak}}}, \bibinfo {author} {\bibfnamefont {S.}~\bibnamefont {{Kaspi}}},
  \bibinfo {author} {\bibfnamefont {C.-U.}\ \bibnamefont {{Lee}}}, \bibinfo
  {author} {\bibfnamefont {F.}~\bibnamefont {{Mallia}}}, \bibinfo {author}
  {\bibfnamefont {A.}~\bibnamefont {{Maury}}}, \bibinfo {author} {\bibfnamefont
  {L.~A.~G.}\ \bibnamefont {{Monard}}}, \bibinfo {author} {\bibfnamefont
  {D.}~\bibnamefont {{Moorhouse}}}, \bibinfo {author} {\bibfnamefont
  {N.}~\bibnamefont {{Morgan}}}, \bibinfo {author} {\bibfnamefont
  {T.}~\bibnamefont {{Natusch}}}, \bibinfo {author} {\bibfnamefont {E.~O.}\
  \bibnamefont {{Ofek}}}, \bibinfo {author} {\bibfnamefont {B.-G.}\
  \bibnamefont {{Park}}}, \bibinfo {author} {\bibfnamefont {R.~W.}\
  \bibnamefont {{Pogge}}}, \bibinfo {author} {\bibfnamefont {D.}~\bibnamefont
  {{Polishook}}}, \bibinfo {author} {\bibfnamefont {R.}~\bibnamefont
  {{Santallo}}}, \bibinfo {author} {\bibfnamefont {A.}~\bibnamefont
  {{Shporer}}}, \bibinfo {author} {\bibfnamefont {O.}~\bibnamefont
  {{Spector}}}, \bibinfo {author} {\bibfnamefont {G.}~\bibnamefont
  {{Thornley}}}, \bibinfo {author} {\bibfnamefont {J.~C.}\ \bibnamefont
  {{Yee}}}, \bibinfo {author} {\bibnamefont {{MiNDSTEp Consortium}}}, \bibinfo
  {author} {\bibfnamefont {V.}~\bibnamefont {{Bozza}}}, \bibinfo {author}
  {\bibfnamefont {P.}~\bibnamefont {{Browne}}}, \bibinfo {author}
  {\bibfnamefont {M.}~\bibnamefont {{Dominik}}}, \bibinfo {author}
  {\bibfnamefont {S.}~\bibnamefont {{Dreizler}}}, \bibinfo {author}
  {\bibfnamefont {F.}~\bibnamefont {{Finet}}}, \bibinfo {author} {\bibfnamefont
  {M.}~\bibnamefont {{Glitrup}}}, \bibinfo {author} {\bibfnamefont
  {F.}~\bibnamefont {{Grundahl}}}, \bibinfo {author} {\bibfnamefont
  {K.}~\bibnamefont {{Harps{\o}e}}}, \bibinfo {author} {\bibfnamefont {F.~V.}\
  \bibnamefont {{Hessman}}}, \bibinfo {author} {\bibfnamefont {T.~C.}\
  \bibnamefont {{Hinse}}}, \bibinfo {author} {\bibfnamefont {M.}~\bibnamefont
  {{Hundertmark}}}, \bibinfo {author} {\bibfnamefont {U.~G.}\ \bibnamefont
  {{J{\o}rgensen}}}, \bibinfo {author} {\bibfnamefont {C.}~\bibnamefont
  {{Liebig}}}, \bibinfo {author} {\bibfnamefont {G.}~\bibnamefont {{Maier}}},
  \bibinfo {author} {\bibfnamefont {L.}~\bibnamefont {{Mancini}}}, \bibinfo
  {author} {\bibfnamefont {M.}~\bibnamefont {{Mathiasen}}}, \bibinfo {author}
  {\bibfnamefont {S.}~\bibnamefont {{Rahvar}}}, \bibinfo {author}
  {\bibfnamefont {D.}~\bibnamefont {{Ricci}}}, \bibinfo {author} {\bibfnamefont
  {G.}~\bibnamefont {{Scarpetta}}}, \bibinfo {author} {\bibfnamefont
  {J.}~\bibnamefont {{Southworth}}}, \bibinfo {author} {\bibfnamefont
  {J.}~\bibnamefont {{Surdej}}}, \bibinfo {author} {\bibfnamefont
  {F.}~\bibnamefont {{Zimmer}}}, \bibinfo {author} {\bibnamefont {{RoboNet
  Collaboration}}}, \bibinfo {author} {\bibfnamefont {A.}~\bibnamefont
  {{Allan}}}, \bibinfo {author} {\bibfnamefont {D.~M.}\ \bibnamefont
  {{Bramich}}}, \bibinfo {author} {\bibfnamefont {C.}~\bibnamefont
  {{Snodgrass}}}, \bibinfo {author} {\bibfnamefont {I.~A.}\ \bibnamefont
  {{Steele}}}, \ and\ \bibinfo {author} {\bibfnamefont {R.~A.}\ \bibnamefont
  {{Street}}},\ }\href {\doibase 10.1051/0004-6361/201016111} {\bibfield
  {journal} {\bibinfo  {journal} {\aap}\ }\textbf {\bibinfo {volume} {529}},\
  \bibinfo {eid} {A102} (\bibinfo {year} {2011})},\ \Eprint
  {http://arxiv.org/abs/1102.0558} {arXiv:1102.0558 [astro-ph.EP]} \BibitemShut
  {NoStop}%
\bibitem [{\citenamefont {{Muraki}}\ \emph {et~al.}(2011)\citenamefont
  {{Muraki}}, \citenamefont {{Han}}, \citenamefont {{Bennett}}, \citenamefont
  {{Suzuki}}, \citenamefont {{Monard}}, \citenamefont {{Street}}, \citenamefont
  {{Jorgensen}}, \citenamefont {{Kundurthy}}, \citenamefont {{Skowron}},
  \citenamefont {{Becker}}, \citenamefont {{Albrow}}, \citenamefont
  {{Fouqu{\'e}}}, \citenamefont {{Heyrovsk{\'y}}}, \citenamefont {{Barry}},
  \citenamefont {{Beaulieu}}, \citenamefont {{Wellnitz}}, \citenamefont
  {{Bond}}, \citenamefont {{Sumi}}, \citenamefont {{Dong}}, \citenamefont
  {{Gaudi}}, \citenamefont {{Bramich}}, \citenamefont {{Dominik}},
  \citenamefont {{Abe}}, \citenamefont {{Botzler}}, \citenamefont {{Freeman}},
  \citenamefont {{Fukui}}, \citenamefont {{Furusawa}}, \citenamefont
  {{Hayashi}}, \citenamefont {{Hearnshaw}}, \citenamefont {{Hosaka}},
  \citenamefont {{Itow}}, \citenamefont {{Kamiya}}, \citenamefont {{Korpela}},
  \citenamefont {{Kilmartin}}, \citenamefont {{Lin}}, \citenamefont {{Ling}},
  \citenamefont {{Makita}}, \citenamefont {{Masuda}}, \citenamefont
  {{Matsubara}}, \citenamefont {{Miyake}}, \citenamefont {{Nishimoto}},
  \citenamefont {{Ohnishi}}, \citenamefont {{Perrott}}, \citenamefont
  {{Rattenbury}}, \citenamefont {{Saito}}, \citenamefont {{Skuljan}},
  \citenamefont {{Sullivan}}, \citenamefont {{Sweatman}}, \citenamefont
  {{Tristram}}, \citenamefont {{Wada}}, \citenamefont {{Yock}}, \citenamefont
  {{MOA Collaboration}}, \citenamefont {{Christie}}, \citenamefont {{DePoy}},
  \citenamefont {{Gorbikov}}, \citenamefont {{Gould}}, \citenamefont {{Kaspi}},
  \citenamefont {{Lee}}, \citenamefont {{Mallia}}, \citenamefont {{Maoz}},
  \citenamefont {{McCormick}}, \citenamefont {{Moorhouse}}, \citenamefont
  {{Natusch}}, \citenamefont {{Park}}, \citenamefont {{Pogge}}, \citenamefont
  {{Polishook}}, \citenamefont {{Shporer}}, \citenamefont {{Thornley}},
  \citenamefont {{Yee}}, \citenamefont {{{$\mu$}FUN Collaboration}},
  \citenamefont {{Allan}}, \citenamefont {{Browne}}, \citenamefont {{Horne}},
  \citenamefont {{Kains}}, \citenamefont {{Snodgrass}}, \citenamefont
  {{Steele}}, \citenamefont {{Tsapras}}, \citenamefont {{RoboNet
  Collaboration}}, \citenamefont {{Batista}}, \citenamefont {{Bennett}},
  \citenamefont {{Brillant}}, \citenamefont {{Caldwell}}, \citenamefont
  {{Cassan}}, \citenamefont {{Cole}}, \citenamefont {{Corrales}}, \citenamefont
  {{Coutures}}, \citenamefont {{Dieters}}, \citenamefont {{Dominis Prester}},
  \citenamefont {{Donatowicz}}, \citenamefont {{Greenhill}}, \citenamefont
  {{Kubas}}, \citenamefont {{Marquette}}, \citenamefont {{Martin}},
  \citenamefont {{Menzies}}, \citenamefont {{Sahu}}, \citenamefont {{Waldman}},
  \citenamefont {{Williams}}, \citenamefont {{Zub}}, \citenamefont {{PLANET
  Collaboration}}, \citenamefont {{Bourhrous}}, \citenamefont {{Matsuoka}},
  \citenamefont {{Nagayama}}, \citenamefont {{Oi}}, \citenamefont
  {{Randriamanakoto}}, \citenamefont {{IRSF Observers}}, \citenamefont
  {{Bozza}}, \citenamefont {{Burgdorf}}, \citenamefont {{Calchi Novati}},
  \citenamefont {{Dreizler}}, \citenamefont {{Finet}}, \citenamefont
  {{Glitrup}}, \citenamefont {{Harps{\o}e}}, \citenamefont {{Hinse}},
  \citenamefont {{Hundertmark}}, \citenamefont {{Liebig}}, \citenamefont
  {{Maier}}, \citenamefont {{Mancini}}, \citenamefont {{Mathiasen}},
  \citenamefont {{Rahvar}}, \citenamefont {{Ricci}}, \citenamefont
  {{Scarpetta}}, \citenamefont {{Skottfelt}}, \citenamefont {{Surdej}},
  \citenamefont {{Southworth}}, \citenamefont {{Wambsganss}}, \citenamefont
  {{Zimmer}}, \citenamefont {{MiNDSTEp Consortium}}, \citenamefont {{Udalski}},
  \citenamefont {{Poleski}}, \citenamefont {{Wyrzykowski}}, \citenamefont
  {{Ulaczyk}}, \citenamefont {{Szyma{\'n}ski}}, \citenamefont {{Kubiak}},
  \citenamefont {{Pietrzy{\'n}ski}}, \citenamefont {{Soszy{\'n}ski}},\ and\
  \citenamefont {{OGLE Collaboration}}}]{muraki}%
  \BibitemOpen
  \bibfield  {author} {\bibinfo {author} {\bibfnamefont {Y.}~\bibnamefont
  {{Muraki}}}, \bibinfo {author} {\bibfnamefont {C.}~\bibnamefont {{Han}}},
  \bibinfo {author} {\bibfnamefont {D.~P.}\ \bibnamefont {{Bennett}}}, \bibinfo
  {author} {\bibfnamefont {D.}~\bibnamefont {{Suzuki}}}, \bibinfo {author}
  {\bibfnamefont {L.~A.~G.}\ \bibnamefont {{Monard}}}, \bibinfo {author}
  {\bibfnamefont {R.}~\bibnamefont {{Street}}}, \bibinfo {author}
  {\bibfnamefont {U.~G.}\ \bibnamefont {{Jorgensen}}}, \bibinfo {author}
  {\bibfnamefont {P.}~\bibnamefont {{Kundurthy}}}, \bibinfo {author}
  {\bibfnamefont {J.}~\bibnamefont {{Skowron}}}, \bibinfo {author}
  {\bibfnamefont {A.~C.}\ \bibnamefont {{Becker}}}, \bibinfo {author}
  {\bibfnamefont {M.~D.}\ \bibnamefont {{Albrow}}}, \bibinfo {author}
  {\bibfnamefont {P.}~\bibnamefont {{Fouqu{\'e}}}}, \bibinfo {author}
  {\bibfnamefont {D.}~\bibnamefont {{Heyrovsk{\'y}}}}, \bibinfo {author}
  {\bibfnamefont {R.~K.}\ \bibnamefont {{Barry}}}, \bibinfo {author}
  {\bibfnamefont {J.-P.}\ \bibnamefont {{Beaulieu}}}, \bibinfo {author}
  {\bibfnamefont {D.~D.}\ \bibnamefont {{Wellnitz}}}, \bibinfo {author}
  {\bibfnamefont {I.~A.}\ \bibnamefont {{Bond}}}, \bibinfo {author}
  {\bibfnamefont {T.}~\bibnamefont {{Sumi}}}, \bibinfo {author} {\bibfnamefont
  {S.}~\bibnamefont {{Dong}}}, \bibinfo {author} {\bibfnamefont {B.~S.}\
  \bibnamefont {{Gaudi}}}, \bibinfo {author} {\bibfnamefont {D.~M.}\
  \bibnamefont {{Bramich}}}, \bibinfo {author} {\bibfnamefont {M.}~\bibnamefont
  {{Dominik}}}, \bibinfo {author} {\bibfnamefont {F.}~\bibnamefont {{Abe}}},
  \bibinfo {author} {\bibfnamefont {C.~S.}\ \bibnamefont {{Botzler}}}, \bibinfo
  {author} {\bibfnamefont {M.}~\bibnamefont {{Freeman}}}, \bibinfo {author}
  {\bibfnamefont {A.}~\bibnamefont {{Fukui}}}, \bibinfo {author} {\bibfnamefont
  {K.}~\bibnamefont {{Furusawa}}}, \bibinfo {author} {\bibfnamefont
  {F.}~\bibnamefont {{Hayashi}}}, \bibinfo {author} {\bibfnamefont {J.~B.}\
  \bibnamefont {{Hearnshaw}}}, \bibinfo {author} {\bibfnamefont
  {S.}~\bibnamefont {{Hosaka}}}, \bibinfo {author} {\bibfnamefont
  {Y.}~\bibnamefont {{Itow}}}, \bibinfo {author} {\bibfnamefont
  {K.}~\bibnamefont {{Kamiya}}}, \bibinfo {author} {\bibfnamefont {A.~V.}\
  \bibnamefont {{Korpela}}}, \bibinfo {author} {\bibfnamefont {P.~M.}\
  \bibnamefont {{Kilmartin}}}, \bibinfo {author} {\bibfnamefont
  {W.}~\bibnamefont {{Lin}}}, \bibinfo {author} {\bibfnamefont {C.~H.}\
  \bibnamefont {{Ling}}}, \bibinfo {author} {\bibfnamefont {S.}~\bibnamefont
  {{Makita}}}, \bibinfo {author} {\bibfnamefont {K.}~\bibnamefont {{Masuda}}},
  \bibinfo {author} {\bibfnamefont {Y.}~\bibnamefont {{Matsubara}}}, \bibinfo
  {author} {\bibfnamefont {N.}~\bibnamefont {{Miyake}}}, \bibinfo {author}
  {\bibfnamefont {K.}~\bibnamefont {{Nishimoto}}}, \bibinfo {author}
  {\bibfnamefont {K.}~\bibnamefont {{Ohnishi}}}, \bibinfo {author}
  {\bibfnamefont {Y.~C.}\ \bibnamefont {{Perrott}}}, \bibinfo {author}
  {\bibfnamefont {N.~J.}\ \bibnamefont {{Rattenbury}}}, \bibinfo {author}
  {\bibfnamefont {T.}~\bibnamefont {{Saito}}}, \bibinfo {author} {\bibfnamefont
  {L.}~\bibnamefont {{Skuljan}}}, \bibinfo {author} {\bibfnamefont {D.~J.}\
  \bibnamefont {{Sullivan}}}, \bibinfo {author} {\bibfnamefont {W.~L.}\
  \bibnamefont {{Sweatman}}}, \bibinfo {author} {\bibfnamefont {P.~J.}\
  \bibnamefont {{Tristram}}}, \bibinfo {author} {\bibfnamefont
  {K.}~\bibnamefont {{Wada}}}, \bibinfo {author} {\bibfnamefont {P.~C.~M.}\
  \bibnamefont {{Yock}}}, \bibinfo {author} {\bibnamefont {{MOA
  Collaboration}}}, \bibinfo {author} {\bibfnamefont {G.~W.}\ \bibnamefont
  {{Christie}}}, \bibinfo {author} {\bibfnamefont {D.~L.}\ \bibnamefont
  {{DePoy}}}, \bibinfo {author} {\bibfnamefont {E.}~\bibnamefont {{Gorbikov}}},
  \bibinfo {author} {\bibfnamefont {A.}~\bibnamefont {{Gould}}}, \bibinfo
  {author} {\bibfnamefont {S.}~\bibnamefont {{Kaspi}}}, \bibinfo {author}
  {\bibfnamefont {C.-U.}\ \bibnamefont {{Lee}}}, \bibinfo {author}
  {\bibfnamefont {F.}~\bibnamefont {{Mallia}}}, \bibinfo {author}
  {\bibfnamefont {D.}~\bibnamefont {{Maoz}}}, \bibinfo {author} {\bibfnamefont
  {J.}~\bibnamefont {{McCormick}}}, \bibinfo {author} {\bibfnamefont
  {D.}~\bibnamefont {{Moorhouse}}}, \bibinfo {author} {\bibfnamefont
  {T.}~\bibnamefont {{Natusch}}}, \bibinfo {author} {\bibfnamefont {B.-G.}\
  \bibnamefont {{Park}}}, \bibinfo {author} {\bibfnamefont {R.~W.}\
  \bibnamefont {{Pogge}}}, \bibinfo {author} {\bibfnamefont {D.}~\bibnamefont
  {{Polishook}}}, \bibinfo {author} {\bibfnamefont {A.}~\bibnamefont
  {{Shporer}}}, \bibinfo {author} {\bibfnamefont {G.}~\bibnamefont
  {{Thornley}}}, \bibinfo {author} {\bibfnamefont {J.~C.}\ \bibnamefont
  {{Yee}}}, \bibinfo {author} {\bibnamefont {{{$\mu$}FUN Collaboration}}},
  \bibinfo {author} {\bibfnamefont {A.}~\bibnamefont {{Allan}}}, \bibinfo
  {author} {\bibfnamefont {P.}~\bibnamefont {{Browne}}}, \bibinfo {author}
  {\bibfnamefont {K.}~\bibnamefont {{Horne}}}, \bibinfo {author} {\bibfnamefont
  {N.}~\bibnamefont {{Kains}}}, \bibinfo {author} {\bibfnamefont
  {C.}~\bibnamefont {{Snodgrass}}}, \bibinfo {author} {\bibfnamefont
  {I.}~\bibnamefont {{Steele}}}, \bibinfo {author} {\bibfnamefont
  {Y.}~\bibnamefont {{Tsapras}}}, \bibinfo {author} {\bibnamefont {{RoboNet
  Collaboration}}}, \bibinfo {author} {\bibfnamefont {V.}~\bibnamefont
  {{Batista}}}, \bibinfo {author} {\bibfnamefont {C.~S.}\ \bibnamefont
  {{Bennett}}}, \bibinfo {author} {\bibfnamefont {S.}~\bibnamefont
  {{Brillant}}}, \bibinfo {author} {\bibfnamefont {J.~A.~R.}\ \bibnamefont
  {{Caldwell}}}, \bibinfo {author} {\bibfnamefont {A.}~\bibnamefont
  {{Cassan}}}, \bibinfo {author} {\bibfnamefont {A.}~\bibnamefont {{Cole}}},
  \bibinfo {author} {\bibfnamefont {R.}~\bibnamefont {{Corrales}}}, \bibinfo
  {author} {\bibfnamefont {C.}~\bibnamefont {{Coutures}}}, \bibinfo {author}
  {\bibfnamefont {S.}~\bibnamefont {{Dieters}}}, \bibinfo {author}
  {\bibfnamefont {D.}~\bibnamefont {{Dominis Prester}}}, \bibinfo {author}
  {\bibfnamefont {J.}~\bibnamefont {{Donatowicz}}}, \bibinfo {author}
  {\bibfnamefont {J.}~\bibnamefont {{Greenhill}}}, \bibinfo {author}
  {\bibfnamefont {D.}~\bibnamefont {{Kubas}}}, \bibinfo {author} {\bibfnamefont
  {J.-B.}\ \bibnamefont {{Marquette}}}, \bibinfo {author} {\bibfnamefont
  {R.}~\bibnamefont {{Martin}}}, \bibinfo {author} {\bibfnamefont
  {J.}~\bibnamefont {{Menzies}}}, \bibinfo {author} {\bibfnamefont {K.~C.}\
  \bibnamefont {{Sahu}}}, \bibinfo {author} {\bibfnamefont {I.}~\bibnamefont
  {{Waldman}}}, \bibinfo {author} {\bibfnamefont {A.}~\bibnamefont
  {{Williams}}}, \bibinfo {author} {\bibfnamefont {M.}~\bibnamefont {{Zub}}},
  \bibinfo {author} {\bibnamefont {{PLANET Collaboration}}}, \bibinfo {author}
  {\bibfnamefont {H.}~\bibnamefont {{Bourhrous}}}, \bibinfo {author}
  {\bibfnamefont {Y.}~\bibnamefont {{Matsuoka}}}, \bibinfo {author}
  {\bibfnamefont {T.}~\bibnamefont {{Nagayama}}}, \bibinfo {author}
  {\bibfnamefont {N.}~\bibnamefont {{Oi}}}, \bibinfo {author} {\bibfnamefont
  {Z.}~\bibnamefont {{Randriamanakoto}}}, \bibinfo {author} {\bibnamefont
  {{IRSF Observers}}}, \bibinfo {author} {\bibfnamefont {V.}~\bibnamefont
  {{Bozza}}}, \bibinfo {author} {\bibfnamefont {M.~J.}\ \bibnamefont
  {{Burgdorf}}}, \bibinfo {author} {\bibfnamefont {S.}~\bibnamefont {{Calchi
  Novati}}}, \bibinfo {author} {\bibfnamefont {S.}~\bibnamefont {{Dreizler}}},
  \bibinfo {author} {\bibfnamefont {F.}~\bibnamefont {{Finet}}}, \bibinfo
  {author} {\bibfnamefont {M.}~\bibnamefont {{Glitrup}}}, \bibinfo {author}
  {\bibfnamefont {K.}~\bibnamefont {{Harps{\o}e}}}, \bibinfo {author}
  {\bibfnamefont {T.~C.}\ \bibnamefont {{Hinse}}}, \bibinfo {author}
  {\bibfnamefont {M.}~\bibnamefont {{Hundertmark}}}, \bibinfo {author}
  {\bibfnamefont {C.}~\bibnamefont {{Liebig}}}, \bibinfo {author}
  {\bibfnamefont {G.}~\bibnamefont {{Maier}}}, \bibinfo {author} {\bibfnamefont
  {L.}~\bibnamefont {{Mancini}}}, \bibinfo {author} {\bibfnamefont
  {M.}~\bibnamefont {{Mathiasen}}}, \bibinfo {author} {\bibfnamefont
  {S.}~\bibnamefont {{Rahvar}}}, \bibinfo {author} {\bibfnamefont
  {D.}~\bibnamefont {{Ricci}}}, \bibinfo {author} {\bibfnamefont
  {G.}~\bibnamefont {{Scarpetta}}}, \bibinfo {author} {\bibfnamefont
  {J.}~\bibnamefont {{Skottfelt}}}, \bibinfo {author} {\bibfnamefont
  {J.}~\bibnamefont {{Surdej}}}, \bibinfo {author} {\bibfnamefont
  {J.}~\bibnamefont {{Southworth}}}, \bibinfo {author} {\bibfnamefont
  {J.}~\bibnamefont {{Wambsganss}}}, \bibinfo {author} {\bibfnamefont
  {F.}~\bibnamefont {{Zimmer}}}, \bibinfo {author} {\bibnamefont {{MiNDSTEp
  Consortium}}}, \bibinfo {author} {\bibfnamefont {A.}~\bibnamefont
  {{Udalski}}}, \bibinfo {author} {\bibfnamefont {R.}~\bibnamefont
  {{Poleski}}}, \bibinfo {author} {\bibfnamefont {{\L}.}~\bibnamefont
  {{Wyrzykowski}}}, \bibinfo {author} {\bibfnamefont {K.}~\bibnamefont
  {{Ulaczyk}}}, \bibinfo {author} {\bibfnamefont {M.~K.}\ \bibnamefont
  {{Szyma{\'n}ski}}}, \bibinfo {author} {\bibfnamefont {M.}~\bibnamefont
  {{Kubiak}}}, \bibinfo {author} {\bibfnamefont {G.}~\bibnamefont
  {{Pietrzy{\'n}ski}}}, \bibinfo {author} {\bibfnamefont {I.}~\bibnamefont
  {{Soszy{\'n}ski}}}, \ and\ \bibinfo {author} {\bibnamefont {{OGLE
  Collaboration}}},\ }\href {\doibase 10.1088/0004-637X/741/1/22} {\bibfield
  {journal} {\bibinfo  {journal} {\apj}\ }\textbf {\bibinfo {volume} {741}},\
  \bibinfo {eid} {22} (\bibinfo {year} {2011})},\ \Eprint
  {http://arxiv.org/abs/1106.2160} {arXiv:1106.2160 [astro-ph.EP]} \BibitemShut
  {NoStop}%
\bibitem [{\citenamefont {Afonso}(2001)}]{Afonso:2001gh}%
  \BibitemOpen
  \bibfield  {author} {\bibinfo {author} {\bibfnamefont {C.}~\bibnamefont
  {Afonso}},\ }\href {\doibase 10.1051/0004-6361:20011204} {\bibfield
  {journal} {\bibinfo  {journal} {Astron.Astrophys.}\ }\textbf {\bibinfo
  {volume} {378}},\ \bibinfo {pages} {1014} (\bibinfo {year} {2001})},\ \Eprint
  {http://arxiv.org/abs/astro-ph/0106231} {arXiv:astro-ph/0106231 [astro-ph]}
  \BibitemShut {NoStop}%
%%CITATION = ASTRO-PH/0106231;%%
\bibitem [{\citenamefont {Gould}(2001)}]{Gould:2001bg}%
  \BibitemOpen
  \bibfield  {author} {\bibinfo {author} {\bibfnamefont {A.}~\bibnamefont
  {Gould}},\ }\href {\doibase 10.1086/322149} {\bibfield  {journal} {\bibinfo
  {journal} {Publ.Astron.Soc.Pac.}\ }\textbf {\bibinfo {volume} {113}},\
  \bibinfo {pages} {903} (\bibinfo {year} {2001})},\ \Eprint
  {http://arxiv.org/abs/astro-ph/0103516} {arXiv:astro-ph/0103516 [astro-ph]}
  \BibitemShut {NoStop}%
%%CITATION = ASTRO-PH/0103516;%%
\bibitem [{\citenamefont {Abe}\ \emph {et~al.}(2003)\citenamefont {Abe},
  \citenamefont {Bennett}, \citenamefont {Bond}, \citenamefont {Calitz},
  \citenamefont {Claret} \emph {et~al.}}]{Abe:2003xb}%
  \BibitemOpen
  \bibfield  {author} {\bibinfo {author} {\bibfnamefont {F.}~\bibnamefont
  {Abe}}, \bibinfo {author} {\bibfnamefont {D.}~\bibnamefont {Bennett}},
  \bibinfo {author} {\bibfnamefont {I.}~\bibnamefont {Bond}}, \bibinfo {author}
  {\bibfnamefont {J.}~\bibnamefont {Calitz}}, \bibinfo {author} {\bibfnamefont
  {A.}~\bibnamefont {Claret}},  \emph {et~al.},\ }\href {\doibase
  10.1051/0004-6361:20031602} {\bibfield  {journal} {\bibinfo  {journal}
  {Astron.Astrophys.}\ }\textbf {\bibinfo {volume} {411}},\ \bibinfo {pages}
  {L493} (\bibinfo {year} {2003})},\ \Eprint
  {http://arxiv.org/abs/astro-ph/0310410} {arXiv:astro-ph/0310410 [astro-ph]}
  \BibitemShut {NoStop}%
%%CITATION = ASTRO-PH/0310410;%%
\bibitem [{\citenamefont {Fields}\ \emph {et~al.}(2003)\citenamefont {Fields}
  \emph {et~al.}}]{Fields:2003zx}%
  \BibitemOpen
  \bibfield  {author} {\bibinfo {author} {\bibfnamefont {D.~L.}\ \bibnamefont
  {Fields}} \emph {et~al.} (\bibinfo {collaboration} {PLANET Collaboration}),\
  }\href {\doibase 10.1086/378196} {\bibfield  {journal} {\bibinfo  {journal}
  {Astrophys.J.}\ }\textbf {\bibinfo {volume} {596}},\ \bibinfo {pages} {1305}
  (\bibinfo {year} {2003})},\ \Eprint {http://arxiv.org/abs/astro-ph/0303638}
  {arXiv:astro-ph/0303638 [astro-ph]} \BibitemShut {NoStop}%
%%CITATION = ASTRO-PH/0303638;%%
\bibitem [{\citenamefont {{Sajadian}}\ and\ \citenamefont
  {{Rahvar}}(2015)}]{sajadian}%
  \BibitemOpen
  \bibfield  {author} {\bibinfo {author} {\bibfnamefont {S.}~\bibnamefont
  {{Sajadian}}}\ and\ \bibinfo {author} {\bibfnamefont {S.}~\bibnamefont
  {{Rahvar}}},\ }\href {\doibase 10.1093/mnras/stv2297} {\bibfield  {journal}
  {\bibinfo  {journal} {\mnras}\ }\textbf {\bibinfo {volume} {454}},\ \bibinfo
  {pages} {4429} (\bibinfo {year} {2015})},\ \Eprint
  {http://arxiv.org/abs/1510.01856} {arXiv:1510.01856 [astro-ph.SR]}
  \BibitemShut {NoStop}%
\bibitem [{\citenamefont {{Moniez}}\ \emph {et~al.}(2017)\citenamefont
  {{Moniez}}, \citenamefont {{Sajadian}}, \citenamefont {{Karami}},
  \citenamefont {{Rahvar}},\ and\ \citenamefont {{Ansari}}}]{moniez}%
  \BibitemOpen
  \bibfield  {author} {\bibinfo {author} {\bibfnamefont {M.}~\bibnamefont
  {{Moniez}}}, \bibinfo {author} {\bibfnamefont {S.}~\bibnamefont
  {{Sajadian}}}, \bibinfo {author} {\bibfnamefont {M.}~\bibnamefont
  {{Karami}}}, \bibinfo {author} {\bibfnamefont {S.}~\bibnamefont {{Rahvar}}},
  \ and\ \bibinfo {author} {\bibfnamefont {R.}~\bibnamefont {{Ansari}}},\
  }\href {\doibase 10.1051/0004-6361/201730488} {\bibfield  {journal} {\bibinfo
   {journal} {\aap}\ }\textbf {\bibinfo {volume} {604}},\ \bibinfo {eid} {A124}
  (\bibinfo {year} {2017})},\ \Eprint {http://arxiv.org/abs/1701.07006}
  {arXiv:1701.07006} \BibitemShut {NoStop}%
\bibitem [{\citenamefont {{Wambsganss}}(2001)}]{2001ASPC..237..185W}%
  \BibitemOpen
  \bibfield  {author} {\bibinfo {author} {\bibfnamefont {J.}~\bibnamefont
  {{Wambsganss}}},\ }in\ \href@noop {} {\emph {\bibinfo {booktitle}
  {Gravitational Lensing: Recent Progress and Future Go}}},\ \bibinfo {series}
  {Astronomical Society of the Pacific Conference Series}, Vol.\ \bibinfo
  {volume} {237},\ \bibinfo {editor} {edited by\ \bibinfo {editor}
  {\bibfnamefont {T.~G.}\ \bibnamefont {{Brainerd}}}\ and\ \bibinfo {editor}
  {\bibfnamefont {C.~S.}\ \bibnamefont {{Kochanek}}}}\ (\bibinfo {year}
  {2001})\ p.\ \bibinfo {pages} {185}\BibitemShut {NoStop}%
\bibitem [{\citenamefont {{Zackrisson}}\ and\ \citenamefont
  {{Riehm}}(2007{\natexlab{a}})}]{2007A&A...475..453Z}%
  \BibitemOpen
  \bibfield  {author} {\bibinfo {author} {\bibfnamefont {E.}~\bibnamefont
  {{Zackrisson}}}\ and\ \bibinfo {author} {\bibfnamefont {T.}~\bibnamefont
  {{Riehm}}},\ }\href {\doibase 10.1051/0004-6361:20066707} {\bibfield
  {journal} {\bibinfo  {journal} {\aap}\ }\textbf {\bibinfo {volume} {475}},\
  \bibinfo {pages} {453} (\bibinfo {year} {2007}{\natexlab{a}})},\ \Eprint
  {http://arxiv.org/abs/0709.1571} {arXiv:0709.1571} \BibitemShut {NoStop}%
\bibitem [{\citenamefont {{Rahvar}}\ \emph {et~al.}(2014)\citenamefont
  {{Rahvar}}, \citenamefont {{Baghram}},\ and\ \citenamefont
  {{Afshordi}}}]{rahvar2014}%
  \BibitemOpen
  \bibfield  {author} {\bibinfo {author} {\bibfnamefont {S.}~\bibnamefont
  {{Rahvar}}}, \bibinfo {author} {\bibfnamefont {S.}~\bibnamefont {{Baghram}}},
  \ and\ \bibinfo {author} {\bibfnamefont {N.}~\bibnamefont {{Afshordi}}},\
  }\href {\doibase 10.1103/PhysRevD.89.063001} {\bibfield  {journal} {\bibinfo
  {journal} {\prd}\ }\textbf {\bibinfo {volume} {89}},\ \bibinfo {eid} {063001}
  (\bibinfo {year} {2014})},\ \Eprint {http://arxiv.org/abs/1310.5412}
  {arXiv:1310.5412} \BibitemShut {NoStop}%
\bibitem [{\citenamefont {{Mediavilla}}\ \emph {et~al.}(2017)\citenamefont
  {{Mediavilla}}, \citenamefont {{Jim{\'e}nez-Vicente}}, \citenamefont
  {{Mu{\~n}oz}}, \citenamefont {{Vives-Arias}},\ and\ \citenamefont
  {{Calder{\'o}n-Infante}}}]{pbh2017}%
  \BibitemOpen
  \bibfield  {author} {\bibinfo {author} {\bibfnamefont {E.}~\bibnamefont
  {{Mediavilla}}}, \bibinfo {author} {\bibfnamefont {J.}~\bibnamefont
  {{Jim{\'e}nez-Vicente}}}, \bibinfo {author} {\bibfnamefont {J.~A.}\
  \bibnamefont {{Mu{\~n}oz}}}, \bibinfo {author} {\bibfnamefont
  {H.}~\bibnamefont {{Vives-Arias}}}, \ and\ \bibinfo {author} {\bibfnamefont
  {J.}~\bibnamefont {{Calder{\'o}n-Infante}}},\ }\href {\doibase
  10.3847/2041-8213/aa5dab} {\bibfield  {journal} {\bibinfo  {journal} {\apjl}\
  }\textbf {\bibinfo {volume} {836}},\ \bibinfo {eid} {L18} (\bibinfo {year}
  {2017})},\ \Eprint {http://arxiv.org/abs/1702.00947} {arXiv:1702.00947}
  \BibitemShut {NoStop}%
\bibitem [{\citenamefont {{Mediavilla}}\ \emph {et~al.}(2009)\citenamefont
  {{Mediavilla}}, \citenamefont {{Mu{\~n}oz}}, \citenamefont {{Falco}},
  \citenamefont {{Motta}}, \citenamefont {{Guerras}}, \citenamefont
  {{Canovas}}, \citenamefont {{Jean}}, \citenamefont {{Oscoz}},\ and\
  \citenamefont {{Mosquera}}}]{meld}%
  \BibitemOpen
  \bibfield  {author} {\bibinfo {author} {\bibfnamefont {E.}~\bibnamefont
  {{Mediavilla}}}, \bibinfo {author} {\bibfnamefont {J.~A.}\ \bibnamefont
  {{Mu{\~n}oz}}}, \bibinfo {author} {\bibfnamefont {E.}~\bibnamefont
  {{Falco}}}, \bibinfo {author} {\bibfnamefont {V.}~\bibnamefont {{Motta}}},
  \bibinfo {author} {\bibfnamefont {E.}~\bibnamefont {{Guerras}}}, \bibinfo
  {author} {\bibfnamefont {H.}~\bibnamefont {{Canovas}}}, \bibinfo {author}
  {\bibfnamefont {C.}~\bibnamefont {{Jean}}}, \bibinfo {author} {\bibfnamefont
  {A.}~\bibnamefont {{Oscoz}}}, \ and\ \bibinfo {author} {\bibfnamefont
  {A.~M.}\ \bibnamefont {{Mosquera}}},\ }\href {\doibase
  10.1088/0004-637X/706/2/1451} {\bibfield  {journal} {\bibinfo  {journal}
  {\apj}\ }\textbf {\bibinfo {volume} {706}},\ \bibinfo {pages} {1451}
  (\bibinfo {year} {2009})},\ \Eprint {http://arxiv.org/abs/0910.3645}
  {arXiv:0910.3645} \BibitemShut {NoStop}%
\bibitem [{\citenamefont {{Giannini}}\ \emph {et~al.}(2017)\citenamefont
  {{Giannini}}, \citenamefont {{Schmidt}}, \citenamefont {{Wambsganss}},
  \citenamefont {{Alsubai}}, \citenamefont {{Andersen}}, \citenamefont
  {{Anguita}}, \citenamefont {{Bozza}}, \citenamefont {{Bramich}},
  \citenamefont {{Browne}}, \citenamefont {{Calchi Novati}}, \citenamefont
  {{Damerdji}}, \citenamefont {{Diehl}}, \citenamefont {{Dodds}}, \citenamefont
  {{Dominik}}, \citenamefont {{Elyiv}}, \citenamefont {{Fang}}, \citenamefont
  {{Figuera Jaimes}}, \citenamefont {{Finet}}, \citenamefont {{Gerner}},
  \citenamefont {{Gu}}, \citenamefont {{Hardis}}, \citenamefont {{Harps{\o}e}},
  \citenamefont {{Hinse}}, \citenamefont {{Hornstrup}}, \citenamefont
  {{Hundertmark}}, \citenamefont {{Jessen-Hansen}}, \citenamefont
  {{J{\o}rgensen}}, \citenamefont {{Juncher}}, \citenamefont {{Kains}},
  \citenamefont {{Kerins}}, \citenamefont {{Korhonen}}, \citenamefont
  {{Liebig}}, \citenamefont {{Lund}}, \citenamefont {{Lundkvist}},
  \citenamefont {{Maier}}, \citenamefont {{Mancini}}, \citenamefont {{Masi}},
  \citenamefont {{Mathiasen}}, \citenamefont {{Penny}}, \citenamefont
  {{Proft}}, \citenamefont {{Rabus}}, \citenamefont {{Rahvar}}, \citenamefont
  {{Ricci}}, \citenamefont {{Scarpetta}}, \citenamefont {{Sahu}}, \citenamefont
  {{Sch{\"a}fer}}, \citenamefont {{Sch{\"o}nebeck}}, \citenamefont
  {{Skottfelt}}, \citenamefont {{Snodgrass}}, \citenamefont {{Southworth}},
  \citenamefont {{Surdej}}, \citenamefont {{Tregloan-Reed}}, \citenamefont
  {{Vilela}}, \citenamefont {{Wertz}},\ and\ \citenamefont
  {{Zimmer}}}]{Giannini}%
  \BibitemOpen
  \bibfield  {author} {\bibinfo {author} {\bibfnamefont {E.}~\bibnamefont
  {{Giannini}}}, \bibinfo {author} {\bibfnamefont {R.~W.}\ \bibnamefont
  {{Schmidt}}}, \bibinfo {author} {\bibfnamefont {J.}~\bibnamefont
  {{Wambsganss}}}, \bibinfo {author} {\bibfnamefont {K.}~\bibnamefont
  {{Alsubai}}}, \bibinfo {author} {\bibfnamefont {J.~M.}\ \bibnamefont
  {{Andersen}}}, \bibinfo {author} {\bibfnamefont {T.}~\bibnamefont
  {{Anguita}}}, \bibinfo {author} {\bibfnamefont {V.}~\bibnamefont {{Bozza}}},
  \bibinfo {author} {\bibfnamefont {D.~M.}\ \bibnamefont {{Bramich}}}, \bibinfo
  {author} {\bibfnamefont {P.}~\bibnamefont {{Browne}}}, \bibinfo {author}
  {\bibfnamefont {S.}~\bibnamefont {{Calchi Novati}}}, \bibinfo {author}
  {\bibfnamefont {Y.}~\bibnamefont {{Damerdji}}}, \bibinfo {author}
  {\bibfnamefont {C.}~\bibnamefont {{Diehl}}}, \bibinfo {author} {\bibfnamefont
  {P.}~\bibnamefont {{Dodds}}}, \bibinfo {author} {\bibfnamefont
  {M.}~\bibnamefont {{Dominik}}}, \bibinfo {author} {\bibfnamefont
  {A.}~\bibnamefont {{Elyiv}}}, \bibinfo {author} {\bibfnamefont
  {X.}~\bibnamefont {{Fang}}}, \bibinfo {author} {\bibfnamefont
  {R.}~\bibnamefont {{Figuera Jaimes}}}, \bibinfo {author} {\bibfnamefont
  {F.}~\bibnamefont {{Finet}}}, \bibinfo {author} {\bibfnamefont
  {T.}~\bibnamefont {{Gerner}}}, \bibinfo {author} {\bibfnamefont
  {S.}~\bibnamefont {{Gu}}}, \bibinfo {author} {\bibfnamefont {S.}~\bibnamefont
  {{Hardis}}}, \bibinfo {author} {\bibfnamefont {K.}~\bibnamefont
  {{Harps{\o}e}}}, \bibinfo {author} {\bibfnamefont {T.~C.}\ \bibnamefont
  {{Hinse}}}, \bibinfo {author} {\bibfnamefont {A.}~\bibnamefont
  {{Hornstrup}}}, \bibinfo {author} {\bibfnamefont {M.}~\bibnamefont
  {{Hundertmark}}}, \bibinfo {author} {\bibfnamefont {J.}~\bibnamefont
  {{Jessen-Hansen}}}, \bibinfo {author} {\bibfnamefont {U.~G.}\ \bibnamefont
  {{J{\o}rgensen}}}, \bibinfo {author} {\bibfnamefont {D.}~\bibnamefont
  {{Juncher}}}, \bibinfo {author} {\bibfnamefont {N.}~\bibnamefont {{Kains}}},
  \bibinfo {author} {\bibfnamefont {E.}~\bibnamefont {{Kerins}}}, \bibinfo
  {author} {\bibfnamefont {H.}~\bibnamefont {{Korhonen}}}, \bibinfo {author}
  {\bibfnamefont {C.}~\bibnamefont {{Liebig}}}, \bibinfo {author}
  {\bibfnamefont {M.~N.}\ \bibnamefont {{Lund}}}, \bibinfo {author}
  {\bibfnamefont {M.~S.}\ \bibnamefont {{Lundkvist}}}, \bibinfo {author}
  {\bibfnamefont {G.}~\bibnamefont {{Maier}}}, \bibinfo {author} {\bibfnamefont
  {L.}~\bibnamefont {{Mancini}}}, \bibinfo {author} {\bibfnamefont
  {G.}~\bibnamefont {{Masi}}}, \bibinfo {author} {\bibfnamefont
  {M.}~\bibnamefont {{Mathiasen}}}, \bibinfo {author} {\bibfnamefont
  {M.}~\bibnamefont {{Penny}}}, \bibinfo {author} {\bibfnamefont
  {S.}~\bibnamefont {{Proft}}}, \bibinfo {author} {\bibfnamefont
  {M.}~\bibnamefont {{Rabus}}}, \bibinfo {author} {\bibfnamefont
  {S.}~\bibnamefont {{Rahvar}}}, \bibinfo {author} {\bibfnamefont
  {D.}~\bibnamefont {{Ricci}}}, \bibinfo {author} {\bibfnamefont
  {G.}~\bibnamefont {{Scarpetta}}}, \bibinfo {author} {\bibfnamefont
  {K.}~\bibnamefont {{Sahu}}}, \bibinfo {author} {\bibfnamefont
  {S.}~\bibnamefont {{Sch{\"a}fer}}}, \bibinfo {author} {\bibfnamefont
  {F.}~\bibnamefont {{Sch{\"o}nebeck}}}, \bibinfo {author} {\bibfnamefont
  {J.}~\bibnamefont {{Skottfelt}}}, \bibinfo {author} {\bibfnamefont
  {C.}~\bibnamefont {{Snodgrass}}}, \bibinfo {author} {\bibfnamefont
  {J.}~\bibnamefont {{Southworth}}}, \bibinfo {author} {\bibfnamefont
  {J.}~\bibnamefont {{Surdej}}}, \bibinfo {author} {\bibfnamefont
  {J.}~\bibnamefont {{Tregloan-Reed}}}, \bibinfo {author} {\bibfnamefont
  {C.}~\bibnamefont {{Vilela}}}, \bibinfo {author} {\bibfnamefont
  {O.}~\bibnamefont {{Wertz}}}, \ and\ \bibinfo {author} {\bibfnamefont
  {F.}~\bibnamefont {{Zimmer}}},\ }\href {\doibase 10.1051/0004-6361/201527422}
  {\bibfield  {journal} {\bibinfo  {journal} {\aap}\ }\textbf {\bibinfo
  {volume} {597}},\ \bibinfo {eid} {A49} (\bibinfo {year} {2017})},\ \Eprint
  {http://arxiv.org/abs/1610.03732} {arXiv:1610.03732} \BibitemShut {NoStop}%
\bibitem [{\citenamefont {{Kimball}}\ \emph {et~al.}(2015)\citenamefont
  {{Kimball}}, \citenamefont {{Lacy}}, \citenamefont {{Lonsdale}},\ and\
  \citenamefont {{Macquart}}}]{2015MNRAS.452...88K}%
  \BibitemOpen
  \bibfield  {author} {\bibinfo {author} {\bibfnamefont {A.~E.}\ \bibnamefont
  {{Kimball}}}, \bibinfo {author} {\bibfnamefont {M.}~\bibnamefont {{Lacy}}},
  \bibinfo {author} {\bibfnamefont {C.~J.}\ \bibnamefont {{Lonsdale}}}, \ and\
  \bibinfo {author} {\bibfnamefont {J.-P.}\ \bibnamefont {{Macquart}}},\ }\href
  {\doibase 10.1093/mnras/stv1160} {\bibfield  {journal} {\bibinfo  {journal}
  {\mnras}\ }\textbf {\bibinfo {volume} {452}},\ \bibinfo {pages} {88}
  (\bibinfo {year} {2015})},\ \Eprint {http://arxiv.org/abs/1505.05262}
  {arXiv:1505.05262} \BibitemShut {NoStop}%
\bibitem [{\citenamefont {Venemans}\ \emph {et~al.}(2016)\citenamefont
  {Venemans}, \citenamefont {Walter}, \citenamefont {Zschaechner},
  \citenamefont {Decarli}, \citenamefont {De~Rosa}, \citenamefont {Findlay},
  \citenamefont {McMahon},\ and\ \citenamefont
  {Sutherland}}]{Venemans:2015hyr}%
  \BibitemOpen
  \bibfield  {author} {\bibinfo {author} {\bibfnamefont {B.~P.}\ \bibnamefont
  {Venemans}}, \bibinfo {author} {\bibfnamefont {F.}~\bibnamefont {Walter}},
  \bibinfo {author} {\bibfnamefont {L.}~\bibnamefont {Zschaechner}}, \bibinfo
  {author} {\bibfnamefont {R.}~\bibnamefont {Decarli}}, \bibinfo {author}
  {\bibfnamefont {G.}~\bibnamefont {De~Rosa}}, \bibinfo {author} {\bibfnamefont
  {J.~R.}\ \bibnamefont {Findlay}}, \bibinfo {author} {\bibfnamefont {R.~G.}\
  \bibnamefont {McMahon}}, \ and\ \bibinfo {author} {\bibfnamefont {W.~J.}\
  \bibnamefont {Sutherland}},\ }\href {\doibase 10.3847/0004-637X/816/1/37}
  {\bibfield  {journal} {\bibinfo  {journal} {Astrophys. J.}\ }\textbf
  {\bibinfo {volume} {816}},\ \bibinfo {pages} {37} (\bibinfo {year} {2016})},\
  \Eprint {http://arxiv.org/abs/1511.07432} {arXiv:1511.07432 [astro-ph.GA]}
  \BibitemShut {NoStop}%
%%CITATION = ARXIV:1511.07432;%%
\bibitem [{\citenamefont {{Nagai}}\ \emph {et~al.}(2016)\citenamefont
  {{Nagai}}, \citenamefont {{Nakanishi}}, \citenamefont {{Paladino}},
  \citenamefont {{Hull}}, \citenamefont {{Cortes}}, \citenamefont
  {{Moellenbrock}}, \citenamefont {{Fomalont}}, \citenamefont {{Asada}},\ and\
  \citenamefont {{Hada}}}]{2016ApJ...824..132N}%
  \BibitemOpen
  \bibfield  {author} {\bibinfo {author} {\bibfnamefont {H.}~\bibnamefont
  {{Nagai}}}, \bibinfo {author} {\bibfnamefont {K.}~\bibnamefont
  {{Nakanishi}}}, \bibinfo {author} {\bibfnamefont {R.}~\bibnamefont
  {{Paladino}}}, \bibinfo {author} {\bibfnamefont {C.~L.~H.}\ \bibnamefont
  {{Hull}}}, \bibinfo {author} {\bibfnamefont {P.}~\bibnamefont {{Cortes}}},
  \bibinfo {author} {\bibfnamefont {G.}~\bibnamefont {{Moellenbrock}}},
  \bibinfo {author} {\bibfnamefont {E.}~\bibnamefont {{Fomalont}}}, \bibinfo
  {author} {\bibfnamefont {K.}~\bibnamefont {{Asada}}}, \ and\ \bibinfo
  {author} {\bibfnamefont {K.}~\bibnamefont {{Hada}}},\ }\href {\doibase
  10.3847/0004-637X/824/2/132} {\bibfield  {journal} {\bibinfo  {journal}
  {\apj}\ }\textbf {\bibinfo {volume} {824}},\ \bibinfo {eid} {132} (\bibinfo
  {year} {2016})},\ \Eprint {http://arxiv.org/abs/1605.00051}
  {arXiv:1605.00051} \BibitemShut {NoStop}%
\bibitem [{\citenamefont {Ohanian}(1983)}]{ohanian:1983}%
  \BibitemOpen
  \bibfield  {author} {\bibinfo {author} {\bibfnamefont {H.~C.}\ \bibnamefont
  {Ohanian}},\ }\href@noop {} {\bibfield  {journal} {\bibinfo  {journal} {ApJ}\
  }\textbf {\bibinfo {volume} {271}},\ \bibinfo {pages} {551} (\bibinfo {year}
  {1983})}\BibitemShut {NoStop}%
%%CITATION = ARXIV:0910.3922;%%
\bibitem [{\citenamefont {Jaroszynski}\ and\ \citenamefont
  {Paczynski}(1995)}]{Jaroszynski:1995cd}%
  \BibitemOpen
  \bibfield  {author} {\bibinfo {author} {\bibfnamefont {M.}~\bibnamefont
  {Jaroszynski}}\ and\ \bibinfo {author} {\bibfnamefont {B.}~\bibnamefont
  {Paczynski}},\ }\href {\doibase 10.1086/176593} {\bibfield  {journal}
  {\bibinfo  {journal} {Astrophys. J.}\ }\textbf {\bibinfo {volume} {455}},\
  \bibinfo {pages} {443} (\bibinfo {year} {1995})},\ \Eprint
  {http://arxiv.org/abs/astro-ph/9503043} {arXiv:astro-ph/9503043 [astro-ph]}
  \BibitemShut {NoStop}%
%%CITATION = ASTRO-PH/9503043;%%
\bibitem [{\citenamefont {Heyl}(2010)}]{Heyl:2009av}%
  \BibitemOpen
  \bibfield  {author} {\bibinfo {author} {\bibfnamefont {J.}~\bibnamefont
  {Heyl}},\ }\href {\doibase 10.1111/j.1745-3933.2009.00795.x} {\bibfield
  {journal} {\bibinfo  {journal} {Mon.Not.Roy.Astron.Soc.}\ }\textbf {\bibinfo
  {volume} {402}},\ \bibinfo {pages} {39} (\bibinfo {year} {2010})},\ \Eprint
  {http://arxiv.org/abs/0910.3922} {arXiv:0910.3922 [astro-ph.EP]} \BibitemShut
  {NoStop}%
%%CITATION = ARXIV:0910.3922;%%
\bibitem [{\citenamefont {Heyl}(2011{\natexlab{a}})}]{Heyl:2011b}%
  \BibitemOpen
  \bibfield  {author} {\bibinfo {author} {\bibfnamefont {J.}~\bibnamefont
  {Heyl}},\ }\href {\doibase doi:10.1111/j.1365-2966.2010.17806.x} {\bibfield
  {journal} {\bibinfo  {journal} {Mon.Not.Roy.Astron.Soc.}\ }\textbf {\bibinfo
  {volume} {411}},\ \bibinfo {pages} {1780} (\bibinfo {year}
  {2011}{\natexlab{a}})}\BibitemShut {NoStop}%
%%CITATION = ARXIV:0910.3922;%%
\bibitem [{\citenamefont {Heyl}(2011{\natexlab{b}})}]{Heyl:2010hm}%
  \BibitemOpen
  \bibfield  {author} {\bibinfo {author} {\bibfnamefont {J.~S.}\ \bibnamefont
  {Heyl}},\ }\href {\doibase 10.1111/j.1365-2966.2010.17814.x} {\bibfield
  {journal} {\bibinfo  {journal} {Mon.Not.Roy.Astron.Soc.}\ }\textbf {\bibinfo
  {volume} {411}},\ \bibinfo {pages} {1787} (\bibinfo {year}
  {2011}{\natexlab{b}})},\ \Eprint {http://arxiv.org/abs/1003.0250}
  {arXiv:1003.0250 [astro-ph.GA]} \BibitemShut {NoStop}%
%%CITATION = ARXIV:1003.0250;%%
\bibitem [{\citenamefont {Mehrabi}\ and\ \citenamefont
  {Rahvar}(2013)}]{Mehrabi:2012dy}%
  \BibitemOpen
  \bibfield  {author} {\bibinfo {author} {\bibfnamefont {A.}~\bibnamefont
  {Mehrabi}}\ and\ \bibinfo {author} {\bibfnamefont {S.}~\bibnamefont
  {Rahvar}},\ }\href {\doibase 10.1093/mnras/stt243} {\bibfield  {journal}
  {\bibinfo  {journal} {Mon.Not.Roy.Astron.Soc.}\ }\textbf {\bibinfo {volume}
  {431}},\ \bibinfo {pages} {1264} (\bibinfo {year} {2013})},\ \Eprint
  {http://arxiv.org/abs/1207.4034} {arXiv:1207.4034 [astro-ph.EP]} \BibitemShut
  {NoStop}%
%%CITATION = ARXIV:1207.4034;%%
\bibitem [{\citenamefont {Schneider}(1985)}]{Schneider1985}%
  \BibitemOpen
  \bibfield  {author} {\bibinfo {author} {\bibfnamefont {P.}~\bibnamefont
  {Schneider}},\ }\href@noop {} {\bibfield  {journal} {\bibinfo  {journal}
  {A\&A}\ }\textbf {\bibinfo {volume} {143}},\ \bibinfo {pages} {413} (\bibinfo
  {year} {1985})}\BibitemShut {NoStop}%
%%CITATION = ARXIV:0901.0115;%%
\bibitem [{\citenamefont {Schneider}\ \emph {et~al.}(1992)\citenamefont
  {Schneider}, \citenamefont {Ehlers},\ and\ \citenamefont
  {Falco}}]{Falco:1992}%
  \BibitemOpen
  \bibfield  {author} {\bibinfo {author} {\bibfnamefont {P.}~\bibnamefont
  {Schneider}}, \bibinfo {author} {\bibfnamefont {J.}~\bibnamefont {Ehlers}}, \
  and\ \bibinfo {author} {\bibfnamefont {E.~E.}\ \bibnamefont {Falco}},\
  }\href@noop {} {\emph {\bibinfo {title} {Gravitational Lenses}}}\ (\bibinfo
  {publisher} {Springer-Verlag},\ \bibinfo {address} {Berlin},\ \bibinfo {year}
  {1992})\BibitemShut {NoStop}%
\bibitem [{\citenamefont {{Peth}}\ \emph {et~al.}(2011)\citenamefont {{Peth}},
  \citenamefont {{Ross}},\ and\ \citenamefont
  {{Schneider}}}]{2011AJ....141..105P}%
  \BibitemOpen
  \bibfield  {author} {\bibinfo {author} {\bibfnamefont {M.~A.}\ \bibnamefont
  {{Peth}}}, \bibinfo {author} {\bibfnamefont {N.~P.}\ \bibnamefont {{Ross}}},
  \ and\ \bibinfo {author} {\bibfnamefont {D.~P.}\ \bibnamefont
  {{Schneider}}},\ }\href {\doibase 10.1088/0004-6256/141/4/105} {\bibfield
  {journal} {\bibinfo  {journal} {\aj}\ }\textbf {\bibinfo {volume} {141}},\
  \bibinfo {eid} {105} (\bibinfo {year} {2011})},\ \Eprint
  {http://arxiv.org/abs/1012.4187} {arXiv:1012.4187 [astro-ph.CO]} \BibitemShut
  {NoStop}%
\bibitem [{\citenamefont {Pâris}\ \emph {et~al.}(2017)\citenamefont {Pâris}
  \emph {et~al.}}]{Paris:2016xdm}%
  \BibitemOpen
  \bibfield  {author} {\bibinfo {author} {\bibfnamefont {I.}~\bibnamefont
  {Pâris}} \emph {et~al.},\ }\href {\doibase 10.1051/0004-6361/201527999}
  {\bibfield  {journal} {\bibinfo  {journal} {Astron. Astrophys.}\ }\textbf
  {\bibinfo {volume} {597}},\ \bibinfo {pages} {A79} (\bibinfo {year}
  {2017})},\ \Eprint {http://arxiv.org/abs/1608.06483} {arXiv:1608.06483
  [astro-ph.GA]} \BibitemShut {NoStop}%
%%CITATION = ARXIV:1608.06483;%%
\bibitem [{\citenamefont {{Sedgwick}}\ \emph {et~al.}(2017)\citenamefont
  {{Sedgwick}}, \citenamefont {{Serjeant}}, \citenamefont {{Pearson}},
  \citenamefont {{Yamamura}}, \citenamefont {{Makiuti}}, \citenamefont
  {{Ikeda}}, \citenamefont {{Fukuda}}, \citenamefont {{Oyabu}}, \citenamefont
  {{Koga}}, \citenamefont {{Amber}},\ and\ \citenamefont
  {{White}}}]{2017PKAS...32..305S}%
  \BibitemOpen
  \bibfield  {author} {\bibinfo {author} {\bibfnamefont {C.}~\bibnamefont
  {{Sedgwick}}}, \bibinfo {author} {\bibfnamefont {S.}~\bibnamefont
  {{Serjeant}}}, \bibinfo {author} {\bibfnamefont {C.}~\bibnamefont
  {{Pearson}}}, \bibinfo {author} {\bibfnamefont {I.}~\bibnamefont
  {{Yamamura}}}, \bibinfo {author} {\bibfnamefont {S.}~\bibnamefont
  {{Makiuti}}}, \bibinfo {author} {\bibfnamefont {N.}~\bibnamefont {{Ikeda}}},
  \bibinfo {author} {\bibfnamefont {Y.}~\bibnamefont {{Fukuda}}}, \bibinfo
  {author} {\bibfnamefont {S.}~\bibnamefont {{Oyabu}}}, \bibinfo {author}
  {\bibfnamefont {T.}~\bibnamefont {{Koga}}}, \bibinfo {author} {\bibfnamefont
  {S.}~\bibnamefont {{Amber}}}, \ and\ \bibinfo {author} {\bibfnamefont
  {G.~J.}\ \bibnamefont {{White}}},\ }\href {\doibase
  10.5303/PKAS.2017.32.1.305} {\bibfield  {journal} {\bibinfo  {journal}
  {Publication of Korean Astronomical Society}\ }\textbf {\bibinfo {volume}
  {32}},\ \bibinfo {pages} {305} (\bibinfo {year} {2017})}\BibitemShut
  {NoStop}%
\bibitem [{\citenamefont {Whiting}\ \emph {et~al.}(2002)\citenamefont
  {Whiting}, \citenamefont {Oshlack},\ and\ \citenamefont
  {Webster}}]{Whiting:2001vj}%
  \BibitemOpen
  \bibfield  {author} {\bibinfo {author} {\bibfnamefont {M.}~\bibnamefont
  {Whiting}}, \bibinfo {author} {\bibfnamefont {A.}~\bibnamefont {Oshlack}}, \
  and\ \bibinfo {author} {\bibfnamefont {R.~L.}\ \bibnamefont {Webster}},\
  }\href {\doibase 10.1071/AS01083} {\bibfield  {journal} {\bibinfo  {journal}
  {Publ. Astron. Soc. Austral.}\ }\textbf {\bibinfo {volume} {19}},\ \bibinfo
  {pages} {222} (\bibinfo {year} {2002})},\ \Eprint
  {http://arxiv.org/abs/astro-ph/0111171} {arXiv:astro-ph/0111171 [astro-ph]}
  \BibitemShut {NoStop}%
%%CITATION = ASTRO-PH/0111171;%%
\bibitem [{\citenamefont {{Ricci}}\ \emph {et~al.}(2011)\citenamefont
  {{Ricci}}, \citenamefont {{Poels}}, \citenamefont {{Elyiv}}, \citenamefont
  {{Finet}}, \citenamefont {{Sprimont}}, \citenamefont {{Anguita}},
  \citenamefont {{Bozza}}, \citenamefont {{Browne}}, \citenamefont
  {{Burgdorf}}, \citenamefont {{Calchi Novati}}, \citenamefont {{Dominik}},
  \citenamefont {{Dreizler}}, \citenamefont {{Glitrup}}, \citenamefont
  {{Grundahl}}, \citenamefont {{Harps{\o}e}}, \citenamefont {{Hessman}},
  \citenamefont {{Hinse}}, \citenamefont {{Hornstrup}}, \citenamefont
  {{Hundertmark}}, \citenamefont {{J{\o}rgensen}}, \citenamefont {{Liebig}},
  \citenamefont {{Maier}}, \citenamefont {{Mancini}}, \citenamefont {{Masi}},
  \citenamefont {{Mathiasen}}, \citenamefont {{Rahvar}}, \citenamefont
  {{Scarpetta}}, \citenamefont {{Skottfelt}}, \citenamefont {{Snodgrass}},
  \citenamefont {{Southworth}}, \citenamefont {{Teuber}}, \citenamefont
  {{Th{\"o}ne}}, \citenamefont {{Wambsgan{\ss}}}, \citenamefont {{Zimmer}},
  \citenamefont {{Zub}},\ and\ \citenamefont {{Surdej}}}]{Ricci2011}%
  \BibitemOpen
  \bibfield  {author} {\bibinfo {author} {\bibfnamefont {D.}~\bibnamefont
  {{Ricci}}}, \bibinfo {author} {\bibfnamefont {J.}~\bibnamefont {{Poels}}},
  \bibinfo {author} {\bibfnamefont {A.}~\bibnamefont {{Elyiv}}}, \bibinfo
  {author} {\bibfnamefont {F.}~\bibnamefont {{Finet}}}, \bibinfo {author}
  {\bibfnamefont {P.~G.}\ \bibnamefont {{Sprimont}}}, \bibinfo {author}
  {\bibfnamefont {T.}~\bibnamefont {{Anguita}}}, \bibinfo {author}
  {\bibfnamefont {V.}~\bibnamefont {{Bozza}}}, \bibinfo {author} {\bibfnamefont
  {P.}~\bibnamefont {{Browne}}}, \bibinfo {author} {\bibfnamefont
  {M.}~\bibnamefont {{Burgdorf}}}, \bibinfo {author} {\bibfnamefont
  {S.}~\bibnamefont {{Calchi Novati}}}, \bibinfo {author} {\bibfnamefont
  {M.}~\bibnamefont {{Dominik}}}, \bibinfo {author} {\bibfnamefont
  {S.}~\bibnamefont {{Dreizler}}}, \bibinfo {author} {\bibfnamefont
  {M.}~\bibnamefont {{Glitrup}}}, \bibinfo {author} {\bibfnamefont
  {F.}~\bibnamefont {{Grundahl}}}, \bibinfo {author} {\bibfnamefont
  {K.}~\bibnamefont {{Harps{\o}e}}}, \bibinfo {author} {\bibfnamefont
  {F.}~\bibnamefont {{Hessman}}}, \bibinfo {author} {\bibfnamefont {T.~C.}\
  \bibnamefont {{Hinse}}}, \bibinfo {author} {\bibfnamefont {A.}~\bibnamefont
  {{Hornstrup}}}, \bibinfo {author} {\bibfnamefont {M.}~\bibnamefont
  {{Hundertmark}}}, \bibinfo {author} {\bibfnamefont {U.~G.}\ \bibnamefont
  {{J{\o}rgensen}}}, \bibinfo {author} {\bibfnamefont {C.}~\bibnamefont
  {{Liebig}}}, \bibinfo {author} {\bibfnamefont {G.}~\bibnamefont {{Maier}}},
  \bibinfo {author} {\bibfnamefont {L.}~\bibnamefont {{Mancini}}}, \bibinfo
  {author} {\bibfnamefont {G.}~\bibnamefont {{Masi}}}, \bibinfo {author}
  {\bibfnamefont {M.}~\bibnamefont {{Mathiasen}}}, \bibinfo {author}
  {\bibfnamefont {S.}~\bibnamefont {{Rahvar}}}, \bibinfo {author}
  {\bibfnamefont {G.}~\bibnamefont {{Scarpetta}}}, \bibinfo {author}
  {\bibfnamefont {J.}~\bibnamefont {{Skottfelt}}}, \bibinfo {author}
  {\bibfnamefont {C.}~\bibnamefont {{Snodgrass}}}, \bibinfo {author}
  {\bibfnamefont {J.}~\bibnamefont {{Southworth}}}, \bibinfo {author}
  {\bibfnamefont {J.}~\bibnamefont {{Teuber}}}, \bibinfo {author}
  {\bibfnamefont {C.~C.}\ \bibnamefont {{Th{\"o}ne}}}, \bibinfo {author}
  {\bibfnamefont {J.}~\bibnamefont {{Wambsgan{\ss}}}}, \bibinfo {author}
  {\bibfnamefont {F.}~\bibnamefont {{Zimmer}}}, \bibinfo {author}
  {\bibfnamefont {M.}~\bibnamefont {{Zub}}}, \ and\ \bibinfo {author}
  {\bibfnamefont {J.}~\bibnamefont {{Surdej}}},\ }\href {\doibase
  10.1051/0004-6361/201016188} {\bibfield  {journal} {\bibinfo  {journal}
  {\aap}\ }\textbf {\bibinfo {volume} {528}},\ \bibinfo {eid} {A42} (\bibinfo
  {year} {2011})},\ \Eprint {http://arxiv.org/abs/1101.3664} {arXiv:1101.3664}
  \BibitemShut {NoStop}%
\bibitem [{\citenamefont {{Ricci}}\ \emph {et~al.}(2013)\citenamefont
  {{Ricci}}, \citenamefont {{Elyiv}}, \citenamefont {{Finet}}, \citenamefont
  {{Wertz}}, \citenamefont {{Alsubai}}, \citenamefont {{Anguita}},
  \citenamefont {{Bozza}}, \citenamefont {{Browne}}, \citenamefont
  {{Burgdorf}}, \citenamefont {{Calchi Novati}}, \citenamefont {{Dodds}},
  \citenamefont {{Dominik}}, \citenamefont {{Dreizler}}, \citenamefont
  {{Gerner}}, \citenamefont {{Glitrup}}, \citenamefont {{Grundahl}},
  \citenamefont {{Hardis}}, \citenamefont {{Harps{\o}e}}, \citenamefont
  {{Hinse}}, \citenamefont {{Hornstrup}}, \citenamefont {{Hundertmark}},
  \citenamefont {{J{\o}rgensen}}, \citenamefont {{Kains}}, \citenamefont
  {{Kerins}}, \citenamefont {{Liebig}}, \citenamefont {{Maier}}, \citenamefont
  {{Mancini}}, \citenamefont {{Masi}}, \citenamefont {{Mathiasen}},
  \citenamefont {{Penny}}, \citenamefont {{Proft}}, \citenamefont {{Rahvar}},
  \citenamefont {{Scarpetta}}, \citenamefont {{Sahu}}, \citenamefont
  {{Sch{\"a}fer}}, \citenamefont {{Sch{\"o}nebeck}}, \citenamefont {{Schmidt}},
  \citenamefont {{Skottfelt}}, \citenamefont {{Snodgrass}}, \citenamefont
  {{Southworth}}, \citenamefont {{Th{\"o}ne}}, \citenamefont {{Wambsganss}},
  \citenamefont {{Zimmer}}, \citenamefont {{Zub}},\ and\ \citenamefont
  {{Surdej}}}]{Ricci}%
  \BibitemOpen
  \bibfield  {author} {\bibinfo {author} {\bibfnamefont {D.}~\bibnamefont
  {{Ricci}}}, \bibinfo {author} {\bibfnamefont {A.}~\bibnamefont {{Elyiv}}},
  \bibinfo {author} {\bibfnamefont {F.}~\bibnamefont {{Finet}}}, \bibinfo
  {author} {\bibfnamefont {O.}~\bibnamefont {{Wertz}}}, \bibinfo {author}
  {\bibfnamefont {K.}~\bibnamefont {{Alsubai}}}, \bibinfo {author}
  {\bibfnamefont {T.}~\bibnamefont {{Anguita}}}, \bibinfo {author}
  {\bibfnamefont {V.}~\bibnamefont {{Bozza}}}, \bibinfo {author} {\bibfnamefont
  {P.}~\bibnamefont {{Browne}}}, \bibinfo {author} {\bibfnamefont
  {M.}~\bibnamefont {{Burgdorf}}}, \bibinfo {author} {\bibfnamefont
  {S.}~\bibnamefont {{Calchi Novati}}}, \bibinfo {author} {\bibfnamefont
  {P.}~\bibnamefont {{Dodds}}}, \bibinfo {author} {\bibfnamefont
  {M.}~\bibnamefont {{Dominik}}}, \bibinfo {author} {\bibfnamefont
  {S.}~\bibnamefont {{Dreizler}}}, \bibinfo {author} {\bibfnamefont
  {T.}~\bibnamefont {{Gerner}}}, \bibinfo {author} {\bibfnamefont
  {M.}~\bibnamefont {{Glitrup}}}, \bibinfo {author} {\bibfnamefont
  {F.}~\bibnamefont {{Grundahl}}}, \bibinfo {author} {\bibfnamefont
  {S.}~\bibnamefont {{Hardis}}}, \bibinfo {author} {\bibfnamefont
  {K.}~\bibnamefont {{Harps{\o}e}}}, \bibinfo {author} {\bibfnamefont {T.~C.}\
  \bibnamefont {{Hinse}}}, \bibinfo {author} {\bibfnamefont {A.}~\bibnamefont
  {{Hornstrup}}}, \bibinfo {author} {\bibfnamefont {M.}~\bibnamefont
  {{Hundertmark}}}, \bibinfo {author} {\bibfnamefont {U.~G.}\ \bibnamefont
  {{J{\o}rgensen}}}, \bibinfo {author} {\bibfnamefont {N.}~\bibnamefont
  {{Kains}}}, \bibinfo {author} {\bibfnamefont {E.}~\bibnamefont {{Kerins}}},
  \bibinfo {author} {\bibfnamefont {C.}~\bibnamefont {{Liebig}}}, \bibinfo
  {author} {\bibfnamefont {G.}~\bibnamefont {{Maier}}}, \bibinfo {author}
  {\bibfnamefont {L.}~\bibnamefont {{Mancini}}}, \bibinfo {author}
  {\bibfnamefont {G.}~\bibnamefont {{Masi}}}, \bibinfo {author} {\bibfnamefont
  {M.}~\bibnamefont {{Mathiasen}}}, \bibinfo {author} {\bibfnamefont
  {M.}~\bibnamefont {{Penny}}}, \bibinfo {author} {\bibfnamefont
  {S.}~\bibnamefont {{Proft}}}, \bibinfo {author} {\bibfnamefont
  {S.}~\bibnamefont {{Rahvar}}}, \bibinfo {author} {\bibfnamefont
  {G.}~\bibnamefont {{Scarpetta}}}, \bibinfo {author} {\bibfnamefont
  {K.}~\bibnamefont {{Sahu}}}, \bibinfo {author} {\bibfnamefont
  {S.}~\bibnamefont {{Sch{\"a}fer}}}, \bibinfo {author} {\bibfnamefont
  {F.}~\bibnamefont {{Sch{\"o}nebeck}}}, \bibinfo {author} {\bibfnamefont
  {R.}~\bibnamefont {{Schmidt}}}, \bibinfo {author} {\bibfnamefont
  {J.}~\bibnamefont {{Skottfelt}}}, \bibinfo {author} {\bibfnamefont
  {C.}~\bibnamefont {{Snodgrass}}}, \bibinfo {author} {\bibfnamefont
  {J.}~\bibnamefont {{Southworth}}}, \bibinfo {author} {\bibfnamefont {C.~C.}\
  \bibnamefont {{Th{\"o}ne}}}, \bibinfo {author} {\bibfnamefont
  {J.}~\bibnamefont {{Wambsganss}}}, \bibinfo {author} {\bibfnamefont
  {F.}~\bibnamefont {{Zimmer}}}, \bibinfo {author} {\bibfnamefont
  {M.}~\bibnamefont {{Zub}}}, \ and\ \bibinfo {author} {\bibfnamefont
  {J.}~\bibnamefont {{Surdej}}},\ }\href {\doibase 10.1051/0004-6361/201118755}
  {\bibfield  {journal} {\bibinfo  {journal} {\aap}\ }\textbf {\bibinfo
  {volume} {551}},\ \bibinfo {eid} {A104} (\bibinfo {year} {2013})},\ \Eprint
  {http://arxiv.org/abs/1302.0766} {arXiv:1302.0766 [astro-ph.CO]} \BibitemShut
  {NoStop}%
\bibitem [{\citenamefont {{Zackrisson}}\ and\ \citenamefont
  {{Riehm}}(2007{\natexlab{b}})}]{Zackrisson}%
  \BibitemOpen
  \bibfield  {author} {\bibinfo {author} {\bibfnamefont {E.}~\bibnamefont
  {{Zackrisson}}}\ and\ \bibinfo {author} {\bibfnamefont {T.}~\bibnamefont
  {{Riehm}}},\ }\href {\doibase 10.1051/0004-6361:20066707} {\bibfield
  {journal} {\bibinfo  {journal} {\aap}\ }\textbf {\bibinfo {volume} {475}},\
  \bibinfo {pages} {453} (\bibinfo {year} {2007}{\natexlab{b}})},\ \Eprint
  {http://arxiv.org/abs/0709.1571} {arXiv:0709.1571} \BibitemShut {NoStop}%
\bibitem [{\citenamefont {Soldi}\ \emph {et~al.}(2008)\citenamefont {Soldi}
  \emph {et~al.}}]{Soldi:2008ev}%
  \BibitemOpen
  \bibfield  {author} {\bibinfo {author} {\bibfnamefont {S.}~\bibnamefont
  {Soldi}} \emph {et~al.},\ }\href {\doibase 10.1051/0004-6361:200809947}
  {\bibfield  {journal} {\bibinfo  {journal} {Astron. Astrophys.}\ }\textbf
  {\bibinfo {volume} {486}},\ \bibinfo {pages} {411} (\bibinfo {year}
  {2008})},\ \Eprint {http://arxiv.org/abs/0805.3411} {arXiv:0805.3411
  [astro-ph]} \BibitemShut {NoStop}%
%%CITATION = ARXIV:0805.3411;%%
\end{thebibliography}%

\label{lastpage}

\end{document}